%% file: thesis.tex
\documentclass[11pt, twoside]{book}
\usepackage{mycommand, amsmath, latexsym, amssymb, amscd, amsfonts, amsthm}
\usepackage[english]{babel}

\DeclareFixedFont{\sfracFont}{U}{euf}{b}{n}{7pt}
\newtheoremstyle{mydefi}
  {15pt}
  {15pt}
  {}
  {}
  {\bfseries}
  {:}
  {.5em}
  {}
\newtheoremstyle{mytheo}
  {15pt}
  {15pt}
  {\slshape}
  {}
  {\bfseries}
  {:}
  {.5em}
  {}

\theoremstyle{mytheo}
\newtheorem{stel}{Theorem}[section]
\newtheorem{lem}[stel]{Lemma}

\theoremstyle{mydefi}
\newtheorem{de}[stel]{Definition}

\begin{document}
\title{Filtering and Control in Quantum Optics}
\author{Luc Bouten}
\date{}
\maketitle
\thispagestyle{empty}
\cleardoublepage

\pagenumbering{roman}
\input{ack.tex}
\tableofcontents
\newpage

\pagenumbering{arabic}
\input{introduction.tex}

\input{davies.tex}

\input{ssequation.tex}
\input{sec.tex}
\input{summary.tex}

\bibliography{thesis}
\end{document}

%% file: ack.tex
{\pagestyle{empty}\headsep=80pt

\begin{center}
{\huge \textbf{Filtering and Control in Quantum Optics} }

\vspace{1.5cm}

{\large een wetenschappelijke proeve op het gebied van de \\ 
Natuurwetenschappen, Wiskunde en Informatica} 

\vspace{1.5cm}

{\large \textbf{Proefschrift}}

\vspace{1.5cm}

{\large ter verkrijging van de graad van doctor \\ 
aan de Radboud Universiteit Nijmegen\\ 
op gezag van de Rector Magnificus prof.~dr.~C.W.P.M.~Blom,\\
volgens het besluit van het College van Decanen\\
in het openbaar te verdedigen op maandag 29 november 2004\\
des morgens om 10.30 precies\\
door}

\vspace{1.5cm}

{\large \textbf{Lucas Martinus Bouten}}

\vspace{1.5cm}

{\large geboren op 11 oktober 1976\\ 
te Helden}

\end{center}

\newpage
\headsep=0 pt
{\Large\textbf{Promotor:}
Prof.~Dr.~G.J.~Heckman}\\

{\Large\textbf{Copromotor:}
Dr.~J.D.M.~Maassen\\}

{\Large \textbf{Manuscriptcommissie:}\\
\\
Prof.~Dr.~V.P.~Belavkin (University of Nottingham)\\
Prof.~Dr.~R.D.~Gill (University of Utrecht)\\
Prof.~Dr.~B.~K\"ummerer (University of Darmstadt)}

\vspace{14.2cm}

ISBN 90-9018790-1

\newpage

\section*{Acknowledgments}

The theme for this thesis, applying the tools of quantum 
probability to problems in quantum optics, was provided 
by Hans Maassen, who has been my supervisor for the past 
four years. Apart from sharing his deep insight in quantum 
mechanics, I am grateful to Hans for his patience and 
encouragement in finding my own way in research. I would 
like to extend these words of thanks to M\u{a}d\u{a}lin 
Gu\c{t}\u{a}, my colleague in Nijmegen in the first 
two years, and a good friend in close scientific contact
afterwards. I have benefited a lot from the seminar on 
operator algebras organised by our little group in Nijmegen
for the first two years. It grew to become the 
"QRandom seminar" in collaboration with 
Richard Gill's groups in Eindhoven and 
Utrecht afterwards.   

The last four years I have been in very pleasant and 
friendly contact with Gert Heckman, who has advised 
and encouraged me on numerous occasions. I am grateful to 
Gert for the many enriching discussions on mathematics
we have had over the years. Gert has also encouraged 
me to attend the seminar on noncommutative geometry
organised by Klaas Landsman and Eric Opdam in Amsterdam, 
which was a very stimulating experience.  

In the second year Hans sent me to Oregon (USA) on a 
three month visit to Howard Carmichael. This visit 
proved to be a turning point for me. Howard 
introduced me to the stochastic Schr\"odinger 
equations and he also provided me with literature 
pointing me in the direction of quantum filtering 
theory.
Apart from showing me a new direction in research, 
I am grateful to him for the many things he taught 
me about quantum optics and for his kind hospitality.  

In the last year I met Slava Belavkin who invited 
me to Nottingham. I am thankful to him for the 
many discussions we had on quantum filtering theory. 
I am also grateful to him for introducing me to 
optimal control theory in a quantum setting, 
a line of research in which I would like to 
continue in the future. I thank Slava and Martin 
Lindsay for arranging the nice time spent in 
Nottingham. 

I am indebted to Andreas Buchleitner for organising
the many workshops I attended at the Max Planck Institut 
in Dresden. I also thank him for his advice in rewriting 
the introduction to the paper that makes up 
Chapter \ref{ch sseq} of this thesis. I am thankful 
to Burkhard K\"ummerer for his inspiring lectures 
at the "Coherent Evolution in Noisy Environments" summer school 
in Dresden, and for his interest and encouragement 
in the work I have done.  

I thank Joost and Stefan, with whom I shared an 
office, and many friends for the pleasant atmosphere
of the past four years. I thank my parents 
for their support.

\cleardoublepage

}

%% file: introduction.tex
\chapter{Introduction}

Sixty years after its invention by the Japanese mathematician 
Kiyosi It\^o, stochastic analysis has found a wide range 
of applications varying from pure mathematics to physics, 
enginering, biology and economics. Stochastic differential equations
are an important tool for solving problems in pricing options at
the stock market, finding solutions to boundary value problems,
filtering signals in electrical enginering and for solving many 
other problems. In 1984 Hudson and Parthasarathy published a 
paper \cite{HuP} extending the definition of It\^o's stochastic 
integral and its subsequent calculus to the non-commutative
world of quantum mechanics. The goal of this thesis is to apply 
this generalized stochastic calculus to problems in quantum 
optics.

\section{Some spectral theory and quantum probability}\label{sec qp}

The question of the mathematical foundations of 
quantum mechanics was settled by von Neumann in 
a series of papers written between 1927 and 1932,
culminating in his book "Mathematische Grundlagen 
der Quantenmechanik" \cite{vNe}. In this series of articles 
von Neumann developed spectral theory for normal operators
on a Hilbert space, i.e.\ operators that commute 
with their adjoint. Avoiding tedious considerations
regarding domains of operators we will now first 
focus on spectral theory for bounded normal operators 
on a separable Hilbert space $\BB{H}$. 

Denote by $\BC(\BB{H})$ the algebra of all bounded operators
on $\BB{H}$ and let $\SC$ be a subset of $\BC(\BB{H})$. We call the 
set $\SC' := \{R \in \BC(\BB{H});\ SR = RS\ \forall S \in \SC\}$
the \emph{commutant} of $\SC$ in $\BC(\BB{H})$. 
A \emph{von Neumann algebra} $\AC$ on $\BB{H}$ is a $^*$-subalgebra of 
$\BC(\BB{H})$ such that $\AC$ equals its double commutant,
i.e.\ $\AC'' = \AC$. It follows from von Neumann's double commutant 
theorem, cf.\ \cite{KaR0}, that a von Neumann algebra is 
a $C^*$-subalgebra, i.e. a $^*$-subalgebra closed in the 
operator norm topology, of $\BC(\BB{H})$ that is 
closed even in the weak operator topology. 

It immediately follows from $\AC'' = \AC$ that the identity 
$\I \in \BC(\BB{H})$ is an element of the von Neumann algebra 
$\AC$. A \emph{state} on $\AC$ is a 
linear map $\rho:\ \AC \to \BB{C}$ such that $\rho$
is \emph{positive} in the sense that $\rho(A^*A) \ge 0$ for
all $A \in \AC$ and such that $\rho$ is \emph{normalised} $\rho(\I) = 1$.
A state is called \emph{normal} if it is weak operator 
continuous on the unit ball of $\AC$. The following
theorem, see \cite{KaR} sections 9.3, 9.4 and 9.5 for 
a proof, is at the heart of spectral theory. 
  \begin{stel}\label{thm spectral}
  Let $\CC$ be a commutative von Neumann algebra and $\rho$
  a normal state on $\CC$. Then there is a probability space
  $(\Omega, \Sigma, \BB{P})$ such that $\CC$ is $^*$-isomorphic
  to $L^\infty(\Omega, \Sigma, \BB{P})$, the space of all bounded
  measurable functions on $\Omega$. Furthermore, if we denote the
  $^*$-isomorphism between $\CC$ and $L^\infty(\Omega, \Sigma, \BB{P})$  
  by $C \mapsto C_\bullet$ we have
    \begin{equation*}
    \rho(C) = \int_\Omega C_\omega\BB{P}(d\omega), \ \ \ \ C \in \CC,
    \end{equation*}
  where $C_\omega$ denotes the function $C_\bullet$ evaluated at $\omega$,
  i.e.\ $C_\omega = C_\bullet(\omega)$. 
  \end{stel}    
Since a normal operator, i.e.\ an operator that commutes
with its adjoint, generates a commutative von Neumann algebra
the above theorem can be applied. The simplest example of 
a normal operator is a Hermitian operator $A$ on a finite 
dimensional Hilbert space. The above theorem then states that 
$A$ is equivalent with a function $A_\bullet$ and that the algebra generated
by $A$ is isomorphic to the algebra generated by this function $A_\bullet$. 
Indeed, after diagonalisation, $A$ is equivalent with the function 
that maps $i$, standing for diagonal entry number $i$, to its 
corresponding eigenvalue. Note that the above theorem also covers 
simultaneous diagonalisation of several commuting Hermitian 
operators.

Given a probability space $(\Omega, \Sigma, \BB{P})$, we can study 
the commutative von Neumann algebra $\BC := L^\infty(\Omega, \Sigma, \BB{P})$, 
acting on the Hilbert space $L^2(\Omega, \Sigma, \BB{P})$ by pointwise 
multiplication, equipped with the normal state $\rho$ given by 
expectation with respect to the measure $\BB{P}$. The pair $(\BC, \rho)$
faithfully encodes the probability space $(\Omega, \Sigma, \BB{P})$
\cite{Maa03}. Indeed, the $\sigma$-algebra $\Sigma$ can be reconstructed
(up to equivalence of sets with symmetric difference $0$, a point on
which we will not dwell here) as the set of projections in $\BC$, i.e.\
the set of characteristic functions of sets in $\Sigma$, 
and the probability measure is given by acting with the state 
$\rho$ on this set of projections. We conclude that studying 
commutative von Neumann algebras equipped with normal states
is the same as studying probability spaces. This motivates 
the definition of a \emph{non-commutative} or \emph{quantum}
probability space as a von Neumann algebra equipped with a 
normal state.    

From the point of view of physics it is violation of Bell's 
inequalities \cite{Bell} in for instance the Aspect experiment \cite{Asp}, 
that motivates the study of non-commutative probability. 
See for instance \cite{KuM} for a clear exposition of the point that 
this violation can not be accounted for by a local classical 
probabilistic model. The way out if one wants to preserve locality 
is to weaken Kolmogorov's axioms of probability ending up 
with a non-commutative probability theory, namely quantum mechanics. 

Let $(\BC, \rho)$ be a quantum probability space. The events are
given by the projections in $\BC$, i.e. elements satisfying
$E^2 = E = E^*$. Two events $E$ and $F$ are called \emph{compatible}
if $EF$ is again an event, which is equivalent to $EF = FE$.
The interpretation is such that only events that are compatible 
can occur simultaneously. Note that in classical probability, 
i.e.\ $\BC$ is commutative, all events are compatible. If $E$ 
and $F$ are compatible events then $EF$ stands for the occurence 
of both $E$ and $F$ and $E\vee F := E +F - EF$ stands for the 
occurence of $E$ or $F$ or both. When an event $E$ has occurred we 
have to update the state $\rho$ to $\rho_E$ with $\rho_E(A) = 
\rho(EAE)/\rho(E)$,  the state conditioned on $E$. 

In classical probability theory a random variable $X$ is 
a measurable map from a probability space $(\Omega, \Sigma, \BB{P})$
to some other measure space $(\Omega', \Sigma')$. The probability
distribution of $X$ is given by 
  \begin{equation*}
  \BB{P}_X:\ \Sigma' \to [0,1]: S \mapsto \BB{P}\big(X^{-1}(S)\big). 
  \end{equation*} 
Algebraically $X$ is completely determined by the pull back
  \begin{equation*}
  j_X:\ L^\infty(\Omega', \Sigma', \BB{P}_X) \to L^\infty(\Omega,\Sigma,\BB{P}):\
  f \mapsto f\circ X.
  \end{equation*}
This motivates the definition of a \emph{random variable} on a quantum 
probability space $(\BC, \rho)$ as a $^*$-homomorphism $j$ from some
other von Neumann algebra $\AC$ to $\BC$ mapping $\I_\AC$ to $\I_\BC$.
The probability distribution of $j$ is then given by $\sigma := \rho\circ j$
on $\AC$. A \emph{quantum stochastic process} \cite{AFL} is a family
$\{j_t\}_{t \in \BB{T}}$ of random variables indexed by time $\BB{T}$
which is a linearly ordered set such as $\BB{N}, \BB{Z}, \BB{R}$ or $\BB{R}_+$.
See \cite{Kum4} for theory on non-commutative Markov processes.

Let $j:\ \AC \to \BC$ be a random variable on $(\BC,\rho)$. If 
$\AC = L^\infty(\BB{R})$ then $j$ is called a \emph{real valued} 
random variable. The following brief exposition of spectral 
theory for unbounded selfadjoint operators shows that there 
is a one-to-one correspondence between real valued random variables
and selfadjoint operators. Given $j$ we can define a projection 
valued measure on the Borel sigma-algebra $\Sigma$ of $\BB{R}$ by
  \begin{equation*}
  E(S) := j(\chi_S), \ \ \ \ S \in \Sigma,
  \end{equation*}
where $\chi_{S}$ stands for the characteristic function of $S$, i.e.\ the
function that is $1$ on $S$ and $0$ elsewhere. We can now define 
a selfadjoint operator $A$ by
  \begin{equation}\label{eq spectral}
  A = \int_\BB{R} \lambda E(d\lambda).
  \end{equation}  
Conversely, since the spectrum of a selfadjoint 
operator $A$ is real, we can define bounded operators 
$T_+ := (A+i\I)^{-1}$ and $T_- := (A-i\I)^{-1}$. The 
operators $T_+$ and $T_-$ generate a commutative von Neumann
algebra $\AC$. From Theorem \ref{thm spectral} it follows 
that there exist a measurable space 
$(\Omega, \Sigma')$ and a $^*$-isomorphism 
$i:\ L^\infty(\Omega,\Sigma') \to \AC$. Define a measurable
function $A_\bullet$ from $\Omega$  to the extended 
reals $\overline{\BB{R}}$ by
  \begin{equation}\label{eq defAdot}
  A_\bullet := \frac{\big(i^{-1}(T_+)\big)^{-1} + \big(i^{-1}(T_-)\big)^{-1}}{2}.
  \end{equation}
Then we can define a projection valued measure on the Borel sigma algebra
of $\BB{R}$, generated by the sets $(-\infty, \lambda]$ with 
$\lambda \in \BB{R}$, by
  \begin{equation*}
  E\big((-\infty, \lambda]\big):= i(\chi_{[A_\bullet \le \lambda]}),  
  \end{equation*}
and in this way equation \eqref{eq spectral} reappears. Furthermore the 
spectral measure, in its turn, uniquely determines a real valued random variable 
$j:\ L^\infty(\BB{R}) \to \BC$. The above expositions shows three 
equivalent ways of characterising real valued random variables 
or \emph{observables}, i.e.\ by a selfadjoint operator $A$,
by a projection valued measure $E$,
and by a $^*$-homomorphism $j$.
A fourth way of looking at random variables is provided by 
Stone's theorem, see for instance \cite{KaR0}.
\begin{stel}\label{thm stone}\textbf{(Stone's theorem)}
There is a one-to-one correspondence between strongly continuous
unitary representations $t \mapsto U_t$ of the abelian group 
$\BB{R}$ into a von Neumann algebra $\AC$ and selfadjoint 
operators $A$ affiliated to $\AC$, i.e.\ having all its spectral 
projections in $\AC$, such that
  \begin{equation*}
  U_t = \exp(itA) := \int_\BB{R}e^{it\lambda}E(d\lambda).
  \end{equation*}
\end{stel}

Let $\AC$ and $\BC$ be von Neumann algebras. We denote their 
Banach space duals by $\AC^*$ and $\BC^*$, respectively. A linear map $T:\
\AC \to \BC$ determines a dual map $T^*:\ \BC^* \to \AC^*$
by $\rho \mapsto \rho \circ T$. An element $A$ of a $C^*$-algebra
is called \emph{positive} if it can be written as $A=B^*B$ for 
some $B$ in that $C^*$-algebra. Recall that von Neumann algebras 
are a special class of $C^*$-algebras. The map $T^*$ maps states 
on $\BC$ to states on $\AC$ if and only if $T$ is \emph{positive}, 
in the sense that it maps positive elements of $\AC$ 
into positive elements of $\BC$, and $T(\I_\AC) = \I_\BC$.
The map $T$ is said to be \emph{completely positive} if 
for all $n \in \BB{N}$ the map $T\ten\Id_n: \AC \ten M_n(\BB{C})
\to \BC \ten M_n(\BB{C})$ is positive. An \emph{operation} $T$
is a completely positive map such that $T(\I_\AC) = \I_\BC$ and its 
dual $T^*$ is also called an \emph{operation}. An operation 
$T^*: \BC^*\to \AC^*$ represent some physical procedure which 
takes as an input a state on a quantum system described by $\BC$
and turns out a state on the quantum system $\AC$. The \emph{complete}
positivity ensures that operations on a quantum system $\AC$
can always be extended to operations on $\AC\ten \WC$, the 
quantum system $\AC$ with its environment $\WC$.

The operational approach to quantum mechanics was 
pioneered by Davies and Lewis \cite{DaL}, \cite{Dav}, 
\cite{EvL}. In this approach all physical procedures 
performed on quantum systems are given by operations 
between their algebras of observables. An example 
of an operation is a random variable, i.e.\ a $^*$-homomorphism
$i$ from an algebra $\AC$ to $\BC$. Suppose $\BC = \AC \ten \WC$
and $i = \Id \ten \I_\WC$, then the dual of $i$ is what is
known in physics as a partial trace, i.e.\ restriction of the 
state to the smaller system $\AC\ten\I_\WC$. A second example is 
adjoining with a unitary operator or a projection, i.e.\ time
evolution or conditioning on a measurement result, respectively.
A third example of an operation is a \emph{conditional 
expectation}, the topic of the next paragraph. 

Let $\BC$ be a von Neumann subalgebra of a von Neumann 
algebra $\AC$. A \emph{conditional expectation} of 
$\AC$ onto $\BC$ is a linear surjective map 
$\EC:\ \AC \to \BC$ such that $\EC^2 = \EC$ and
$\p \EC\p = 1$. We are interested in a conditional 
expectation $\EC_\rho:\ \AC \to \BC$ that leaves a 
certain state $\rho$ on $\AC$ invariant, i.e. 
$\rho\circ\EC_\rho = \rho$. However, such a conditional 
expectation does not always exist, but if it exists 
it is unique \cite{Tak}. The interpretation is as 
follows, if we start with a quantum system 
$\AC$ in a state $\rho$ and we gain (for example by 
observation) the additional information that 
$\BC$ is in a state $\sigma$, then $\sigma \circ \EC_\rho$ 
is the updated state of $\AC$ \cite{Kum3}. It follows from 
\cite{Tom} that $\EC_\rho$ is an operation and that it 
satisfies the \emph{module property}: $\EC_\rho(B_1AB_2) = B_1\EC_\rho(A)B_2$
for all $B_1,B_2 \in \BC$ and $A \in \AC$. An example
of a conditonal expectation is given by taking $\AC = \BC \ten \WC$
and $\EC = \Id \ten \phi$ with $\phi$ a state on $\WC$.
If $\AC$ is a commutative algebra then we retrieve the 
conditional expectation of classical probability theory.

\section{Stochastic calculus on Fock spaces}\label{sec scFs}

Let $\BB{H}$ be a Hilbert space and think of its elements 
as the pure states of one particle in quantum mechanics. The 
particles we are interested in are photons. Since photons 
are bosons they have to be described by symmetrised wave functions.
This motivates the definition of the \emph{symmetric Fock 
space over $\BB{H}$} as
  \begin{equation*}
  \FC(\BB{H}) := \BB{C} \oplus \bigoplus_{n=1}^\infty \BB{H}^{\ten_s n}. 
  \end{equation*}
It describes situations where the number of 
particles present is arbitrary. For every $f \in \BB{H}$
we define the \emph{exponential vector} 
$e(f) \in \FC(\BB{H})$ by
  \begin{equation*}
  e(f) := 1 \oplus\bigoplus_{n=1}^\infty \frac{1}{\sqrt{n!}}f^{\ten n}.  
  \end{equation*}
The linear span $\DC$ of all exponential vectors is a dense subspace of
$\FC(\BB{H})$. On the dense domain $\DC$ we define for all $f \in \BB{H}$
an operator $W(f)$ by
  \begin{equation}\label{eq defWeyl}
  W(f)e(g) := \exp\big(-\langle f, g\rangle - \frac{1}{2}\p f\p^2\big)e(f+g),
  \ \ \ \ g \in \BB{H},
  \end{equation} 
which is isometric and therefore uniquely extends to a unitary operator 
on $\FC(\BB{H})$. The operators $W(f): \FC(\BB{H}) \to \FC(\BB{H})$ are 
called \emph{Weyl operators} and they satisfy the following \emph{Weyl relations}
  \begin{equation}\begin{split}\label{eq Weylrel} 
  &1.\ \ \ \ \  W(f)^* = W(-f),\ \ \ \ \ f \in \BB{H},                        \\
  &2.\ \ \ \ \  W(f)W(g) = \exp\big(-i\mbox{Im}\langle f,g\rangle\big)W(f+g),
          \ \ \ \ \ f,g \in \BB{H}.
  \end{split}\end{equation} 

For all $f \in \BB{H}$ the family $\{W(tf)\}_{t \in \BB{R}}$ forms a 
one-parameter group, continuous in the strong operator topology. 
Therefore it follows from Stone's theorem \ref{thm stone} that for 
all $f \in \BB{H}$ there exists a selfadjoint $B(f)$ such that
  \begin{equation*}
  W(tf) = \exp\big(itB(f)\big).
  \end{equation*}
The operators $B(f)$ are called \emph{field operators}, see 
also Section \ref{sec qo}.
The domain of the operator $B(f_k)\ldots B(f_1)$
contains $\DC$ for every $f_1,\ldots f_k \in \BB{H}$ and
$k \in \BB{N}$ (cf.\ \cite{Pet}). For $f,g \in \BB{H}$ and $t \in \BB{R}$
it follows from the Weyl relations that on the domain $\DC$
  \begin{equation}\label{eq prop fields}\begin{split}
  &1.\ \ \   B(tf) = tB(f),          \\
  &2.\ \ \   B(f+g) = B(f) + B(g),          \\
  &3.\ \ \   [B(f), B(g)] = 2 i\mbox{Im}\langle f, g\rangle.
  \end{split}\end{equation}

Let us take $\BB{H} = L^2(\BB{R})$ and let us consider it
in the canonical way as a real space, then 
the pair $(\BB{H},\mbox{Im}\langle\cdot, \cdot\rangle)$ forms a 
symplectic space. Denote by $H_0$ the real subspace of $\BB{H}$ given by 
$\{f = (f_1, f_2)\in H;\ f_2 =0\}$. From 
(\ref{eq prop fields}.3) it immediately follows that the 
family of operators $\{B(f);\ f \in H_0\}$ is commutative.

As in the previous section we can consider the bounded operators 
$T(f)_+ := (B(f)+i\I)^{-1}$ and $T(f)_- := (B(f)-i\I)^{-1}$
for $f \in H_0$. They generate a commutative von Neumann algebra $\CC$. 
We denote by $\phi$ the state on $\CC$ given by 
  \begin{equation*}
  \phi(C) := \langle \Phi, C\Phi\rangle,\ \ \mbox{with}\ \ 
  \Phi = 1 \oplus 0 \oplus 0 \ldots \in \FC(\BB{H}).
  \end{equation*}
Using Theorem \ref{thm spectral} we see that $(\CC, \phi)$ is 
isomorphic to $L^\infty(\Omega, \Sigma, \BB{P})$ for some
probability space $(\Omega, \Sigma, \BB{P})$. We denote this
isomorphism from $\CC$ to $L^\infty(\Omega, \Sigma, \BB{P})$
by $C \mapsto C_\bullet$. As in the previous
section, see equation \eqref{eq defAdot}, we can let the 
operator $B(f)$ $(f\in H_0)$, affiliated with $\CC$, correspond to 
a measurable function $B(f)_\bullet$ on $\Omega$ taking values in 
the extended reals $\overline{\BB{R}}$.

The joint characteristic function of the random variables 
$B(f_1)_\bullet, B(f_2)_\bullet,\ldots,B(f_k)_\bullet$ is 
for $x_1,\ldots,x_k \in \BB{R}$ given by
  \begin{equation}\label{eq charfunc}\begin{split}
  &\BB{E}\big[e^{ix_1B(f_1)_\bullet}e^{ix_2B(f_2)_\bullet}\ldots
  e^{ix_kB(f_k)_\bullet}\big] =
  \int_\Omega e^{ix_1B(f_1)_\omega}e^{ix_2B(f_2)_\omega}\ldots
  e^{ix_kB(f_k)_\omega}\BB{P}(d\omega) = \\
  &\Big\langle\Phi, e^{ix_1B(f_1)}e^{ix_2B(f_2)}\ldots
  e^{ix_kB(f_k)}\Phi\Big\rangle = 
  \Big\langle\Phi, \exp\big(i\sum_{i=1}^k x_i B(f_i)\big)\Phi\Big\rangle = \\
  &\Big\langle\Phi, W\Big(\sum_{i=1}^k x_i f_i\Big)\Phi\Big\rangle = 
  \exp\Big(-\frac{1}{2}\sum_{i,j=1}^kx_ix_j\langle f_i,f_j\rangle\Big),
  \end{split}\end{equation}
where we used Theorem \ref{thm spectral} in the second step and 
equation (\ref{eq prop fields}.1) and (\ref{eq prop fields}.2) in the
fourth step. For $t \in \BB{R}$ we define a random variable 
$B_t := B(\chi_{[0,t)})_\bullet$
where $\chi_{[0,t)}$ stands for the function that is $1$ on the interval 
$[0,t)$ and $0$ elsewhere. For $s_1 \le t_1 \le s_2 \le t_2$ it immediately 
follows from \eqref{eq charfunc} that the joint characteristic function 
of the increments $B_{t_1}-B_{s_1} = B(\chi_{[s_1,t_1)})_\bullet$ 
and $B_{t_2}-B_{s_2} = B(\chi_{[s_2,t_2)})_\bullet$ factorizes
  \begin{equation*}\begin{split}
  &\BB{E}\big[e^{ix(B_{t_1}-B_{s_1})}e^{iy(B_{t_2}-B_{s_2})}\big] = 
  \exp\Big(-\frac{1}{2}\big(x^2(t_1-s_1) + y^2(t_2-s_2)\big)\Big) = \\
  &\exp\Big(-\frac{x^2(t_1-s_1)}{2}\Big)\exp\Big(-\frac{y^2(t_2-s_2)}{2}\Big) = 
  \BB{E}\big[e^{ix(B_{t_1}-B_{s_1})}\big]\BB{E}\big[e^{iy(B_{t_2}-B_{s_2})}\big],
  \end{split}\end{equation*} 
i.e.\ $\{B_t\}_{t \ge 0}$ is a process with independent increments. 
Furthermore, from its characteristic function it follows that 
the increments $B_t-B_s$ are normally distributed with mean zero
and variance $t-s$. Summarizing, the process $\{B_t\}_{t\ge 0}$ is a 
\emph{Brownian motion}. 

The idea to simultaneously diagonalise
the fields in the family $\{B(\chi_{[0,t)});\ t \ge 0\}$ is 
implicit is some of the earliest work in quantum field theory. 
However, Segal \cite{Seg} in the 1950s was the first to emphasise the 
connection with probability theory. See also \cite{Sim} for a
nice review of the importance of these ideas for Euclidean
quantum field theory. 

Fix $\alpha$ in the interval $[0, \pi)$. Then, as in the 
previous paragraph, the family of operators 
$B\big(e^{i\alpha}\chi_{[0,t)}\big)$ is 
commutative. This means there is a probability space
$(\Omega_\alpha, \Sigma_\alpha, \BB{P}_\alpha)$ on 
which the operators $B\big(e^{i\alpha}\chi_{[0,t)}\big)$ 
are realised as random variables 
$B\big(e^{i\alpha}\chi_{[0,t)}\big)_\bullet =: B^\alpha_t$.
In a similar fashion as in the previous paragraph it 
can be shown that the process $\{B^\alpha_t\}_{t\ge 0}$ is 
a Brownian motion. However, for two different alphas, i.e.\
$\alpha_1$ and $\alpha_2$, it follows from equation 
(\ref{eq prop fields}.3) that the Brownian motions $B^{\alpha_1}_t$
and $B^{\alpha_2}_t$ do not commute. For instance we have
  \begin{equation*}
  [B^0_s, B^{\pi/2}_t] = 2i\mbox{min}\{s,t\}.
  \end{equation*}

Expression (\ref{eq prop fields}.3) is known as the 
\emph{canonical commutation relation}. It is clear that 
the Weyl relation (\ref{eq Weylrel}.2) is an exponentiated
form of the canonical commutation relation. Weyl operators
have the advantage that they are bounded, so that there 
are no domain problems involved. That is why we
abstractly define the \emph{$C^*$-algebra of canonical commutation
relations $CCR(H,\sigma)$} with respect to some 
symplectic space $(H,\sigma)$ as the $C^*$-algebra generated
by abstract elements $W(f)\ (f\in H)$ satisfying the relations
in \eqref{eq Weylrel} where $\BB{H}$ and 
$\mbox{Im}\langle\cdot, \cdot\rangle$ should be replaced by 
$H$ and $\sigma$, respectively. It follows from \cite{Sla} that this
$C^*$-algebra exists and moreover that it is unique up 
to isomorphism.

Equation \eqref{eq defWeyl} defines a representation of 
$CCR(\BB{H}, \mbox{Im}\langle\cdot,\cdot\rangle)$ on the Fock 
space $\FC(\BB{H})$. This representation is the 
GNS-representation (cf.\ \cite{Pet}) with respect to the vacuum
state, i.e.\ the state that maps $W(f)$ to $\exp(-1/2\p f\p^2)$.
In Chapter \ref{ch sec} we will encounter GNS-representations
of $CCR(\BB{H}, \mbox{Im}\langle\cdot,\cdot\rangle)$ with
respect to other states than the vacuum. Given a representation
of $CCR(\BB{H}, \mbox{Im}\langle\cdot,\cdot\rangle)$, 
we can define the von Neumann algebra 
$\WC$ generated by the $W(f)\ (f\in \BB{H})$ represented 
as operators on the representation Hilbert space. 
For the vacuum GNS-representation it turns out
that $\WC$ is the whole algebra of bounded operators on 
$\FC(\BB{H})$ cf.\ \cite{Pet}. The algebra 
$\WC = \BC\big(\FC(\BB{H})\big)$ equipped with the 
vacuum state is the quantum probability space with which 
we will describe an electromagnetic field in the vacuum state, 
see also Section \ref{sec qo}. More information about 
$CCR$-algebras can be found in \cite{BrR1}, \cite{BrR2} and 
\cite{Pet}.

Apart from the Brownian motions $B^\alpha_t$ the 
algebra $\WC= \BC\big(\FC\big(L^2(\BB{R})\big)\big)$ contains 
Poisson processes \cite{FrM}, a point that we will investigate below. 
Let $P_t:\ L^2(\BB{R})\to L^2(\BB{R})$
denote the projection given by $f \mapsto \chi_{[0,t)}f$.
The \emph{second quantisation} of an operator 
$A \in \BC\big(L^2(\BB{R})\big)$ is the operator in 
$\BC\big(\FC\big(L^2(\BB{R})\big)\big)$ given by
  \begin{equation*}
  \Gamma(A) := 1 \oplus \bigoplus_{n=1}^\infty A^{\ten n}.
  \end{equation*} 
Then for $A,B \in \BC\big(L^2(\BB{R})\big)$ we have 
$\Gamma(AB) =\Gamma(A)\Gamma(B)$. This means that the 
one-parameter group of unitaries generated by 
$P_t$, i.e.\ $\exp(isP_t)$, leads 
to a one-parameter group of unitaries
$\Gamma\big(\exp(isP_t)\big)$.  
Stone's theorem \ref{thm stone} asserts the 
existence of a selfadjoint operator $\Lambda(t)$ on 
$\FC\big(L^2(\BB{R})\big)$ such that for all $s\in \BB{R}$
  \begin{equation*}
  \Gamma\big(\exp(isP_t)\big) = \exp\big(is\Lambda(t)\big).
  \end{equation*}
The operator $\Lambda(t)$ is interpreted as the random variable
that counts how many particles, i.e. photons, are present in the 
interval $[0,t)$.

The family of selfadjoint operators $\{\Lambda(t);\ t \ge 0\}$ 
is commutative, i.e.\ the bounded operators $T^t_+ := (\Lambda(t)+i)^{-1}$
and $T^t_- := (\Lambda(t)-i)^{-1}$ generate a commutative von Neumann
algebra $\CC$ to which the operators $\Lambda(t)$ are affiliated. 
For $f\in L^2(\BB{R})$ a state $\rho$ on $\CC$ is defined by
  \begin{equation*}
  \rho(C) := \langle\psi(f),C\psi(f)\rangle,
  \end{equation*}
where $\psi(f)\in \FC\big(L^2(\BB{R})\big)$ is the \emph{coherent vector}
defined as the exponential vector $e(f)$ normalised to unity
  \begin{equation*}
  \psi(f) := \exp\big(-\frac{1}{2} \p f\p^2\big)e(f).
  \end{equation*}
The state given by the coherent vector $\psi(f)$ is the state of
the field in a laser beam. Spectral theory provides 
a probability space $(\Omega, \Sigma, \BB{P})$ 
such that $(\CC,\rho)$ is isomorphic to 
$L^\infty(\Omega, \Sigma, \BB{P})$. As before 
we can realise the selfadjoint operators $\Lambda(t)$ as 
measurable functions $\Lambda(t)_\bullet$ on $\Omega$ taking 
values in $\overline{\BB{R}}$.

For $s_1\le t_1\le s_2 \le t_2$ the characteristic
function of the increments $\Lambda(t_1)_\bullet-\Lambda(s_1)_\bullet$
and $\Lambda(t_2)_\bullet-\Lambda(s_2)_\bullet$ for $x, y \in \BB{R}$
is given by
  \begin{equation*}\begin{split}
  &\BB{E}\Big[e^{ix\big(\Lambda(t_1)_\bullet-\Lambda(s_1)_\bullet\big)}
  e^{iy\big(\Lambda(t_2)_\bullet-\Lambda(s_2)_\bullet\big)}\Big] =
  \int_\Omega e^{ix\big(\Lambda(t_1)_\omega-\Lambda(s_1)_\omega\big)}
  e^{iy\big(\Lambda(t_2)_\omega-\Lambda(s_2)_\omega\big)}\BB{P}(d\omega) = \\
  &\Big\langle\psi(f),e^{ix\big(\Lambda(t_1)_\omega-\Lambda(s_1)_\omega\big)}
  e^{iy\big(\Lambda(t_2)_\omega-\Lambda(s_2)_\omega\big)}\psi(f)\Big\rangle = \\
  &e^{-\p f\p^2}\Big\langle e(f),\Gamma\Big(e^{ix(P_{t_1}-P_{s_1})}\Big)
  \Gamma\Big(e^{iy(P_{t_2}-P_{s_2})}\Big)e(f)\Big\rangle = \\
  &e^{-\p f\p^2}\Big\langle e(f),e\big(e^{ix(P_{t_1}-P_{s_1})}
  e^{iy(P_{t_2}-P_{s_2})}f\big)\Big\rangle = 
  e^{\big\langle f,\big(e^{ix(P_{t_1}-P_{s_1})}
  e^{iy(P_{t_2}-P_{s_2})}-1\big) f\big\rangle} = \\
  &\exp\big(\int_{s_1}^{t_1}(e^{ix}-1)|f|^2d\lambda\big)
  \exp\big(\int_{s_2}^{t_2}(e^{iy}-1)|f|^2d\lambda\big),
  \end{split}\end{equation*}
i.e.\ $\{\Lambda(t)_\bullet\}_{t\ge 0}$ is a process with independent
increments. Moreover, the characteristic function shows that it is 
a Poisson process with intensity measure $|f|^2d\lambda$, where $\lambda$
stands for the Lebesgue measure. Summarizing, when photons are counted in 
a laser beam they arrive Poisson distributed.
Since $\psi(f) = W(f)e(0)= W(f)\Phi$, we can just as well 
study the commutative von Neumann algebra $W(f)^*\CC W(f)$, 
generated by the operators $W(f)^*\Lambda(t)W(f)$, 
equipped with the vacuum state to arrive again at a 
Poisson process with intensity measure $|f|^2d\lambda$.
  
The above exposition shows that the quantum probability 
space $(\WC, \phi)$ contains many processes that do 
not commute with each other. Hudson and 
Parthasarathy showed \cite{HuP} that it is possible 
to extend the definition of It\^o's stochastic integrals 
to deal with all these processes simultaneously. 
To this end we introduce \emph{annihilation} $A(t)$ 
and \emph{creation operators} $A^*(t)$ by
  \begin{equation*}\label{eq ancr}
  A(t) := \frac{1}{2}\big(B(i\chi_{[0,t)})-iB(\chi_{[0,t)})\big)\ \ \ \mbox{and} \ \ \
  A^*(t) := \frac{1}{2}\big(B(i\chi_{[0,t)})+iB(\chi_{[0,t)})\big).
  \end{equation*}
Let $M_t$ stand for one of the three 
processes $A(t), A^*(t)$ or $\Lambda(t)$, all 
restricted to the domain $\DC$.
The stochastic integrals will be defined with  
respect to $M_t$.  

Let us write $L^2(\BB{R})$ as the direct sum 
$L^2\big((-\infty,t)\big)\oplus L^2\big([t,\infty)\big)$, 
then $\FC\big(L^2(\BB{R})\big)$ is unitarily equivalent 
with $\FC\big(L^2\big((-\infty,t)\big)\big)\ten\FC\big(L^2\big([t,\infty)\big)\big)$
through the identification $e(f)\cong e(f_{t)})\ten e(f_{[t})$
with $f_{t)} := f\chi_{(-\infty,t)}$ and $f_{[t} := f\chi_{[t,\infty)}$.
We will also use the notation $f_{[s,t)}$ for $f\chi_{[s,t)}$ and 
omit the tensor product signs between exponential vectors.
Furthermore, the algebra $\WC = \BC\big(\FC\big(L^2(\BB{R})\big)\big)$ splits
as $\WC = \WC_{t)}\ten\WC_{[t}$ where 
$\WC_{t)} = \BC\big(\FC\big(L^2((-\infty,t))\big)\big)$ and 
$W_{[t} =\BC\big(\FC\big(L^2([t,\infty))\big)\big)$. The following
factorisation property \cite{HuP}, \cite{Par} makes the 
definition of stochastic integration with respect to $M_t$ possible
  \begin{equation*}
  (M_t - M_s)e(f) = e(f_{s)})\big\{(M_t-M_s) e(f_{[s,t)})\big\}e(f_{[t}), 
  \end{equation*} 
with $(M_t-M_s) e(f_{[s,t)}) \in \FC\big(L^2([s,t))\big)$. 
We first define the stochastic integral for the so called  
{\it simple} operator processes with values in $\BC\ten\WC$,
where $\BC$ is an $n$-dimensional von Neumann algebra called
the \emph{initial system}.

\begin{de} 
Let $\{L_s\}_{0 \le s \le t}$ be an adapted 
(i.e.\ $L_s \in \BC \ten \WC_{s)}$ for all $0 \le s \le t$) simple 
process with respect to the partition $\{s_0=0, s_1,\dots, s_p= t\}$ in the sense that 
$L_s = L_{s_j}$ whenever $s_j \le s < s_{j+1}$. Then the stochastic integral of 
$L$ with respect to $M$ on $\BB{C}^n \ten \DC$ is given by \cite{HuP}, \cite{Par}:  
  \begin{equation*}
  \int_0^t L_s dM_s  ~f e(u) := 
\sum_{j=0}^{p-1} \big(L_{s_j}fe(u_{s_j)})\big)\big((M_{s_{j+1}}- M_{s_{j}})
   e(u_{[s_j, s_{j+1})})\big)e(u_{[s_{j+1}}).
  \end{equation*}
\end{de}
By the usual approximation by simple processes the definition of the 
stochastic integral can be extended 
to a large class of {stochastically integrable processes} \cite{HuP}, \cite{Par}. 
The notation is simplified by writing $dX_t = L_tdM_t$ for 
$X_t = X_0 + \int_0^t L_sdM_s$.

The following theorem of Hudson and Parthasarathy \cite{HuP} 
extends the It\^o rule of classical probability theory. 
\begin{stel}\label{thm Ito}\textbf{(Quantum It\^o rule \cite{HuP}, \cite{Par})}
Let $M_1$ and $M_2$ each be one of the processes $A(t), A^*(t)$ or $\Lambda(t)$. Then 
$M_1M_2$ is an adapted process satisfying the relation:
  \begin{equation*}
  d(M_1M_2) = M_1dM_2 + M_2dM_1 + dM_1dM_2,
  \end{equation*}
where $dM_1dM_2$ is given by the quantum It\^o table:
\begin{center}
{\large \begin{tabular} {l|lll}
$dM_1\backslash dM_2$ & $dA^*(t)$ & $d\Lambda(t)$ & $dA(t)$ \\
\hline 
$dA^*(t)$ & $0$ & $0$ & $0$ \\
$d\Lambda(t)$ & $dA^*(t)$ & $d\Lambda(t)$ & $0$  \\
$dA(t)$ & $dt$ & $dA(t)$ & $0$ 
\end{tabular} }
\end{center}
\end{stel}
This theorem will prove to be much more useful in 
calculations than the actual definition of the stochastic
integral. It reduces hard questions regarding analysis 
to algebraic manipulations with increments $dA^*(t), dA(t)$ and 
$d\Lambda(t)$.  

For a fixed $\alpha \in [0,\pi)$ we saw that the 
fields $B(e^{i\alpha}\chi_{[0,t)})$ define a Brownian 
motion. Note that from the definition of the creation 
and annihilation operator \eqref{eq ancr} it 
follows that $B(e^{i\alpha}\chi_{[0,t)}) = 
ie^{-i\alpha}A(t)-ie^{i\alpha}A^*(t)$, i.e.\
$dB(e^{i\alpha}\chi_{[0,t)}) = ie^{-i\alpha}dA(t)-ie^{i\alpha}dA^*(t)$. 
Therefore, using Theorem \ref{thm Ito}, 
$\big(dB(e^{i\alpha}\chi_{[0,t)})\big)^2 = dt$, which
is exactly what would be expected for a Brownian motion. 
Note that in the non-commutative theory we can also
calculate products of increments of two non-commuting
processes. For instance for $\alpha_1, \alpha_2 \in [0,\pi)$,
$dB(e^{i\alpha_1}\chi_{[0,t)})dB(e^{i\alpha_2}\chi_{[0,t)})
= e^{i(\alpha_2-\alpha_1)}dt$.

Let $f$ be an element of $L^2(\BB{R})$. It is not hard 
to see (cf.\ \cite{Par}) that the Weyl operator $W(f_{t)}) = 
\exp\big(A^*(f_{t)})-A(f_{t)})\big)$
satisfies the quantum stochastic differential equation
  \begin{equation}\label{eq W}
  dW(f_{t)}) = \big\{f(t)dA^*(t) - \overline{f}(t)dA(t) - \frac{1}{2}|f(t)|^2dt\big\}W(f_{t)}).
  \end{equation}
Denote $N_t := W(f)^*\Lambda(t)W(f) = W(f_{t)})^*\Lambda(t)W(f_{t)})$
then it follows from Theorem \ref{thm Ito} and equation \eqref{eq W} 
that the Poisson process $N_t$ satisfies \cite{FrM}
  \begin{equation*}
  dN_t = d\Lambda(t) + f(t)dA^*(t) + \overline{f}(t)dA(t)+ |f(t)|^2dt. 
  \end{equation*}
It easily follows that $dN_t^2 = dN_t$, which is exactly what would 
be expected for a Poisson process.

\section{Quantum optics}\label{sec qo}

Quantum optics deals with the interaction between 
quantum systems and the quantized electromagnetic 
field. In this section we will point out that in 
some Markovian approximation this 
interaction is governed by a stochastic differential 
equation in the sense of Hudson and 
Parthasarathy. Indeed, the Markovian approximation is justified 
if the time scale of the field evolution can be considered 
to be extremely much faster than the time scale 
of the quantum system with which it interacts. 
In this way the field can be considered as a 
(non-commutative) noise acting on the quantum system.  

A convenient starting point (cf.\ \cite{WaM}, \cite{GaZ}) for quantizing the classical
free electromagnetic field is the vector potential 
$\mbox{\boldmath $A$}(\mbox{\boldmath $r$},t)$ in the Coulomb gauge, 
i.e.\ $\mbox{\boldmath $\nabla$} \cdot \mbox{\boldmath $A$} = 0$. The magnetic 
field $\mbox{\boldmath $B$}$ and the electric field $\mbox{\boldmath $E$}$
are determined by $\mbox{\boldmath $B$} = \mbox{\boldmath $\nabla$} \times \mbox{\boldmath $A$}$
and $\mbox{\boldmath $E$} = -\frac{\partial\mbox{\boldmath $A$}}{\partial t}$, respectively.  
The Hamiltonian of the free field is given by
  \begin{equation*}
  H = \frac{1}{2}\int \Big(\varepsilon_0 \mbox{\boldmath $E$}^2 + \frac{1}{\mu_0}
  \mbox{\boldmath $B$}^2\Big)d^3\mbox{\boldmath $r$},
  \end{equation*}
where $\mu_0$ and $\varepsilon_0$ are the magnetic permeability 
and electric permittivity of free space, respectively. 
Then a Fourier expansion of $\mbox{\boldmath $A$}
(\mbox{\boldmath $r$},t)$ is made 
and $\mbox{\boldmath $E$}(\mbox{\boldmath $r$},t)$ and 
$\mbox{\boldmath $B$}(\mbox{\boldmath $r$},t)$ are expressed 
in terms of this Fourier expansion. After filling these expressions
for $\mbox{\boldmath $E$}(\mbox{\boldmath $r$},t)$ and 
$\mbox{\boldmath $B$}(\mbox{\boldmath $r$},t)$ into the 
Hamiltonian one easily recognizes the Hamiltonian for 
an assembly of independent harmonic oscillators,
see \cite{WaM} and \cite{GaZ}. An assembly of \emph{quantum} 
harmonic oscillators is described by its algebra of 
creation and annihilation operators obeying the usual commutation
relations, which justifies 
our description of the quantized electromagnetic field 
by a $CCR$-algebra. In this description the electric 
field of a mode $g \in L^2(\BB{R})$ is given by 
a polarisation vector times the operator $i\big(A(g)-A^*(g)\big)$, 
which is just the operator $B(g)$ from the previous section, 
see also \cite{WaM}, \cite{GaZ}.    

Let $\BC$ stand for the algebra of observables of an $n$-dimensional
quantum system, i.e.\ the elements of $\BC$ are operators on $\BB{C}^n$. 
Let $\WC$ be the von Neumann algebra of observables of the 
electromagnetic field, i.e.\ elements of 
$\WC$ are operators on some representation Hilbert space for the 
$CCR$-algebra. In the theory of open quantum systems the field $\WC$ is 
often referred to as the reservoir. 
The time evolution of the system $\BC$ and the 
electromagnetic field $\WC$ together is governed by the
Hamiltonian $H$ from quantum electrodynamics, cf.\ \cite{Fey}. 
Using shorthand notation $H_s+H_r := H_s\ten\I_r + \I_s\ten H_r$, 
we can write this Hamiltonian as
  \begin{equation}\label{eq QEDHam}
  H = H_s + H_r + H_{sr}, 
  \end{equation}
where the subscripts $s$ and $r$ stand for the system $\BC$ and the 
reservoir $\WC$, respectively. 
The interaction between system and field is given by 
the interaction Hamiltonian $H_{sr}$. 

Let $\rho$ and $\gamma$ be states on $\BC$ and
$\WC$, respectively. We can define a time evolved 
state on $\BC\ten\WC$ by
  \begin{equation*}
  \chi^t(Z) := \rho\ten\gamma\Big(\exp\big(\frac{it}{\hbar}H\big)
  Z \exp\big(-\frac{it}{\hbar}H\big)\Big), \ \ \ \ Z \in \BC\ten\WC,\ t\ge0.
  \end{equation*}
This leads for times $t\ge 0$ to the following reduced state on system $\BC$ 
  \begin{equation*}
  \rho^t(X) := \chi^t(X\ten\I_r), \ \ \ \ X \in \BC,\ t\ge0,
  \end{equation*}
and it defines an operation $T_t^*:\ \BC^* \to \BC^*$ for $t \ge 0$ 
by $T_t^*(\rho) := \rho^t$. The evolution $T_t$ is still rather 
complicated, for instance it is not a semigroup, i.e.\ 
we do not have $T_{t+s} = T_tT_s$ for all $s,t \ge 0$. We proceed
by assuming that $H_{sr}$ is small compared to 
$H_s + H_r$. We separate the slow evolution 
generated by $H_{sr}$ from the rapid evolution generated by $H_s+H_r$
by transforming to the \emph{interaction picture}, i.e.\ we define 
for $Z \in \BC\ten\WC$ and $X \in \BC$
  \begin{equation*}\begin{split}
  &\tilde{\chi}^t(Z) := \chi^t\Big(\exp\big(-\frac{it}{\hbar}(H_s+H_r)\big)
  Z\exp\big(\frac{it}{\hbar}(H_s+H_r)\big)\Big),\\  
  &\tilde{\rho}^t(X) := \tilde{\chi}^t(X\ten\I_r) = 
  \rho^t\Big(\exp\big(-\frac{it}{\hbar}H_s\big)
  X\exp\big(\frac{it}{\hbar}H_s\big)\Big),\ \ \  \mbox{and} \\
  &\tilde{H}_{sr}^t := \exp\big(\frac{it}{\hbar}(H_s+H_r)\big)H_{sr}
  \exp\big(-\frac{it}{\hbar}(H_s+H_r)\big).
  \end{split}\end{equation*}
It easily follows that $\frac{d\tilde{\chi}^t(Z)}{dt} = 
\frac{i}{\hbar}\tilde{\chi}^t\big([\tilde{H}_{sr}^t,Z]\big)$ for 
all $Z \in \BC\ten\WC$. Expanding this up to second order we 
find for all $X \in \BC$
  \begin{equation}\label{eq nonMark}
  \frac{d\tilde{\rho}^t(X)}{dt} = 
  \frac{i}{\hbar}\rho\ten\gamma\big([\tilde{H}_{sr}^t, X\ten\I_r]\big) 
  - \frac{1}{\hbar^2}\int_0^t
  \tilde{\chi}^{t'}\Big(\big[\tilde{H}_{sr}^{t}, [\tilde{H}_{sr}^{t'}, X\ten\I_r]\big]\Big)dt'.
  \end{equation}
We may take $\frac{i}{\hbar}\rho\ten\gamma\big([\tilde{H}_{sr}^t, X\ten\I_r]\big) =0$
in the above equation by including a term $\mbox{Id}\ten\gamma(H_{sr})$
in the system Hamiltonian. 

In the coming two paragraphs a heuristic 
exposition, common in physics, on the so-called \emph{Markov approximation} 
to equation \eqref{eq nonMark}, cf.\ \cite{Car}, is given. 
The approximation consists out of two steps. 
In the first step we replace $\tilde{\chi}^{t'}$ 
by $\tilde{\rho}^{t'}\ten\gamma$,
and in the second step 
we replace $\tilde{\rho}^{t'}$ in its turn 
by $\tilde{\rho}^t$. The first step of the 
approximation is called the \emph{Born approximation}. 
It is justified if the reservoir is 
a very large system, i.e.\ its state is virtually unaffected
by the coupling to $\BC$, and the coupling between 
system and reservoir through $H_{sr}$ is very weak, i.e.\
at all times $\chi^t$ shows only deviations of order $H_{sr}$
from an uncorrelated state, see also \cite{Car}.  
The second step is reasonable if the reservoir correlation
times are much shorter than the time scale of the 
evolution of the system. Then the past history of 
the system, imprinted in the reservoir through the interaction,
can not influence the present state of the system since in the 
reservoir it gets lost very quickly, see also \cite{Car}. 

After identifying the states $\tilde{\rho}^t$ and $\gamma$ with
their density matrices, equation \eqref{eq nonMark} reduces in 
the Markov approximation to
  \begin{equation}\label{eq Mark}
  \frac{d\tilde{\rho}^t}{dt} = 
  - \frac{1}{\hbar^2}\int_0^t
  \mbox{Tr}_r\Big(\big[\tilde{H}_{sr}^{t}, 
  [\tilde{H}_{sr}^{t'}, \tilde{\rho}^t\ten\gamma]\big]\Big)dt',
  \end{equation}
i.e.\ it is of the form $d\tilde{\rho}^t = L^t(\tilde{\rho}^t)dt$. 
In many concrete examples, see \cite{Car}, it can be shown, again
by using the fact that the reservoir correlation functions decay
extremely much faster than the time scale of the evolution 
of the system, that the generator $L^t$ is actually time 
independent. This means we end up with a time evolution 
$T_t^*(\rho) := \tilde{\rho}^t$ that is a one-parameter 
semigroup, i.e. $T_tT_s = T_{t+s}$ for all $t,s\ge 0$. 
The semigroup property reflects that there are no 
memory effects present. The equation $d\rho^t = L(\rho^t)dt$
is called the \emph{master equation}.  
The following theorem of Lindblad characterizes the 
generator of a one-parameter semigroup of operations
on $\BC = M_n(\BB{C})$.

\begin{stel}\textbf{(Lindblad \cite{Lin})}
Let $\{T_t\}_{t\ge 0}$ be a semigroup of completely positive
identity preserving operators on $M_n(\BB{C})$ with generator 
$L$. Then there exist a self-adjoint element $H \in M_n(\BB{C})$
and elements $V_j \in M_n(\BB{C})$ for $j = 1,2,\ldots,k$ with 
$k \le n^2$, such that
  \begin{equation}\label{eq genLin}
  L(X) = i[H,X] + \sum_{j=1}^k V_j^*XV_j - 
  \frac{1}{2}\{V_j^*V_j, X\}, \ \ \ \ X \in M_n(\BB{C}),
  \end{equation}
where $\{X,Y\}$ stands for the anti-commutator $XY+YX$. 
Conversely, every operator $L$ of this form generates a 
semigroup of completely positive identity preserving operators.  
\end{stel}
Lindblad's result \cite{Lin} is actually more general, it 
is valid for norm-continuous semigroups on the algebra 
of bounded operators $\BC(\BB{H})$ for some, possibly infinite
dimensional, Hilbert space $\BB{H}$. The commutator with 
$H$ describes the evolution generated by some system 
Hamiltonian and the part with the $V_j$'s describes the
decay into the field. The $V_j$'s are determined through 
equation \eqref{eq Mark} and the expression for 
$H_{sr}$ in quantum electrodynamics. 

Another, more rigorous, approach to the master equation is via the 
\emph{weak coupling limit}. In this limit the Hamiltonian
$H$ of equation \eqref{eq QEDHam} is replaced by 
  \begin{equation*}
  H^\lambda = H_s + H_r + \lambda H_{sr}.
  \end{equation*}
While the coupling constant $\lambda$ is sent to
$0$, the time variable $t$ in the interaction picture 
has to be scaled to $\tau := \frac{t}{\lambda^2}$
to compensate the slower decay of the system, i.e.\ for $X \in \BC$
and $Z \in \BC \ten \WC$ we define
  \begin{equation*}\begin{split}
  &\tilde{\rho}^t_\lambda(X) := 
  \tilde{\chi}^{\tau}_\lambda(X \ten I_r),\ \  \mbox{where} \ \
  \tilde{\chi}^\tau_\lambda(Z) := \rho\ten\gamma
  \big(U^{\lambda*}_\tau
  ZU^\lambda_\tau\big), \ \  \mbox{with} \\
  &U^\lambda_\tau := 
  \exp\big(\frac{i\tau}{\hbar}(H_s+H_r)\big)
  \exp\big(-\frac{i\tau}{\hbar}H^\lambda\big).
  \end{split}\end{equation*}
Davies \cite{Dav2} showed that under some technical assumption
the time-evolution $T_t^\lambda$, given 
by $T_t^{\lambda*}(\rho) := \tilde{\rho}^t_\lambda$,
uniformly converges to a semigroup $T_t$ when 
$\lambda$ goes to $0$.

Let us assume that the electromagnetic field is initially 
in the vacuum state, i.e.\ $\gamma =\phi$. Let $T_t$
be the semigroup describing the irreversible evolution
of the system in the weak coupling limit. Its generator $L$
can be written in the form of equation \eqref{eq genLin}.
Accardi, Frigerio and Lu showed, see \cite{AFLu}, that 
when $\lambda$ goes to $0$, $U^\lambda_\tau$ converges in distribution 
to the solution $U_t$ of the quantum stochastic differential 
equation
  \begin{equation*}
  dU_t = \big\{V_jdA_j^*(t) - V_j^*dA_j(t) -(iH + \frac{1}{2}V_j^*V_j)dt\big\}U_t, 
  \end{equation*}
where repeated indices are summed. In this context the weak coupling limit 
is also called the \emph{stochastic limit}. Just to support this 
result, let us check that $\mbox{Id}\ten\phi(U_t^*X\ten\I_rU_t)$ is 
the semigroup $T_t$. Using the continuous tensor product structure
of the Fock space on which $U_t$ is defined, it is not hard to see that 
the expression $\mbox{Id}\ten\phi(U_t^*X\ten\I_rU_t)$ defines a semigroup, i.e.\
we only have to show that it is generated by $L$. Since the noises $A_j$
and $A_j^*$ are independent for different $j$ and vacuum expectations
of $A^*_j$ and $A_j$ are zero, it follows from 
Ito's rules that
  \begin{equation*}\begin{split}
  &d\mbox{Id}\ten\phi(U_t^*X\ten\I_rU_t) = 
  \mbox{Id}\ten\phi\big(d(U_t^*X\ten\I_rU_t)\big)= \\
  &\mbox{Id}\ten\phi((dU_t^*)X\ten\I_rU_t) +\mbox{Id}\ten\phi(U_t^*X\ten\I_rdU_t) +
  \mbox{Id}\ten\phi((dU_t^*)X\ten\I_rdU_t) = \\
  &(iHX -\frac{1}{2}V_j^*V_jX)dt + (-iXH -\frac{1}{2}XV_j^*V_j)dt + V_j^*XV_jdt =L(X)dt.
  \end{split}\end{equation*}
This result was already obtained by Hudson and Parthasarathy in \cite{HuP}.

In this thesis we will always take the semigroup and its corresponding 
quantum stochastic differential equation, describing the interaction of 
the system $\BC$ and the electromagnetic field $\WC$ in the vacuum, as the 
starting point. The Markov approximation or the weak coupling limit is 
a very good approximation for many phenomenon in quantum optics. Barchielli 
\cite{Bar} was one of the first to see the relevance of quantum stochastic calculus
for quantum optics. See for instance \cite{Bares} for a description 
of the electron shelving effect with quantum stochastic calculus. In 
\cite{RoM} the Mollow spectrum of fluorescence of a driven two 
level atom is derived using quantum stochastic calculus.

\section{Filtering and control, outline of results}\label{sec fandc}

In this thesis a quantum system 
$\BC$ in interaction with the electromagnetic
field $\WC$ is studied in the weak coupling limit. 
For simplicity we always assume $\BC$ to be $M_n(\BB{C})$ for
some $n \in \BB{N}$. Next, an observable 
$Y_t$ of the field $\WC$ is measured continuously in time. For 
instance in Chapter \ref{ch davies} the 
photons emitted into the field by a laser driven 
two-level atom are counted continuously in time. Since the 
system $\BC$ interacts with the field $\WC$, we gain information
about the system $\BC$ when the observable $Y_t$ in the field 
is measured. For instance, when the field is in the vacuum state 
we can infer immediately after a photon appears in the field and 
is counted that the two-level atom is in the ground state. 
The central question in this thesis is how to condition 
the state of $\BC$ continuously in time on information 
gained by measuring some observable $Y_t$ in the field. 

Chapter \ref{ch davies} introduces the subject by analyzing a 
photon counting experiment, i.e.\ the observable $Y_t$ is the 
photon number operator $\Lambda(t)$ of the field. The system $\BC$
is a two-level atom driven by a laser. We explicitly write down 
the solution of the quantum stochastic differential equation 
using Maassen's integral-sum kernels. The conditioning is done 
by sandwiching with the projection corresponding to an observed event.
In this way a whole family of evolutions for the two-level atom 
is obtained, i.e.\ for every possible observed event $E$ in the 
photon counter there is a map that represents the reduced evolution
of the two-level system conditioned on this event $E$. We prove 
that this family of maps satisfies the axioms of the processes 
studied by Davies \cite{Dav1}, \cite{Dav} in the late sixties
and early seventies. We use his theory for these processes and 
the explicit solution, in terms of Maassen's kernels, of the 
quantum stochastic differential equation to calculate the jump 
operators that describe the evolution when a photon appears and is
detected in the field. This leads to a continuous evolution of
the two-level atom interrupted by jumps at the moments at which photons 
are detected. In quantum optics such an evolution is known as 
a quantum trajectory \cite{Car}. We use the trajectory evolution to 
show that the photons are detected according to a renewal process.     
 
In Chapter \ref{ch davies} the explicit solution of the 
quantum stochastic differential equation is used, i.e.\ 
we leave the strength of It\^o's calculus for increments unused.
In Chapter \ref{ch sseq} the conditioned evolution of the system 
$\BC$ is described infinitesimally, fully exploiting the quantum 
It\^o calculus summarized in Theorem \ref{thm Ito}. Some of 
the ideas of Chapter \ref{ch davies} directly carry over to this 
description. For every time $t \ge 0$ there is a measure 
space $(\Omega_t, \Sigma_t, \BB{P}_t)$ with $\Omega_t$ the 
set of all possible paths of the observed process up to time $t$.    
In both chapters we are interested in the \emph{consistency} of the 
family of measures $\{\BB{P}_t\}_{t \ge 0}$, i.e.\ for all $s \ge t$ and 
$E \in \Sigma_t:\ \BB{P}_s(E) = \BB{P}_t(E)$. Kolmogorov's extension
theorem then states that this family extends to a single probability
measure $\BB{P}$ on the paths observed up to infinity.  

The conditioning by sandwiching with projections in Chapter \ref{ch davies} 
is not that easily carried over to the infinitesimal setting 
of Chapter \ref{ch sseq}. In the third chapter this 
is done using the decomposition of a von Neumann algebra 
$\AC$, equipped with a normal state $\chi$ and represented 
on some Hilbert space $\BB{H}$, over its 
center $\CC := \{C \in \AC;\ CA=AC,\ \forall A \in \AC\}$. 
Theorem \ref{thm spectral} states that there exists
a probability space $(\Omega, \Sigma, \BB{P})$ such that 
$(\CC, \chi) \cong L^\infty(\Omega, \Sigma, \BB{P})$. The  
decomposition of $\AC$ over $\CC$ described below (see \cite{KaR} for 
proofs) is an extension of this result.

The Hilbert space $\BB{H}$ has a direct integral representation 
$\BB{H} = \int_\Omega^\oplus \BB{H}_\omega \BB{P}(d\omega)$ \cite{KaR} in 
the sense that there exists a family of Hilbert spaces 
$\{\BB{H}_\omega\}_{\omega \in \Omega}$ and for all $\psi \in \BB{H}$
there exists a measurable map $\omega \mapsto \psi_\omega \in \BB{H}_\omega$
such that
  \begin{equation*}
  \langle \psi, \phi\rangle = \int_\Omega \langle \psi_\omega, \phi_\omega\rangle
  \BB{P}(d\omega), \ \ \ \ \psi, \phi \in \BB{H}.
  \end{equation*} 
The von Neumann algebra $\AC$ has a \emph{central decomposition}
$\AC = \int_\Omega^\oplus \AC_\omega \BB{P}(d\omega)$ \cite{KaR} in the sense that 
there exists a family $\{\AC_\omega\}_{\omega \in \Omega}$ of 
von Neumann algebras with trivial center, called \emph{factors}, 
and for all $A \in \AC$ there exists a measurable map 
$\omega \mapsto A_\omega \in \AC_\omega$ such that 
$(A\psi)_\omega = A_\omega \psi_\omega$ for all $\psi \in \BB{H}$
and for $\BB{P}$-almost all $\omega \in \Omega$. The state $\chi$ on 
$\AC$ has a decomposition in states $\chi_\omega$ on $\AC_\omega$ \cite{KaR}
such that for all $A \in \AC$ its expectation is given by
  \begin{equation*}
  \chi(A) = \int_\Omega \chi_\omega(A_\omega)\BB{P}(d\omega).
  \end{equation*}
We think of the state $\chi$ and an arbitrary operator $A \in \AC$
as maps $\chi_\bullet:\ \omega \mapsto \chi_\omega$ and 
$A_\bullet:\ \omega \mapsto A_\omega$, respectively.  

Let $t \ge 0$  and let $\rho$ be a state on $\BC$. Define a time evolved 
state $\chi^t(Z) := \rho\ten\phi(U_t^*ZU_t),\ Z \in \BC\ten\WC$ in 
the interaction picture, with $U_t$ the solution of the quantum
stochastic differential equation describing the interaction between
$\BC$ and $\WC$. Define a commutative algebra $\CC_t$ as the 
algebra generated by the process $\{Y_s\}_{s\ge 0}$ up to 
time $t$, i.e.\ $\CC_t$ is generated by the set $\{Y_s;\ 0\le s\le t\}$.
Because of the consistency and Kolmogorov's extension theorem 
there exists an increasing family of $\sigma$-algebras $\{\Sigma_t\}_{t\ge 0}$
such that $(\CC_t,\chi^t) \cong L^\infty(\Omega, \Sigma_t, \BB{P_\rho})$
where $\Omega$ is the set of all paths of the observed process 
in the field $\{Y_s\}_{s\ge 0}$ up to infinity. 

We define a subalgebra $\AC_t$ of $\BC\ten\WC$ as the commutant of $\CC_t$,
i.e.\ $\AC_t := \CC_t'$, and we restrict $\chi^t$ to this algebra $\AC_t$. 
Now $\CC_t$ is the center of $\AC_t$ and we can decompose the state $\chi^t$
over this center. In this way we get a random state $\chi^t_\bullet$ from 
$\Omega$ to the states on $\AC_t$ which for each $t \ge 0$ is 
measurable with respect to $\Sigma_t$.
Since $\BC\ten\I_\WC$ is a subalgebra of $\AC_t$ for all $t\ge 0$ and since
$(X\ten\I_\WC)_\bullet$ is the constant map $\omega \mapsto X$, we 
obtain a random reduced state $\rho^t_\bullet$ from $\Omega$ to 
the states on $\BC$ given by
  \begin{equation*}
  \rho_\omega^t(X) := \chi^t_\omega\big((X\ten\I_\WC)_\omega\big),\ \ \ \
  \omega \in \Omega,\ X \in \BC.
  \end{equation*}  
The state $\rho_\omega^t$ is the state of the system $\BC$ after 
$t$ seconds of time evolution conditioned on having observed the 
first $t$ seconds of the path $\omega$. The state $\rho^t_E$ on 
$\BC$ after $t$ seconds of time evolution conditioned 
on some event $E \in \Sigma_t$ is then given by
  \begin{equation*}
  \rho^t_E(X) = \int_E\rho^t_\omega(X)\BB{P}_\rho(d\omega),\ \ \ \ X \in \BC. 
  \end{equation*}
  
The goal of Chapter \ref{ch sseq} is to derive a 
stochastic differential equation for 
the state $\rho^t_\bullet$ in which the 
stochastic part of the equation is determined 
by the observed process $\{Y_s\}_{s\ge 0}$. 
As an example, the stochastic differential 
equation for the two-level atom of Chapter 
\ref{ch davies} when photons are counted in the 
field is given by
  \begin{equation}\label{eq Belfilt}
  d\rho^t_\bullet(X) = \rho^t_\bullet\big(L(X)\big)dt + 
  \Big(\frac{\rho^t_\bullet\big(\JC(X)\big)}{\rho^t_\bullet\big(\JC(\I)\big)}
  -\rho^t_\bullet(X)\Big)
  \big(dN_t - \rho^t_\bullet\big(\JC(\I)\big)dt\big), \ \ \ \ X\in\BC,
  \end{equation}   
where $L$ is the Lindblad generator of the semigroup evolution
when we would not be counting photons in the field, $N_t$ is 
the random variable from $\Omega \to \BB{N}$ that represents the number 
of photons counted up to time $t$, and $\JC$ is the jump operation 
of Chapter \ref{ch davies}. The stochastic differential equations 
derived in Chapter \ref{ch sseq} are called \emph{Belavkin equations} 
\cite{Bel0}, \cite{Bel}. They are known in quantum optics under the somewhat 
misleading name of \emph{stochastic Schr\"odinger equations} \cite{Car} since 
they are actually a stochastic variant of a master equation. 

Apart from the conditioning for which we use the central decomposition, 
our derivation of the Belavkin equation is 
rather close to the original derivation of Belavkin \cite{Bel}. However,
instead of using the explicit construction of the quantum 
conditional expectation encountered in that derivation, 
our proofs try to exploit its characterizing properties. 
Furthermore, we work in the interaction picture, which 
makes the derivation more accessible.

Let us return to the Belavkin equation, see the 
example in equation \eqref{eq Belfilt}.
It is an equation for the expectation $\rho^t_\bullet(X)$ 
of some system  operator $X$ conditional on the observed 
process $Y_t$ in terms of $dt$ and $dY_t$. We only 
have acces to the process $Y_t$ in the field and since 
the field is coupled to the system $\BC$ we indirectly gain 
information on $X$. This is similar to the situation 
in classical filter theory, developed mainly by 
R.L.~Stratonovich in the late 1950s, 
cf.\ \cite{Str}, \cite{Kal}. There one is 
interested in some system process $X_t$ but 
only has access to a process $Y_t$ which is 
the system process polluted with some noise. 
In classical filtering theory one derives a 
non-linear stochastic differential equation 
for the conditional expectation of $X_t$ on 
the observed process $Y_t$ in terms of $dt$
and $dY_t$, called the filtering equation. 
The Belavkin equation is a non-commutative
analogue of this equation and is therefore some times
called the \emph{quantum filtering equation}.

In Chapter \ref{ch sec} a similar setup with 
a system $\BC$ in interaction with the field $\WC$
is being studied. The problem is to control the state 
of the system $\BC$, i.e.\ we start with an unknown 
initial state $\rho$ of the system $\BC$ and then 
we try to control it in a way that depends on 
observation of some process $Y_t$ in the field, aiming
to keep the state of the system $\BC$ as close as possible 
to $\rho$. First the case is studied where the 
stochastic differential equation governing the interaction
between $\BC$ and $\WC$ contains only commutative 
noise terms. This special situation is called 
\emph{essentially commutative} \cite{KuM0}.

To control the state it is first evolved 
over a time interval $\tau$ according to the 
Belavkin equation for the measurement we are 
performing in the field, i.e.\
  \begin{equation*}
  \tilde{\rho}^\tau_\bullet = \rho^0 + \int_0^\tau d\rho^s_\bullet, 
  \end{equation*}
where the tilde is meant to indicate that $\tilde{\rho}^\tau_\bullet$ is the 
state of $\BC$ \emph{before} control. In the 
time interval $\tau$ some measurement result
has been obtained for the process observed in 
the field. We then correct with a unitary $U^\tau_c$
that depends on this result i.e.\ denoting the 
state of $\BC$ with its density matrix, we have
  \begin{equation*}
  \rho^\tau_\bullet = U^\tau_c \tilde{\rho}^\tau_\bullet U^\tau_c. 
  \end{equation*}
We derive a stochastic differential equation for the
controlled state evolution depending on the stochastic 
measurement process. It turns out that in the 
essentially commutative case it is possible 
to freeze the state evolution, i.e.\ $d\rho^t_\bullet =0$. 
Such a control scheme is said to 
\emph{restore quantum information} \cite{GrW}.

However, the interaction between the system $\BC$ and the field
can in general not be treated in an essentially commutative way. 
For instance, the interaction causing spontaneous decay of a two-level atom into 
a vacuum field is given by a quantum stochastic differential 
equation in which the two non-commuting noises $B^0_t = B(\chi_{[0,t)})$
and $B^{\pi/2}_t = B(i\chi_{[0,t)})$ of Section \ref{sec scFs} appear. 
The strategy when we are not in the essentially 
commutative case is to manipulate the field state 
such that the quantum stochastic differential equation 
resembles the essentially commutative situation more and more. 
This can be done by replacing the vacuum state of the field 
by a \emph{squeezed state}.

In the vacuum state the variances of $B^0_t$ and $B^{\pi/2}_t$
are both equal to $t$. In a squeezed state the variance of one of
these noises is decreased while the other one is increased,
since we still have to satisfy Heisenberg's uncertainty relation. 
The noise with the large variance has the biggest disturbing 
influence on the evolution of the system $\BC$, therefore this noise 
is being observed in the field and the control scheme is 
based on the measurement results of this observation. 
It turns out that in the limit for 
squeezing to infinity the situation where quantum 
information can be restored is refound.

\newpage
\thispagestyle{empty}

%% file: davies.tex
\chapter{The Davies process of resonance fluorescence}\label{ch davies}

\begin{center}
{\large Luc Bouten$^\dagger$ \ \ \ \ \ \ Hans Maassen$^\dagger$ \ \ \ \ \ \ Burkhard K\"ummerer}$^{\dagger\dagger}$\\
\vspace{1cm}
$^\dagger$\emph{Mathematisch Instituut, Katholieke Universiteit Nijmegen \\
Toernooiveld 1, 6526 ED Nijmegen, The Netherlands}\\
\vspace{0.5 cm}
$^{\dagger\dagger}$\emph{Fachbereich Mathematik, Technische Universit\"at Darmstadt \\
Schlo\ss gartenstra\ss e 7, 64289 Darmstadt, Germany}
\end{center}
\vspace{0.3cm}

{\small
\begin{center}\textbf{Abstract\footnote{This chapter is an adapted version of \cite{BMK}.}}\end{center}
Starting point is a given semigroup of completely
positive maps on the $2\times 2$ matrices. This
semigroup describes the irreversible evolution of
a decaying two-level atom.
Using the integral-sum kernel approach 
to quantum stochastic calculus we couple the
two-level atom to an environment, 
the electromagnetic field.
The irreversible evolution of the two-level atom
stems from the reversible time evolution of atom 
and field together. Mathematically speaking, we have
constructed a Markov dilation of the semigroup.\\
Next step is to drive the atom by a laser and
to count the photons emitted into the field by the 
decaying two-level atom. For every possible sequence of 
photon counts a map is constructed that gives the time 
evolution of the two-level atom inferred by that sequence.
This family of maps 
forms a so-called Davies process.
In his book Davies describes the structure 
of these processes, which brings us into the field 
of quantum trajectories. Within our model 
we calculate the jump operators and we briefly 
describe the resulting counting process.}

\section{Introduction}

In this paper we want to illustrate that quantum stochastic calculus together
with the processes studied by Davies in his book \cite{Dav}, and explained in his paper with 
Srinivas \cite{SrD}, form a suitable mathematically rigorous framework for doing 
quantum trajectory theory \cite{Car}. As an example we consider here the case of 
resonance fluorescence.                                                                         

Our starting point is a semigroup of transition operators $\{T_t\}_{t\ge 0}$ on
the algebra $M_2$ of all $2\times 2$-matrices. This semigroup describes the irreversible
evolution of a spontaneously decaying two-level atom in the Heisenberg picture. By
coupling the atom to a quantum noise, we construct a stationary 
quantum Markov process having precisely these transition operators. If we impose the
requirements that the external noise be a Bose field, and the quantum Markov process be
minimal, then the latter is uniquely determined. It is called the \emph{minimal Bose dilation}
of $(M_2, T_t, g)$ \cite{Kum2}, where $g$ is the ground state of the two-level atom.                     

Since this dilation is uniquely determined, any other reversible dynamical model which
couples $(M_2, T_t, g)$ to some Bose field necessarily contains this Bose dilation as a
subsystem. Therefore, without deriving our model from an explicit Schr\"odinger equation
(by performing a Markovian limit) we may safely assume it to be a physically correct way
to describe the interaction of the two-level atom with the electromagnetic field.               

We will couple the two-level atom to the electromagnetic field by using quantum stochastic
calculus \cite{Par}, \cite{Mey}. We use a version of quantum stochastic calculus based on
integral-sum kernels \cite{Maa}, \cite{LiM}, \cite{Mey}, which has the advantage that we
have an explicit construction for the solution of the quantum stochastic differential
equation with which we will describe the coupling of atom and field. 
Having this explicit construction in our hands is important for doing the actual calculations
we encounter later on.                                                                           

To be able to discuss resonance fluorescence a dilation with two
channels in the electromagnetic field is used. On one of them we will put a laser state to drive the 
two-level atom. We will call this field the \emph{forward channel} and the other one the \emph{side channel}.
Then photons are counted in both channels. We need the side channel, because we know
that there all detected photons are fluorescence photons. In the forward channel a detected photon
could just as well be coming directly from the laser.                                              

For every event that can occur in the photon counters a map is constructed that gives the
evolution of the two-level atom inferred by that event. We will see that the family of maps we
obtain, fulfills the axioms for the processes discussed by Davies \cite{Dav}. We have constructed
the Davies process of resonance fluorescence.        
                                               
Using the structure theory for Davies processes \cite{Dav} we can decompose the process into
its trajectories \cite{Car}. Within our model we calculate the expression for the
jump operators and for the time evolution in between jumps. Note that a jump in the system
occurs the moment we detect a photon, since our knowledge concerning the system changes.              
Using the above apparatus we show that the resulting counting process in the 
side channel is a so-called \emph{renewal process}.

\section{The dilation}

Let $M_2$, the algebra of $2 \times 2$-matrices, stand for the algebra of observables of
a two-level atom. On this algebra we are given a (continuous) semigroup $\{T_t\}_{t\ge 0}$
of completely positive maps. This semigroup describes the, generally irreversible,
evolution of the two-level atom. Lindblad's Theorem \cite{Lin} then says that $T_t = \exp{tL}$
where $L: M_2 \to M_2$ can be written as: for $X\in M_2$:
  \begin{equation}
  L(X) = i[H,X] + \sum_{j=1}^k V_j^*XV_j -\frac{1}{2}\{V_j^*V_j, X\}, 
  \end{equation}
where the $V_j$ and $H$ are fixed $2\times2$-matrices, $H$ being Hermitian. In this paper we will 
restrict to the simpler case where $H = 0$ and there are just two $V_j's$. This means there
is dissipation only into two channels, the forward channel described by $V_f$, and the side channel
described by $V_s$. We choose $V_f$ and $V_s$ such that: 
  \begin{equation*}
  V= \begin{pmatrix} 0 & 0  \\ 1 & 0 \end{pmatrix},\mbox{\ \ \ } V_f = \kappa_f V,
  \mbox{\ \ \ }V_s = \kappa_s V, \mbox{\ \ \ } |\kappa_f|^2 + |\kappa_s|^2 = 1.
  \end{equation*}
This exactly gives the time evolution for spontaneous decay to the ground state of the two-level atom
into two decay channels, where the decay rates are given by $|\kappa_f|^2$ and $|\kappa_s|^2$.        

We want to see this irreversible evolution of the two-level atom as stemming from a reversible evolution
of the atom coupled to, in this case, two decay channels in the field.
So let us first construct the algebra of observables for these channels in the field. 
Let $\mathcal{F}$ be the symmetric Fock space over the Hilbert space $L^2(\BB{R})$ of square
integrable wave functions on the real line, i.e.\ $\mathcal{F} := 
\BB{C}\oplus \bigoplus_{n=1}^\infty L^2(\BB{R})^{\ten_s n}$.
The electromagnetic field is given by creation and annihilation operators
on $\mathcal{F}$, generating the algebra of all bounded operators. We need two copies of this 
algebra, which we denote by $\WC_f$, which will be the forward channel, and $\WC_s$, 
which will be the side channel in the field.                                                                 

The evolution over a time $t$ of a free field is given by the second quantization of the
left shift, i.e.\ the second quantization of the operator on $L^2(\BB{R})$ which maps $f(\cdot)$ into
$f(\cdot + t)$. We denote the second quantization of this operator by $S_t$. This means that in
the Heisenberg picture we have an evolution on $\WC_f \ten \WC_s$ mapping $X$ into
$(S_{t}^*\ten S_{t}^*)X(S_{t}\ten S_{t})$ \big(= $(S_{-t}\ten S_{-t})X(S_t\ten S_t)$\big),
also denoted by $\Ad[S_t\ten S_t](X)$.                                                                  

The presence of the atom in the field introduces a perturbation on the evolution of the free
field. We let this perturbation be given by a certain family of unitary operators
$\{U_t\}_{t\in \BB{R}}$ on $\BB{C}^2\ten\mathcal{F}\ten\mathcal{F}$, to be specified later,
that forms a \emph{cocycle} with respect to the shift $S_t\ten S_t$, i.e.\ for all $t,s \in \BB{R}:\
U_{t+s} = (S_{-s}\ten S_{-s})U_t(S_{s}\ten S_{s})U_s$. Given this cocycle, we let the time evolution of
the atom and the field together be given by the following one-parameter group $\{\hat{T}_t\}_{t\in\BB{R}}$
(i.e.\ the evolution is now \emph{reversible}) of $*$-automorphisms on $M_2 \ten \WC_f \ten \WC_s$:
for all $X \in M_2 \ten \WC_f \ten \WC_s$:
  \begin{equation*}
  \hat{T}_t(X) = \left\{ \begin{array}{ll}
  U_t^{-1}(S_{-t}\ten S_{-t})X(S_t\ten S_t)U_t & \mbox{\ \ \ if \ } t \ge 0 \\
  (S_{-t}\ten S_{-t})U_{-t}X U^{-1}_{-t}(S_t\ten S_t) & \mbox{\ \ \ if \ } t < 0 
  \end{array}\right. ,
  \end{equation*}                                                                                      
  
The solution of the following quantum stochastic differential equation 
\cite{HuP}, \cite{Par} provides us with a cocycle of unitaries with respect to the shift:
  \begin{equation}\label{HuPeq}
  dU_t = \{V_f dA^*_{f,t} -V_f^* dA_{f,t} + V_s dA^*_{s,t} - V_s^* dA_{s,t} -\frac{1}{2}V^*Vdt\}U_t,
  \mbox{\ \ \ } U_0 = I.
  \end{equation}
In the next section we will give an explicit construction for the solution $U_t$ of this equation.
It can be shown (\cite{HuP}, \cite{Fri}, \cite{Maa}, \cite{Par}) that if the cocycle satisfies
equation \eqref{HuPeq} we have constructed a so-called \emph{quantum Markov dilation}
$(M_2\ten\WC_f\ten \WC_s, \{\hat{T}\}_{t\in \BB{R}}, \mbox{id}\ten\phi^{\ten 2})$ of the quantum
dynamical system $(M_2, \{T_t\}_{t \ge 0}, g)$ \cite{Kum1}, \cite{Kum2}, where $\phi$ is the
vector state on $\WC_{f,s}$ given by the vacuum vector. This means that the following
\emph{dilation diagram} commutes for all $t \ge 0$ (and that the resulting quantum process is Markov):
    \begin{equation}\label{dil_diag}\begin{CD}
  M_2 @>T_t>> M_2              \\
   @V{\Id \ten I^{\ten 2}}VV        @AA{\Id \ten \phi^{\ten 2}}A      \\
   M_2\ten \WC_f \ten \WC_s @>\hat{T}_t>> M_2\ten\WC_f \ten\WC_s            \\
  \end{CD}\end{equation}
i.e.\ for all $X \in M_2:\ T_t(X) = \big(\Id \ten \phi^{\ten 2}\big) 
\big(\hat{T}_t(X \ten I^{\ten 2})\big)$.  

The diagram can also be read in the Schr\"odinger picture if we reverse the arrows: 
start with a state $\rho$ of the atom $M_2$ in the upper right hand corner, 
then this state undergoes the following sequence of maps:
  \begin{equation*}
  \rho \mapsto  
  \rho \ten \phi^{\ten 2} \mapsto 
  (\rho \ten \phi^{\ten 2})\circ\hat{T}_t =\hat{T}_{t*}(\rho \ten \phi^{\ten 2})
  \mapsto 
  \mbox{Tr}_{\FC_f\ten\FC_s}\big(\hat{T}_{t*}(\rho \ten \phi^{\ten 2})\big).
  \end{equation*} 
This means that at $t=0$, the atom in state $\rho$ is coupled to the $k$ channels in the 
vacuum state, and after $t$ seconds of unitary evolution we take the 
partial trace over the $2$ channels.

\section{Guichardet space and integral-sum kernels}

Let us now turn to giving the explicit construction for the solution of equation \eqref{HuPeq}. 
For this we need the \emph{Guichardet space} $\Omega$ \cite{Gui} of $\BB{R}$, which is
the space of all finite subsets of $\BB{R}$, i.e.\ 
$\Omega := \bigcup_{n\in \BB{N}} \Omega_n$, where $\Omega_n := \{\sigma \subset \BB{R};\ |\sigma| = n \}$.
Let us denote by $\lambda_n$ the Lebesgue measure on $\BB{R}^n$.
If, for $n \in \BB{N}$, we let $j_n: \BB{R}^n \to \Omega_n$ denote the map that maps an $n$-tuple
$(t_1,t_2,\ldots,t_n)$ into the set $\{t_1,t_2, \ldots, t_n\}$, then we can define a measure $\mu_n$
on $\Omega_n$ by: $\mu_n(E) := \frac{1}{n!}\lambda_n\big(j_n^{-1}(E)\big)$ for all $E$ in the sigma field
$\Sigma_n$ of $\Omega_n$ induced by $j_n$ and the Borel sigma field of $\BB{R}^n$. Now we define a measure
$\mu$ on $\Omega$ such that $\mu(\{\emptyset\}) = 1$ and $\mu = \mu_n$ on $\Omega_n$. This means we have
now turned the Guichardet space into the measure space $(\Omega, \Sigma, \mu)$.                          

The key to constructing the solution of equation \eqref{HuPeq} is to identify the symmetric Fock space
$\mathcal{F}$ with the space of all quadratically integrable functions on the Guichardet space
$L^2(\Omega, \mu)$. To see this identification note that $L^2(\Omega_n, \mu_n)$ is, in the canonical
way, unitarily equivalent with the space of all quadratically integrable functions on $\BB{R}^n$
invariant under permutations of coordinates, denoted $L_{\mbox{sym}}^2(\BB{R}^n)$. It is now
obvious how to identify $\mathcal{F} = \BB{C} \oplus \bigoplus_{n = 1}^\infty L^2_{\mbox{sym}}(\BB{R}^n)$
with $L^2(\Omega, \mu) = \BB{C} \oplus \bigoplus_{n=1}^\infty L^2(\Omega_n, \mu_n)$.                        

For every $f \in L^2(\BB{R})$ we define the \emph{exponential vector} $e(f) \in \mathcal{F}$ in the
following way: $e(f) := 1 \oplus f \oplus \frac{1}{\sqrt{2}} f^{\ten 2} \oplus 
\frac{1}{\sqrt{6}} f^{\ten 3} \oplus \dots$. Note that the linear span of all exponential vectors
forms a dense subspace of $\mathcal{F}$. For every $f \in L^2(\BB{R})$ we define the \emph{coherent
vector} $\psi(f)$ to be the exponential vector of $f$ normalised to unity, i.e.\ $\psi(f) =
\exp(-\frac{1}{2}\p f\p^2)e(f)$. Under the above identification of $\mathcal{F}$ with
$L^2(\Omega, \mu)$, the exponential vector (of an $f \in L^2(\BB{R})$) $e(f)$ is mapped into an element
of $L^2(\Omega, \mu)$ which we denote by $\pi(f)$ and which is given by: $\pi(f):\ \Omega \to \BB{C}:\
\omega \mapsto \prod_{s \in \omega} f(s)$, where the empty product $\prod_{s \in \emptyset}f(s)$ is
defined to be $1$. We will often choose for $f$ the \emph{indicator function} of a certain interval
$I \subset \BB{R}$, which we denote by $\chi_I$. This is the function which is $1$ on $I$ and
$0$ elsewhere.                                                                                        

We are now ready to start the construction of the solution $U_t$ of equation $\eqref{HuPeq}$. Define
the \emph{integral-sum kernel} of $U_t$ (name will become apparent in a minute)
to be the map $u_t$ that maps four disjoint finite subsets of $\BB{R}, \sigma_f, \sigma_s, \tau_f, \tau_s$
(where $f$ and $s$ stand for "forward" and "side") to the following $2 \times 2$-matrix, where we write
$\sigma_f \cup \sigma_s\cup \tau_f \cup \tau_s$ also as $\{t_1, t_2, \ldots, t_k\}$ such that
$t_1 < t_2 < \ldots < t_k$ and $k \in \BB{N}$: 
  \begin{equation*}\begin{split}
  u_t(\sigma_f, \sigma_s, \tau_f, \tau_s) := &\pi(\chi_{[0,t]})(\sigma_f \cup \sigma_s\cup \tau_f \cup \tau_s)
  \exp(- \frac{t-t_k}{2}V^*V)V_k \times                                                      \\
  &\exp(- \frac{t_k -t_{k-1}}{2}V^*V)V_{k-1}\ldots V_1 \exp(-\frac{t_1}{2}V^*V),
  \end{split}\end{equation*}
where
  \begin{equation*}
  V_j = \left\{ \begin{array}{ll}
  V_f & \mbox{\ \ if \ } t_j \in \sigma_f \\
  -V_f^* & \mbox{\ \  if \ } t_j \in \tau_f \\
  V_s & \mbox{\ \ if \ } t_j \in \sigma_s \\
  -V_s^* & \mbox{\ \  if \ } t_j \in \tau_s
  \end{array}\right. .
  \end{equation*}
Then we have the following theorem of Maassen, see \cite{Maa}, \cite{LiM}:
\begin{stel}\label{Maassen}
After identifying $\BB{C}^2 \ten \mathcal{F}\ten\mathcal{F}$ with $L^2_{\BB{C}^2}(\Omega\times\Omega, \mu\times\mu)$,
the space of all square integrable functions on $\Omega \times \Omega$ with values in $\BB{C}^2$,
the solution $U_t:\ L^2_{\BB{C}^2}(\Omega\times\Omega, \mu\times\mu) \to L^2_{\BB{C}^2}(\Omega\times\Omega, \mu\times\mu)$
of equation \eqref{HuPeq} is given by:
  \begin{equation*}
  (U_tf)(\omega_f, \omega_s) = \sum_{\substack{\sigma_f\subset\omega_f\\ \sigma_s\subset\omega_s}}
  \int_{\Omega\times\Omega} u_t(\sigma_f,\sigma_s,\tau_f,\tau_s)               
  f\big((\omega_f\backslash\sigma_f)\cup\tau_f, (\omega_s\backslash\sigma_s)\cup\tau_s\big)
  d\tau_f d\tau_s.
  \end{equation*} 
\end{stel}

Now we have an explicit expression for the time evolution $\hat{T}_t = \Ad[\hat{U}_t]$, where $\hat{U}_t$
is given by $(S_t\ten S_t)U_t$ if $t\ge 0$ and $U_{-t}^{-1}(S_t\ten S_t)$ if $t <0$. 
The family $\{\hat{U}_t\}_{t\in \BB{R}}$
forms a group of unitary operators on $\BB{C}^2 \ten \mathcal{F}\ten\mathcal{F}$ describing the time
evolution of the two-level atom and the two channels in the field together. Stone's Theorem says that there must be
a Hamiltonian associated to this time evolution. This Hamiltonian has been calculated recently \cite{Gre1}, \cite{Gre}.

\section{The Davies process}

We now return to the situation in figure \ref{dil_diag}. We wish to make some changes in this diagram
and for this we need to introduce some more notation regarding Guichardet spaces. Let $I \subset \BB{R}$
be an interval. Then the \emph{Guichardet space of $I$} is the set $\Omega(I) =
\bigcup_{n=0}^\infty \Omega_n(I)$, where $\Omega_n(I) = \{\sigma \subset I;\ |\sigma| = n\}$. In
a similar way as for $\Omega$, which is $\Omega(\BB{R})$, we can provide these sets with a measure structure:
$(\Omega(I), \Sigma(I), \mu)$. Given a subset $E$ of $\Omega(I)$ in the sigma field $\Sigma(I)$, we
can construct the projection $M_{\chi_E}:\ L^2(\Omega, \mu) \to L^2(\Omega, \mu):\ f \mapsto \chi_E f$.  

Let $I$ be $[-t,0)$, then the events in $\Sigma\big([-t,0)\big)$, which we abbreviate to $\Sigma_t$,
are events in the output field of the atom up to time $t$. Remember that the evolution of the free
field was given by the left shift and that the atom is sitting in the origin. 
Since the Guichardet space representation corresponds to the photon number picture, we can give concrete
interpretations to the subsets in $\Sigma_t$. For instance, the subsets $\Omega_n\big([-t,0)\big)$,
correspond to the events "there are n photons in the output of the atom into this channel of the 
field up to time t".        

Now back to the situation in figure \ref{dil_diag}. Suppose we have been observing the output in the
forward and side channel of the atom up to time $t$ with two photon counters. Then we are given two
events $E_f$ and $E_s$ in $\Sigma_t$. Since we know the outcome of the measurements we
have to change the time evolution of the two-level atom, i.e.\ we have to project onto the observed events
(see also \cite{BaB}).
This is summarized in the following figure:       
  \begin{equation*}\begin{CD}
  M_2 @>\EC^t_0[E_f, E_s]>> M_2              \\
   @V{\Id \ten \chi_{E_f} \ten \chi_{E_s}}VV        @AA{\Id \ten \phi^{\ten 2}}A      \\
   M_2\ten \WC_f \ten \WC_s @>\hat{T}_t>> M_2\ten\WC_f \ten\WC_s            \\
  \end{CD}\end{equation*}
where we have suppressed the capital letters $M$ in the projections. The map $\EC_0^t[E_f, E_s]:\ M_2 \to M_2:\
X \mapsto  \Id \ten \phi^{\ten 2}\big(\hat{T}_t(X\ten\chi_{E_f} \ten \chi_{E_s})\big)$ is
the unnormalized time evolution of the two-level atom in the Heisenberg picture given that we see
event $E_f$ in the output of the forward channel and event $E_s$ in the output of the side channel.
If we are given a state on $M_2$, i.e.\ a $2 \times 2$ density matrix $\rho$, then the probability
of seeing event $E_f$ in the forward channel and $E_s$ in the side channel after $t$ seconds of
observation is given by: $\BB{P}^t_\rho[(E_f, E_s)] = \mbox{Tr}\big(\rho\EC_0^t[E_f, E_s](I)\big)$.     

The setting is still not complete for describing resonance fluorescence. Since we are not driving the
atom, both the forward and the side channel fields are in the vacuum state, at most one photon can appear 
in the output. We change this by putting on the forward channel a coherent state with
amplitude $z \in \BB{C}$, defined by: $\gamma_{z_t}:\ \WC \to \BB{C}:\ X \mapsto
\exp(-t|z|^2)\big\langle \pi(z\chi_{[0,t]}), X\pi(z\chi_{[0,t]})\big\rangle$. Note that $\gamma_0$ is
the vacuum state. Putting a coherent state on the forward channel mimics a laser driving the atom. 
We have suppressed its oscillations for the sake of simplicity. Now we are ready to do 
resonance fluorescence, i.e.\ the diagram has changed into:
  \begin{equation*}\begin{CD}
  M_2 @>\EC^t_z[E_f, E_s]>> M_2              \\
   @V{\Id \ten \chi_{E_f} \ten \chi_{E_s}}VV        @AA{\Id \ten \gamma_{z_t} 
\ten \gamma_0}A      \\
   M_2\ten \WC_f \ten \WC_s @>\hat{T}_t>> M_2\ten\WC_f \ten\WC_s            \\
  \end{CD}\end{equation*} 
where the map $\EC_z^t[E_f, E_s]:\ M_2 \to M_2$ is now defined by $\EC_z^t[E_f, E_s](X) :=
\Id \ten \gamma_{z_t} \ten \gamma_0 \big(\hat{T}_t(X\ten\chi_{E_f} \ten \chi_{E_s})\big)$. It describes
the unnormalized time evolution of the laser-driven atom given that we see event $E_f$ in the output of
the forward channel and event $E_s$ in the output of the side channel. Given a state $\rho$ of the atom,
the probability of seeing event $E_f$ in the forward channel and $E_s$ in the side channel after $t$ seconds of
observation is now given by: $\BB{P}^t_\rho[(E_f, E_s)] = \mbox{Tr}\big(\rho\EC_z^t[E_f, E_s](I)\big)$.
To make the notation lighter we suppres the $z$ in $\EC^t_z$ in the following.                        

Since $L^2(\Omega, \Sigma, \mu) \ten L^2(\Omega, \Sigma, \mu)$ is canonically isomorphic to
$L^2 (\Omega\times\Omega, \Sigma \ten \Sigma, \mu \times\mu)$ we can simplify our notation even a bit
further. By identifying these spaces we can write $\EC^t[E_f, E_s] = \EC^t[E_f\times E_s]$, where
the righthandside is defined by: for all $E \in \Sigma_t \ten \Sigma_t, X \in M_2, t\ge 0:\
\EC^t[E](X) := \Id \ten \gamma_{z_t,0} \big(\hat{T}_t(X\ten\chi_E)\big)$, where $\gamma_{z_t,0}$
is an abbreviation for $\gamma_{z_t}\ten \gamma_0$. We will now study the properties of the family 
of maps we defined.

\begin{stel}\label{properties} 
The family of maps $\{\EC^t[E]\}_{t\ge0, E \in \Sigma_t \ten \Sigma_t}$ satisfies
the axioms of a \emph{Davies process}, \cite{Dav}: 
\begin{enumerate}
  \item For all $t \ge 0$ and $E \in \Sigma_t \ten \Sigma_t$, $\EC^t[E]$ is completely positive.       
  \item  For all $t \ge 0$ and all countable collections of disjoint sets 
               $\{E_n\}$ in $\Sigma_t \ten \Sigma_t$ \\ 
               and for all $X \in M_2:\
               \EC^t\Big[\bigcup_{n}E_n\Big](X) = \sum_{n} \EC^t[E_n](X)$.
  \item  For all $t \ge 0$ we have 
               $\EC^t\Big[\Omega\big([-t,0)\big)\times\Omega\big([-t,0)\big)\Big] (I) = I$.
  \item  For all $X \in M_2:\ \lim_{t \to 0}
               \EC^t\Big[\Omega\big([-t,0)\big)\times\Omega\big([-t,0)\big)\Big] (X) = X$.
  \item  For all $t,s \ge 0$ and $E \in \Sigma_s \ten \Sigma_s, 
               F \in \Sigma_t \ten \Sigma_t$ and all $X \in M_2$ we have:\\
               $\EC^t[F]\circ\EC^s[E](X) = \EC^{s+t}[F-s \tilde{\cup} E] (X)$, \\
               where $F-s \in \Sigma\big([-t-s, -s) \ten \Sigma\big([-t-s, -s) \big)$
               is given by:                                                              \\
               $F-s = \{(f_f-s,f_s-s);\ (f_f,f_s) \in F)\}$ and $\tilde{\cup}$
               is defined by:                                                    \\
               $A \tilde{\cup} B = \{(\omega_f \cup \sigma_f,\omega_s \cup \sigma_s);\ 
               (\omega_f,\omega_s)\in A, (\sigma_f,\sigma_s) \in B\}$.
\end{enumerate}
\end{stel}

\begin{proof}
The only point where there is really something to prove is point $5$.
Let us first introduce some short notation which we shall only use in this proof.
Let $\pi(z_t, 0)$ denote $\pi(z\chi_{[0,t]}) \ten \pi(0)$ and denote
$S_t \ten S_t$ just by $S_t$. Further we use the notation $\sigma_t(U_s)$ for $S_{-t}U_sS_t$.
Then for all $X \in M_2$, $s,t \ge 0$, $E\in \Sigma_s\ten\Sigma_s$ and $F \in \Sigma_t\ten \Sigma_t$ we have:
  \begin{equation*}\begin{split}
  &\frac{\EC^t[F]\circ\EC^s[E](X)}{\exp(-(s+t)|z|^2)} = 
     \EC^t[F]\Big(\big\langle\pi(z_s,0),\hat{T}_s(X\ten\chi_E)    
      \pi(z_s,0)\big\rangle\Big)\exp(t|z|^2) =                                      \\
  &\Big\langle\pi(z_t,0), \hat{T}_t\Big(\big\langle\pi(z_s,0),\hat{T}_s(X\ten\chi_E)
      \pi(z_s,0)\big\rangle\ten\chi_F\Big)\pi(z_t,0)\Big\rangle =                            \\
  &\Big\langle\pi(z_t,0), U_t^*\big\langle\pi(z_s,0),\hat{T}_s(X\ten\chi_E)    
      \pi(z_s,0)\big\rangle\ten\chi_{F+t} U_t\pi(z_t,0)\Big\rangle =                            \\
  &\Big\langle\pi(z_t,0), U_t^*\big\langle S_{-t}\pi(z_s,0),S_{-t}\hat{T}_s(X\ten\chi_E)
      S_t S_{-t}\pi(z_s,0)\big\rangle\ten\chi_{F+t} U_t\pi(z_t,0)\Big\rangle =                            \\
  &\Big\langle\pi(z_t,0), U_t^*\big\langle S_{-t}\pi(z_s,0),\sigma_t(U_s)^*X\ten\chi_{E+t+s}
      \sigma_t(U_s) S_{-t}\pi(z_s,0)\big\rangle\ten\chi_{F+t} U_t\pi(z_t,0)\Big\rangle.
  \end{split}\end{equation*} 
Now we use the cocycle identity and the continuous tensor product structure of the symmetric Fock
space to obtain:
  \begin{equation*}\begin{split}
  &\frac{\EC^t[F]\circ\EC^s[E](X)}{\exp(-(s+t)|z|^2)} = 
  \Big\langle \pi(z_{t+s},0), (\sigma_t(U_s)U_t)^* 
    X \ten \chi_{F+t \tilde{\cup} E+t+s}\sigma_t(U_s)U_t\pi(z_{t+s},0)\Big\rangle  = \\
  &\Big\langle \pi(z_{t+s},0), U_{t+s}^*
    X \ten \chi_{F+t \tilde{\cup} E+t+s}U_{t+s}\pi(z_{t+s},0)\Big\rangle = \\ 
  &\Big\langle \pi(z_{t+s},0), \hat{T}_{t+s}
    (X \ten \chi_{F-s \tilde{\cup} E})\pi(z_{t+s},0)\Big\rangle = 
  \frac{\EC^{s+t}[F-s \tilde{\cup} E](X)}{\exp(-(s+t)|z|^2)}.
  \end{split}\end{equation*}
\end{proof}

Define maps $Y_t: M_2 \to M_2: X \mapsto \EC^t\big[\{(\emptyset,\emptyset)\}\big](X)$. They represent 
the evolution of the atom when it is observed that no photons entered the decay channels. Then  
for all $t,s \ge 0$ 
  \begin{equation*} 
  Y_tY_s = \EC^t\big[\{(\emptyset,\emptyset)\}\big]\circ
  \EC^s\big[\{(\emptyset, \emptyset)\}\big]= \EC^{t+s}\big[\{(\emptyset, 
  \emptyset)\}-s \tilde{\cup}\{(\emptyset, \emptyset)\}\big] =
  \EC^{t+s}\big[\{(\emptyset, \emptyset)\}\big] = Y_{t+s}, 
  \end{equation*}
i.e.\ the family $\{Y_t\}_{t\ge 0}$ forms
a semigroup. Now observe that for $X \in M_2$ and $t \ge 0$ 
  \begin{equation*}\begin{split}
  &Y_t(X) = \EC^t\big[(\{\emptyset\},\{\emptyset\})\big](X) = 
  \Id\ten \gamma_{z_t}\ten \gamma_0\big(\hat{T}_t(X\ten\chi_{\{\emptyset\}}\ten\chi_{\{\emptyset\}})\big) = \\
  &\big\langle \pi(z_t)\ten\pi(0),U_t^*X\ten\chi_{\{\emptyset\}}\ten\chi_{\{\emptyset\}}
  U_t\pi(z_t)\ten\pi(0)\big\rangle\exp(-t|z|^2) =                                                            \\
  &\big(U_t\pi(z_t)\ten\pi(0)\big)^*(\emptyset,\emptyset)X\big(U_t\pi(z_t)\ten\pi(0)\big)
   (\emptyset,\emptyset)\exp(-t|z|^2),
  \end{split}\end{equation*}
i.e. we can write $Y_t(X) = B_t^*XB_t$ where $B_t$ is 
given by 
  \begin{equation}\label{Bt}
  B_t = 
  \Big(\exp(-\frac{1}{2}t|z|^2)\big(U_t\pi(z_t)\ten\pi(0)\big)\Big)
  (\emptyset,\emptyset).
  \end{equation}
Using Theorem \ref{Maassen} it follows that $B_t$ is the following semigroup
of contractions:
  \begin{equation*}
  B_t = \exp\big(-\frac{1}{2}(|z|^2I_2 + V^*V +2zV_f^*)t\big),
  \end{equation*}
If $Y_t$ can be written as $B_t^*\cdot B_t$ for some semigroup of contractions $B_t$,
the Davies process $\EC^t$ is called \emph{ideal}, see \cite{Dav}. 

\begin{lem}\label{lem bir}
The Davies process $\EC^t$ has \emph{bounded interaction rate} i.e.\ there exists a 
constant $K$ such that for all $t \ge 0$
  \begin{equation*}
  \EC^t\big[\Omega[-t,0)\times \Omega[-t,0)\backslash \{(\emptyset, \emptyset)\}\big] (I) \le tKI.
  \end{equation*}
\end{lem}
\begin{proof}                                                                                       
From Theorem \ref{properties} point $2$ it follows that
  \begin{equation*}
  \EC^t\big[\Omega[-t,0)\times \Omega[-t,0)\backslash \{(\emptyset, \emptyset)\}\big] (I) = I - B_t^*B_t,
  \end{equation*}
with $B_t$ as in equation \eqref{Bt}. Using Theorem \ref{Maassen} we find
  \begin{equation*}
  B_t = \exp(-\frac{1}{2}t|z|^2)\begin{pmatrix}\exp(-\frac{t}{2}) & 
  2z\overline{\kappa}_f\big(\exp(-\frac{t}{2})-1\big) \\ 0 & 1\end{pmatrix}.
  \end{equation*} 
Therefore
  \begin{equation*}\begin{split}
  &\ \ \ \ \ \ \ \  \ \ \ \ \ \ \ \ \ \ \ \ 
   \EC^t\big[\Omega[-t,0)\times \Omega[-t,0)\backslash \{(\emptyset, \emptyset)\}\big] (I) = \\
  &\begin{pmatrix}1-e^{-t(|z|^2+1)} & 
   2z\overline{\kappa}_f e^{-t(|z|^2+\frac{1}{2})}\big(1-e^{-\frac{t}{2}}\big) \\
  2\overline{z}\kappa_fe^{-t(|z|^2+\frac{1}{2})}\big(1-e^{-\frac{t}{2}}\big) & 
  1 -e^{-t|z|^2}\Big(1 + 4|z|^2|\kappa_f|^2\big(1-e^{-\frac{t}{2}}\big)^2\Big)\end{pmatrix}.
  \end{split}\end{equation*}
Since $\EC^t\big[\Omega[-t,0)\times \Omega[-t,0)\backslash \{(\emptyset, \emptyset)\}\big]$ is
completely positive this matrix has to be positive, i.e.\ its eigenvalues are positive. Let us denote
them by $E_+$ and $E_-$ such that $E_+ \ge E_- \ge 0$. Define $\alpha := \exp(-t|z|^2)$ and 
$\beta := \exp(-\frac{t}{2})$. Then for all $t\ge 0$
  \begin{equation*}
  \alpha \ge 0, \ \ \ \ \ \ \ 1-\beta \ge 0, \ \ \ \ \ \ \  
  0 \le 1-\alpha \le |z|^2t, \ \ \ \ \ \ \  0 \le 1-\alpha\beta^2 \le (|z|^2+ 1)t.
  \end{equation*} 
Therefore, for all $t \ge 0$
  \begin{equation*}\begin{split}
  & \EC^t\big[\Omega[-t,0)\times \Omega[-t,0)\backslash \{(\emptyset, \emptyset)\}\big] (I)\ \le \ 
  E_+ I \ \le \\ 
  &(E_+ + E_-) I =                                                        
  \mbox{Tr}\Big(\EC^t\big[\Omega[-t,0)\times \Omega[-t,0)\backslash \{(\emptyset, \emptyset)\}\big] (I)\Big) = \\ 
  &\big((1-\alpha\beta^2) + (1-\alpha) - 4\alpha|z|^2|\kappa_f|^2(\beta-1)^2\big) I  \ \le                  \\
  &\big((1-\alpha\beta^2)+ (1-\alpha)\big)I \ \le \ (2|z|^2+ 1)t I.
  \end{split}\end{equation*}
\end{proof}

\section{Quantum trajectories}

In the seventies Davies studied the structure of what we now call Davies processes \cite{Dav}.
Let us first state his results, as far as relevant, in the context of the process we are studying.
\begin{lem}\label{lemma}\textbf{(Davies \cite{Dav})}
Given any ideal Davies process $\EC^t$ with bounded interaction rate, as defined in the previous
section, we have existence of the following limits:
  \begin{equation*}
  \JC_f := \lim_{t \downarrow 0} \frac{1}{t}\EC^t\big[\Omega_1[-t,0), \{\emptyset\}\big] \mbox{\ \ and \ \ }
  \JC_s := \lim_{t \downarrow 0} \frac{1}{t}\EC^t\big[\{\emptyset\},\Omega_1[-t,0)\big].
  \end{equation*}
\end{lem}  
These completely positive maps represent the action we have to apply on the two-level atom the moment
we see one photon appear in the forward, respectively side channel.They are the \emph{jump operations} for
these channels. We will explicitly calculate these limits later on, but first we turn our attention to
decomposing the Davies process into its trajectories \cite{Car}. For this we need the following
definition. 
\begin{de}\label{W}
Let $Y_t: M_2 \to M_2$ be the maps from the previous section, i.e. $Y_t = \EC^t[\{\emptyset\},\{\emptyset\}]$
and let $\JC_f$ and $\JC_s$ be the maps from lemma \ref{lemma}. Let $\omega_f$ and $\omega_s$ be disjoint
elements of $\Omega[-t,0)$ and denote $\omega_f \cup \omega_s$ also as $\{t_1, \ldots, t_k\}$ where
$-t \le t_1 < t_2 < \ldots < t_k \le 0$ for a $k \in \BB{N}$. Then we define:
  \begin{equation*}
  W_{Y,\JC_f, \JC_s}(\omega_f, \omega_s)
         := Y_{t_1+t}\JC^{t_1}Y_{t_2-t_1}\JC^{t_2}\ldots \JC^{t_k}Y_{-t_k}, 
  \end{equation*}
where $\JC^{t_i}$ denotes $\JC_s$ if $t_i \in \omega_s$ and $\JC_f$ if $t_i \in \omega_f$.
\end{de}
Since $Y_t$ is the time evolution of the system when, both in the forward and the side channels,
no photons are detected and $\JC_f$ and $\JC_s$ are the jump operations that we have to apply when a 
photon in the corresponding channels appears, it is clear that the string of maps 
$Y_{t_1+t}\JC^{t_1}Y_{t_2-t_1}\JC^{t_2}\ldots \JC^{t_k}Y_{-t_k}$ represents the
\emph{trajectory} of an observable $X$ in $M_2$ when we find the outcomes $\omega_f$ in the
forward and $\omega_s$ in the side channel during our counting experiment. The following theorem
of Davies \cite{Dav} shows how to decompose the Davies process into its trajectories.
\begin{stel}\textbf{(Davies \cite{Dav})}\label{deco} 
Given any ideal Davies process $\EC^t$ with bounded interaction rate, as defined in the previous
section, we have for all $t \ge 0$, $E_f, E_s \in \Sigma_t$ and $X \in M_2$:
  \begin{equation*}
  \EC^t[E_f,E_s](X) = \int_{E_f \times E_s} W_{Y,\JC_f, \JC_s}(\omega_f, \omega_s) (X)d\mu(\omega_f)d\mu(\omega_s).
  \end{equation*}
\end{stel}
In the previous section we already found the expression for the time evolution in between jumps: $Y_t$. We
now turn to the calculation of $\JC_f$ and $\JC_s$. For all $X$ in $M_2$ we have:
  \begin{equation*}
  \JC_f(X) = \lim_{t\downarrow 0} \frac{1}{t} \EC^t\big[(\Omega_1[-t,0),\{\emptyset\})\big](X)
  = \lim_{t\downarrow 0} \frac{\int_0^t\Ad\big[U_t\pi(z)\ten\pi(0)(\{s\},\emptyset)\big](X)ds}
       {t\exp(-t|z|^2)}.                                  
  \end{equation*}
Now look at $U_t\pi(z)\ten\pi(0)(\{s\},\emptyset)$, we find by using Theorem \ref{Maassen}:
  \begin{equation*}\begin{split}
  &U_t\pi(z)\ten\pi(0)(\{s\},\emptyset) =                                                  
  \sum_{\sigma\subset\{s\}}\int_{\Omega}u_t(\sigma,\emptyset,\tau,\emptyset)
  z^{1-|\sigma|+|\tau|}d\tau = zu_t(\emptyset,\emptyset,\emptyset,\emptyset)\ +        \\
  &z^2\int_0^tu_t(\emptyset,\emptyset,\{r\},\emptyset)dr\ +                    
  u_t(\{s\},\emptyset,\emptyset,\emptyset) + z\int_0^tu_t(\{s\},\emptyset,\{r\},\emptyset)dr + \\
  &z^2\int_0^t\int_0^{r_2}u_t(\{s\},\emptyset,\{r_1,r_2\},\emptyset)dr_1dr_2 = 
  \begin{pmatrix}z\exp(-\frac{t}{2})&2z^2\overline{\kappa}_f\exp(-\frac{t}{2})- 2z^2\overline{\kappa}_f\\
  \kappa_f\exp(-\frac{s}{2}) & z \end{pmatrix}.  
  \end{split}\end{equation*}
Therefore we get, for all $X \in M_2$:
  \begin{equation*}\begin{split}
  &\JC_f(X) = \lim_{t\downarrow 0} \frac{\int_0^t\Ad\big[U_t\pi(z)\ten\pi(0)(\{s\},\emptyset)\big](X)ds}
       {t\exp(-t|z|^2)} =                                                      
  \Ad\Bigg[\begin{pmatrix}z & 0 \\ \kappa_f & z\end{pmatrix}\Bigg](X) =                             \\
  &\Ad[zI_2+V_f](X).
  \end{split}\end{equation*}

Let us now turn to the calculation of $\JC_s$. We find for all $X \in M_2$:
  \begin{equation*}
  \JC_s(X) = \lim_{t\downarrow 0} \frac{1}{t} \EC^t\big[(\{\emptyset\},\Omega_1[-t,0))\big](X)
  = \lim_{t\downarrow 0} \frac{\int_0^t\Ad\big[U_t\pi(z)\ten\pi(0)(\emptyset,\{s\})\big](X)ds}
       {t\exp(-t|z|^2)}. 
  \end{equation*}
Taking a closer look at $U_t\pi(z)\ten\pi(0)(\emptyset,\{s\})$, applying Theorem \ref{Maassen}:
  \begin{equation*}\begin{split}
  &U_t\pi(z)\ten\pi(0)(\emptyset,\{s\}) =
  \int_{\Omega}u_t(\emptyset,\{s\},\tau,\emptyset)z^{|\tau|}d\tau =    
   u_t(\emptyset,\{s\},\emptyset,\emptyset)\ +                                                \\
  &z\int_0^t u_t(\emptyset,\{s\},\{r\},\emptyset)dr + 
  z^2\int_0^t \int_0^{r_2}u_t(\emptyset,\{s\},\{r_1,r_2\},\emptyset)dr_1dr_2 =
  \begin{pmatrix}0 &0 \\ \kappa_s\exp(-\frac{s}{2}) & 0 \end{pmatrix}. 
  \end{split}\end{equation*}
Therefore we get, for all $X \in M_2$:
  \begin{equation*}\begin{split}
  &\JC_s(X) = \lim_{t\downarrow 0} \frac{\int_0^t\Ad\big[U_t\pi(z)\ten\pi(0)(\emptyset,\{s\})\big](X)ds}
       {t\exp(-t|z|^2)} =                                                      
  \Ad\Bigg[\begin{pmatrix}0 & 0 \\ \kappa_s & 0 \end{pmatrix}\Bigg](X) =                            \\
  & = \Ad[V_s](X).    
  \end{split}\end{equation*}

Since we are driving the atom with a laser now, the time evolution when we do not observe
the side channel nor the forward channel is now given by 
  \begin{equation*}
  T^z_t = \EC^t\big[\Omega[-t,0),\Omega[-t,0)\big]
  \end{equation*}
and no longer by $T_t$. We will now derive the Master equation for this new time evolution.
For this we need the Dyson series: let $L_0$ and $J$ be linear maps from $M_2 \to M_2$, then for all $t \ge 0$:
\begin{equation*}
  \exp\big(t(L_0 +J)\big) = \int_{\Omega[-t,0)} \exp\big((\omega_1+t)L_0\big)J 
  \exp\big((\omega_2-\omega_1)L_0\big)J\ldots J\exp(-\omega_k L_0)d\omega,
\end{equation*}
where we have written $\omega$ as $\{\omega_1,\ldots, \omega_k\}$ 
with $-t \le \omega_1 < \ldots < \omega_k\le 0$.                                        

Now remember that $\{Y_t\}_{t\ge 0}$ is a semigroup, i.e.\ we can write $Y_t = \exp(tL_0)$.
Then, using Theorem \ref{deco} and twice the Dyson series
  \begin{equation*}\begin{split}
  &T_t^z = \EC^t\big[\Omega[-t,0),\Omega[-t,0)\big] = 
  \int_{\Omega[-t,0)\times\Omega[-t,0)} W_{Y, \JC_f, \JC_s}(\omega_f,\omega_s)d\omega_f d\omega_s = \\
  &\exp\big(t(L_0+\JC_f +\JC_s)\big).
  \end{split}\end{equation*}
This means we get the following Master equation:
  \begin{equation}\label{Mastereq}\begin{split}
  &\frac{d}{dt}T^z_t(\,\cdot\,) = (L_0 + \JC_f + \JC_s)T^z_t(\,\cdot\,) = \\
  - \frac{1}{2}\big\{V^*V,\ & T^z_t(\,\cdot\,)\big\} + 
  \big[zV_f^*-\overline{z}V_f,\ T^z_t(\,\cdot\,)\big] +V^* T^z_t(\,\cdot\,)V,
  \end{split}\end{equation}
which is exactly the Master equation for resonance fluorescence (see \cite{Car}) if we take
$z= -i \frac{\Omega}{2\overline{\kappa}_f}$ with $\Omega$, the \emph{Rabi frequency}, real.   

In the quantum optics literature (see for instance \cite{Car}), usually there is no photon counting
measurement done in the forward channel, i.e.\ $E_f = \Omega[-t,0)$. From here on we will do the same,
we define for all $t\ge0$ and $E_s \in \Sigma_t:\ \MC^t[E_s] := \EC^t\big[\Omega[-t,0), E_s\big]$.
In the following we will also suppress the index $s$ on $E_s$. Using the Dyson series and Theorem
\ref{deco} we find, for all $t\ge0$ and $E \in \Sigma_t$:
  \begin{equation}\label{Davside}
  \MC^t[E] = \int_E W_{Z, \JC_s}(\omega)d\mu(\omega),
  \end{equation} 
where the time evolution in between side-channel-jumps $Z_t$ is given by 
$Z_t = \exp\big(t(L_0+ \JC_f)\big)$ and $W_{Z, \JC_s}$ is defined in the obvious way analogous
to Definition \ref{W}. Note that we have found exactly the same jump operator and time evolution in
between jumps as in the usual quantum optics literature, see for instance
\cite{Car}, \cite{Car01}, i.e.\
we have succeeded in constructing the Davies process of resonance fluorescence with quantum 
stochastic calculus.                                                                             \\

\section{A renewal process}

We will now look briefly at some features of the process $\MC^t$ we obtained. It is easily seen
from the fact that $(\JC_s)^2=0$ (i.e.\ $g_2(0) = 0$) that the photons in the side channel arrive
\emph{anti-bunched}: the probability to see two photons immediately after each other is $0$.
We will now show that the photon counting process in the side channel is a so-called
\emph{renewal process}.                                                              

We denote $\Sigma^t := \Sigma[0,t)$ and, via a shift, we let events $E$ in $\Sigma^t$ correspond to events
$E-t$ in the output sigma field $\Sigma_t$. This means that an element $\omega = \{\omega_1, \ldots, \omega_k\}$ 
in $E \in \Sigma^t$ with $0 \le \omega_1 < \ldots < \omega_k < t$ should be interpreted as 
seeing the first photon appear in the side channel at time $\omega_1$, the second at time $\omega_2$
up to the $k$'th photon at time $\omega_k$.                                                       

Given that we start the photon counting measurement in the initial state $\rho$, we define on the 
sigma fields $\Sigma^t$ ($t \ge 0$) probability measures in the natural way:
for $E \in \Sigma^t: \BB{P}^t_\rho[E] := \mbox{Tr}\big(\rho \MC^t[E-t](I)\big)$. The family of sigma fields
$\{\Sigma^t\}_{t\ge0}$ generates a sigma-field $\Sigma^{\infty}$ of $\Omega[0,\infty)$.
Using that $T_s^z(I) = I$, see equation \eqref{Mastereq}, we find for all $E \in \Sigma^t$:
  \begin{equation*}\begin{split}
  &\BB{P}^{t+s}_{\rho}[E] =
  \mbox{Tr}\Big(\rho \MC^{t+s}\Big[\big(E \tilde{\cup} \Omega[t,t+s)\big)-(t+s)\Big](I)\Big) =      \\
  &\mbox{Tr}\Big(\rho \MC^{t+s}\big[E-(t+s) \tilde{\cup} \Omega[-s,0)\big](I)\Big) =
  \mbox{Tr}\Big(\rho\MC^t[E-t]\MC^s\big[\Omega[-s, 0)\big](I)\Big) =                                        \\
  &\mbox{Tr}\big(\rho \MC^t[E-t]T_t^z(I)\big) = 
         \mbox{Tr}\big(\rho \MC^t[E-t](I)\big) = \BB{P}^{t}_{\rho}[E],
  \end{split}\end{equation*} 
i.e.\ $\BB{P}^{t+s}_{\rho}[E]$ does not depend on $s$. Therefore the family 
$\{\BB{P}^t_{\rho}\}_{t\ge 0}$ on the sigma-fields $\{\Sigma^t\}_{t\ge 0}$
is consistent, hence by Kolmogorov's extension theorem it extends to a 
single probability measure $\BB{P}_{\rho}$ on $\Sigma^\infty$.                                        

In the following, when we write $\omega \in \Omega[0,\infty)$ as $\{\omega_{1}, \omega_2, \ldots \}$, we
imply that $0 \le \omega_1 < \omega_2 < \ldots$. For $j =1,2,\ldots$ we define random variables:
  \begin{equation*}
  X_j:\ \Omega[0,\infty) \to \overline{\BB{R}}^+:\ \omega = \{\omega_1, \omega_2,\ldots\}
  \mapsto
  \left\{ \begin{array}{ll}
  \omega_j-\omega_{j-1} & \mbox{\ if\ \ \ }  |\omega| \ge j \\
  \infty & \mbox{\ otherwise\ } \end{array}\right. ,
  \end{equation*}
where we take $\omega_0$ to be $0$. These random variables give the time elapsed between 
the $(j-1)$th and $j$th detection of a photon. To prove that the counting process is a 
\emph{(modified) renewal process} we have to show that for $i= 1,2,\ldots$
the random variables $X_i$ are independent and for $i = 2,3,\ldots$ they are identically distributed. 
This means we have to show that for $i = 2,3,\ldots$ the
distribution functions $F_{X_i}(x) := \BB{P}_\rho[X_i \le x]$ are equal and for $i,j = 1,2,\ldots$
the joint probability distribution function $F_{X_i, X_j}(x,y) := \BB{P}_\rho[X_i \le x \wedge X_j\le y]$
factorizes: $F_{X_i, X_j}(x,y) = F_{X_i}(x)F_{X_j}(y)$.                                                

Let us first introduce some convenient notation. Note that, using equation \eqref{Davside}, we 
have for all $E \in \Sigma^t$:
  \begin{equation*}\begin{split}
  &\BB{P}_\rho[E] = \BB{P}^t_\rho[E] = \mbox{Tr}\Big(\rho\int_{E-t}W_{Z,\JC_s}(\omega)d\mu(\omega)(I)\Big) =  \\
  &\mbox{Tr}\Big(\rho\int_{E}Z_{\omega_1}\JC_s Z_{\omega_2-\omega_1}\JC_s \ldots 
  \JC_s Z_{t-\omega_k}(I)d\mu(\omega)\Big).
  \end{split}\end{equation*}
We will denote: $x_1 := \omega_1, x_2 := \omega_2-\omega_1, \ldots, x_{k+1} := t- \omega_k$, then we can 
write:
  \begin{equation*}
  \BB{P}_\rho[E] = \int_E \mbox{Tr}\Big(\rho Z_{x_1}\JC_s Z_{x_2}\JC_s \ldots 
  \JC_s Z_{x_{k+1}}(I)\Big)d\mu(\omega). 
  \end{equation*}
Let $P$ denote the matrix $\begin{pmatrix}1 & 0 \\ 0 & 0\end{pmatrix}$, then we have: 
  \begin{equation*}\begin{split}
  &\JC_s Z_{x_{k+1}}(I) = \begin{pmatrix}|\kappa_s|^2\big(Z_{x_{k+1}}(I)\big)_{22} & 0 \\
  0 & 0 \end{pmatrix} = |\kappa_s|^2\big(Z_{x_{k+1}}(I)\big)_{22} P,                          \\
  &\JC_s Z_{x_{k}}(P) = \begin{pmatrix}|\kappa_s|^2\big(Z_{x_{k}}(P)\big)_{22} & 0 \\
  0 & 0 \end{pmatrix} = |\kappa_s|^2\big(Z_{x_{k}}(P)\big)_{22} P,  \\
  &\dots\dots\dots,                   \\
  &\JC_s Z_{x_{2}}(P) = \begin{pmatrix}|\kappa_s|^2\big(Z_{x_{2}}(P)\big)_{22} & 0 \\
  0 & 0 \end{pmatrix} = |\kappa_s|^2\big(Z_{x_{2}}(P)\big)_{22} P.
  \end{split}\end{equation*}
Therefore, if we define $z(x) := |\kappa_s|^2\big(Z_x(P)\big)_{22}$,  
$z_{last}(x) := |\kappa_s|^2\big(Z_x(I)\big)_{22}$ and $z_{first}(x) := \mbox{Tr}\big(\rho Z_x (P)\big)$,
we can write (see also \cite{Car01}): 
  \begin{equation}\label{formula}
  \BB{P}_\rho[E] = \int_E
   z_{first}(x_1)\Big(\prod_{l=2}^k z(x_l)\Big)z_{last}(x_{k+1})d\mu(\omega). 
  \end{equation}
We would like to stress that this formula is only valid for events $E \in \Sigma^t$ and not 
for all events in $\Sigma^\infty$.                                                                

For $t \ge 0$ we introduce the following random variables:
  \begin{equation*}
  N_t:\ \Omega[0,\infty) \to \BB{N}:\ \omega \mapsto |\omega \cap [0,t] |,
  \end{equation*}  
counting the number of photons appearing in the side channel up to time $t$. Since, for strictly positive
driving field strengths, i.e.\ $|z| >0$,
the eigenvalues of the generator $L_0 + \JC_f$ of the semigroup $Z_t$ all have strictly negative real parts, 
we have $\lim_{t\to\infty}Z_t = 0$. Using this, formula \eqref{formula} and the fact that 
the event $[N_t =0]$ is an element of $\Sigma^t$, we obtain:
  \begin{equation*}
  \lim_{t\to \infty} \BB{P}_\rho[N_t = 0] = \lim_{t \to \infty} z_{first}(t) = 0. 
  \end{equation*}
Now suppose we have that $\lim_{t\to \infty} \BB{P}_\rho[N_t \le n] = 0$ for a certain $n \in \BB{N}$. 
For $s \le t$ we use: $\BB{P}_\rho[N_t \le n+1] = 
\BB{P}_\rho[N_t \le n+1| N_s \le n ]\BB{P}_\rho[N_s \le n] +
\BB{P}_\rho[N_t \le n+1| N_s > n]\BB{P}_\rho[N_s > n]$. Therefore we have:
  \begin{equation*}\begin{split}
  &\lim_{t\to \infty} \BB{P}_\rho[N_t \le n +1] = \lim_{s\to \infty}\lim_{t\to \infty} 
  \BB{P}_\rho[N_t \le n +1] =                                                                 \\
  &\lim_{s\to \infty}\lim_{t\to \infty}\Big(
  \BB{P}_\rho[N_t \le n+1| N_s \le n ]\BB{P}_\rho[N_s \le n] + 
  \BB{P}_\rho[N_t \le n+1| N_s > n ]\BB{P}_\rho[N_s > n]\Big) =                               \\
  &\lim_{s\to \infty}\lim_{t\to \infty}\BB{P}^t_\rho[N_t \le n+1| N_s > n ] = 
  \lim_{s\to \infty}\lim_{t\to \infty}z_{last}(t-s) = 0.
  \end{split}\end{equation*}
Now using induction, we get for $n \in \BB{N}$:
  \begin{equation*}
  \lim_{t \to \infty} \BB{P}_\rho [N_t \le n] = 0.
  \end{equation*}
We are now ready to calculate the distribution functions $F_{X_i}$ and $F_{X_i, X_j}$. The 
problem is that for instance the event $[X_i \le x] \in \Sigma^\infty$ is not an element of 
$\Sigma^t$ for a $t \in \BB{R}$. We solve this by conditioning on the event $[N_t \ge i]$
and taking the limit for $t$ to infinity:
  \begin{equation*}\begin{split}
  &F_{X_i}(x) = \BB{P}_\rho [X_i \le x] =                                                          \\
  &\lim_{t \to \infty} \Big(\BB{P}_\rho [X_i \le x| N_t \ge i] \BB{P}_\rho[N_t \ge i]+ 
   \BB{P}_\rho [X_i \le x| N_t < i]\BB{P}_\rho[N_t < i]\Big) =                                         \\
  &\lim_{t \to \infty}\BB{P}^t_\rho [X_i \le x \wedge N_t \ge i].
  \end{split}\end{equation*}
Now we use again formula \eqref{formula} to obtain for $i \ge 2$:
  \begin{equation*}\begin{split} 
  &F_{X_i}(x) =  \lim_{t \to \infty} \sum_{k=i}^\infty \int_{\substack{\sum_{l=1}^{k+1} x_l = t \\ x_i \le x}}
  z_{first}(x_1)\Big(\prod_{l=2}^k z(x_l)\Big)z_{last}(x_{k+1})dx_1\ldots dx_{k+1} =                      \\
  &\lim_{t \to \infty} \int_0^x z(x_i)\Bigg(\sum_{k=i}^\infty 
   \int_{\sum_{l \neq i} x_l = t-x_i}
  z_{first}(x_1)dx_1\Big(\prod_{\substack{l =2\\ l\neq i}}^k z(x_l)dx_l\Big)z_{last}(x_{k+1})dx_{k+1}\ 
  \Bigg)dx_i =                                                                                             \\
  &\lim_{t \to \infty} \int_0^x z(x_i) \BB{P}^{t-x_i}_\rho\big[N_{t-x_i} \ge i-1\big] dx_i.
  \end{split}\end{equation*}
Then we use dominated convergence to interchange the limit and the integral to obtain:
  \begin{equation*}
  F_{X_i}(x) = \int_0^x z(x')dx'.
  \end{equation*}
When $i=1$ we can repeat the whole calculation to find the same result when for $z$ we substitute
$z_{first}$. It is now obvious that for $i = 2,3,\ldots$ the random variables $X_i$ are identically
distributed.                                                                                        

In a similar fashion, only extracting two integrals now, we find that for $i,j = 2,3, \ldots:\
F_{X_i,X_j}(x,y) = \int_0^x\int_0^y z(x')z(y') dx'dy'$. If $i$ or $j$ is $1$ we again have to 
substitute $z_{first}$ for $z$. It is now obvious that the random variables $X_i$ and $X_j$ are
independent. We conclude that the family of random variables $\{X_i\}_{i=1,2,\ldots}$ is a 
(modified) renewal process.

%% file: ssequation.tex
\chapter{Stochastic Schr\"odinger equations}\label{ch sseq}

\begin{center}
{\large Luc Bouten$^\dagger$ \ \ \ \ \ \ M\u{a}d\u{a}lin Gu\c{t}\u{a}$^{\dagger\dagger}$ \ \ \ \ \ \  Hans Maassen$^\dagger$}\\
\vspace{1cm}
$^\dagger$\emph{Mathematisch Instituut, Katholieke Universiteit Nijmegen \\
Toernooiveld 1, 6526 ED Nijmegen, The Netherlands}\\
\vspace{0.5 cm}
$^{\dagger\dagger}$\emph{EURANDOM, PO Box 513, 5600 MB Eindhoven, The Netherlands}
\end{center}

{\small
\begin{center}\textbf{Abstract\footnote{This chapter is an adapted version of \cite{BGM}.}}\end{center}
A derivation of Belavkin's stochastic Schr\"odinger equations is 
given using quantum filtering theory. We study an open system in 
contact with its environment, the electromagnetic 
field. Continuous observation of the field yields information on the system: 
it is possible to keep track in real time of the best estimate of the system's
quantum state given the observations made. This estimate satisfies a stochastic
Schr\"odinger equation, which can be derived from the quantum stochastic 
differential equation for the interaction picture evolution of system and 
field together. Throughout the paper we focus on the basic example of resonance 
fluorescence.}

\section{Introduction}\label{Introduction}

It has long been recognized that continuous time measurements can not
be described by the standard projection postulate of quantum mechanics.
In the late 60's, beginning 70's, Davies developed a theory
for continuous time measurement \cite{Dav1} culminating in his book 
\cite{Dav}. His mathematical work became known to the quantum optics 
community through the paper with Srinivas on photon 
counting \cite{SrD}.                                      

The late 80's brought renewed interest to the theory of continuous time 
measurement. For instance the waiting time distribution of fluorescence photons 
of a two-level atom driven by a laser was obtained by associating a continuous
evolution to the atom in between photon detections and jumps at the moments
a photon is detected \cite{Car01}. In this way every record of photon detection
times determines a trajectory in the state space of the atom. Averaging
over all possible detection records leads to the well-known description
of the dissipative evolution of the atom by a master equation.
Advantage of the trajectory approach is the fact that an initially 
pure state will remain pure along the whole trajectory. This 
allows for the use of state vectors instead of density matrices, 
significantly speeding up computer simulations \cite{MCD}, \cite{DCM}, 
\cite{GPZ}, \cite{Car}.                                    

Infinitesimally, the quantum trajectories are solutions of a stochastic 
differential equation with the measurement process as the noise term.
The change in the state is given by the sum of two terms: 
a deterministic one proportional with $dt$ and a stochastic one proportional 
to the number of detected photons $dN_t$ in the interval $dt$. 
For other schemes such as homodyne detection the corresponding
stochastic differential equation is obtained as the diffusive limit
of photon counting where the jumps in the state space 
decrease in size but become increasingly frequent 
\cite{BaB}, \cite{Car}, \cite{WiM}. In this limit the stochastic term 
in the differential equation is replaced by a process with continuous
paths.      

The stochastic Schr\"odinger equations obtained in this way had been 
postulated before by Gisin \cite{Gis}, \cite{Gis1}, \cite{DGHP}, \cite{GPR},
in an attempt to generalize the customary unitary evolution in 
quantum mechanics. The stochastic terms are seen as randomness 
originating from the measurement process. However, in this approach 
the correspondence between the different quantum state diffusion equations 
and the measurements that can be performed is not emphasized.  

Another approach originated from the development of quantum stochastic
calculus \cite{HuP}, \cite{Par}, generalizing the classical It\^o table to quantum 
noises represented by creation and annihilation operators (see Section \ref{QSC}).
Barchielli saw the relevance of this new calculus for quantum optics \cite{Bar}. 
Indeed, in the Markovian approximation the interaction between a quantum system
and the electromagnetic field is governed by a unitary solution of a quantum 
stochastic differential equation in the sense of \cite{HuP}.

Belavkin \cite{Bel2} was the first to see the connection 
between quantum measurement 
theory and classical filtering theory \cite{Kal}, in which one estimates 
a signal or system process when observing a function of the signal in the 
presence of noise. This is done by deriving the filtering equation
which is a stochastic differential equation for the expectation value of the system 
process conditioned on outcomes of the observation process. Belavkin extended
the filtering theory \cite{Bel}, \cite{Bel01} to allow for the quantum noises
of \cite{HuP}. Stochastic Schr\"odinger equations turn out to be examples of the 
quantum filtering or Belavkin equation \cite{Bel0}, \cite{BeS}.       

Aim of this paper is to give an elementary presentation of quantum filtering
theory. We construct the expectation of an observable conditioned on outcomes
of a given measurement process. The differential form of this conditional
expectation is the stochastic Schr\"odinger equation associated with the 
given measurement. At the heart of the derivation lies 
the It\^o table of quantum stochastic calculus enabling a fast computation 
of the equation. The procedure is summarized in a small recipe in Section
\ref{Belavkinslemma}.       

To illustrate the theory we consequently focus on the basic example of 
resonance fluorescence of a two-level atom for which we consider 
photon counting and homodyne detection measurement schemes. The stochastic
Schr\"odinger equations for these examples are derived in two ways, once
via the usual approach using quantum trajectories and a diffusive limit, and
once using quantum filtering theory. In this way we hope to emphasize how 
conceptually different both methods are. 

This paper is organised as follows. Sections \ref{Daviesprocess} and 
\ref{Homodynedetection} serve as an introduction to the guiding example
of this paper: resonance fluorescence of a two-level atom driven by a laser. 
In Section \ref{Daviesprocess} we put the photon counting description of 
resonance fluorescence by Davies \cite{BMK}, \cite{Car01}, \cite{Car0} into 
the form of a stochastic differential equation driven by the counting process. 
In Section \ref{Homodynedetection} we discuss the 
homodyne detection scheme as a diffusive limit of the photon counting 
measurement, arriving at a stochastic differential equation driven by a 
diffusion process. The equations of Sections \ref{Daviesprocess}
and \ref{Homodynedetection} will be rederived later in a more general way 
using quantum filtering theory. 

In Section \ref{conditional expectation} we introduce the concept of
conditional expectation in quantum mechanics by first illustrating it 
in some simple, motivating examples. Section \ref{dilation}  
describes the dissipative evolution of the open system 
within the Markov approximation. The joint evolution of the 
system and its environment, the quantized electromagnetic field, is 
given by unitaries satisfying a quantum stochastic differential equation. 
Given a measurement of some field observables it is shown how to 
condition the state of the system on outcomes of the measurement 
using the construction of Section \ref{conditional expectation}.
Section \ref{QSC} is a short review of quantum stochastic
calculus and its applications to open systems. Sections \ref{dilation}
and \ref{QSC} describe dilation theory and quantum stochastic calculus
in a nutshell. 

Section \ref{Belavkinslemma} contains the derivation of the quantum 
filtering equation, the stochastic differential equation for the 
conditional expectation. This equation is the stochastic Schr\"odinger 
equation for the given measurement. This part ends with a recipe
for computing stochastic Schr\"odinger equations for a large class
of quantum systems and measurements. The end of the article
connects to Sections \ref{Daviesprocess} and \ref{Homodynedetection}
by showing how the recipe works in our main example.

\section{The Davies process}\label{Daviesprocess}

We consider a two-level atom in interaction with the quantized electromagnetic field. 
The state of the atom is described by a $2 \times 2$-density matrix $\rho$, i.e.\ 
$\rho\geq 0$, and $\mbox{Tr}\rho=1$. Atom and field together perform a unitary, thus 
reversible evolution, but by taking a partial trace over the electromagnetic field we 
are left with an irreversible, dissipative evolution of the atom alone. 
In the so called Markov limit it is given by a norm continuous semigroup 
$\{T_t\}_{t \ge 0}$ of completely positive maps. A central example discussed in this paper 
is resonance fluorescence. Here the atom is driven by a laser on the {\it forward} channel, while 
in the {\it side} channel a photon counting measurement is performed. For the time being we will 
suppress the oscillations of the laser for reasons of simplicity. 
In this case the Lindblad generator of $T_t$, or Liouvillian $L$ is given by (cf.\ \cite{Car}):
  \begin{equation}\label{Master}
  \frac{d}{dt}\Big|_{t=0}T_t(\rho) = 
  L(\rho) = -i[H, \rho] + i\frac{\Omega}{2}[V+V^*, \rho] - \frac{1}{2}\{V^*V,\rho\} + V \rho V^*, 
  \end{equation}
where
  \begin{equation*}
  V = \begin{pmatrix} 0 & 0 \\ 1 & 0 \end{pmatrix},
  \end{equation*}
$H := \frac{\omega_0}{2}\sigma_z$ is the Hamiltonian of the atom, and $\Omega$ is the 
\emph{Rabi frequency}.  \\
The master equation \eqref{Master} can be \emph{unravelled} in many ways 
depending on what photon detection measurement is performed. By unravelling the master equation 
we mean writing $L$ as the sum $\LC +\JC$, where $\JC$ represents the instantaneous 
state change taking place when detecting a photon, and $\LC$ describes the smooth state 
variation in between these instants. The unravelling for photon counting in the side
channel is given by \cite{Car}    
  \begin{equation*}
  \LC(\rho) =-i[H, \rho] + i \frac{\Omega}{2}[V+V^*,\rho] - \frac{1}{2}\{V^*V,\rho\} + 
(1-|\kappa_s|^2)V\rho V^* \mbox{\ and \ } \JC(\rho) = |\kappa_s|^2 V\rho V^*,    
  \end{equation*}
with $|\kappa_s|^2$ the decay rate into the side channel.        \\ 
An outcome of the measurement over an arbitrary finite time interval $[0,t)$
is the set of times $\{t_1, t_2, \ldots, t_k\}$ at which photons are detected 
in the side channel of the field. The number of detected photons can be arbitrary, thus the 
space of outcomes is 
 \begin{equation*}
 \Omega\left([0,t)\right) := \bigcup_{n=0}^\infty \Omega_n\left([0,t)\right)=
 \bigcup_{n=0}^\infty  \{\sigma \subset [0,t);\ |\sigma| = n\}
 \end{equation*}
also called the \emph{Guichardet space} \cite{Gui}. 
In order to describe the probability distribution of the 
outcomes we need to make $\Omega\left([0,t)\right)$ into a measure space. 
Let us consider the space of $n$-tuples $[0,t)^n$ with its Borel 
$\sigma$-algebra and the measure $\frac{1}{n!}\lambda_n$ where $\lambda_n$ is the 
Lebesgue measure. Then the map
 \begin{equation*} 
 j_n: [0,t)^n\ni (t_1,\dots, t_n)\to\{t_1,\dots, t_n\} \in \Omega_n\left([0,t)\right)
 \end{equation*}
induces  the $\sigma$-algebra $\Sigma_n\left([0,t)\right)$ and the measure $\mu_n$ on 
$\Omega_n\left([0,t)\right)$.  We define now the measure 
$\mu$ on $\Omega\left([0,t)\right)$ such that $\mu(\{\emptyset\})= 1$ and $\mu = \mu_n$ on 
$\Omega_n\left([0,t)\right)$. 
We will abbreviate $\Omega\big([0,t)\big)$ and $\Sigma\big([0,t)\big)$ to $\Omega^t$ and 
$\Sigma^t$, respectively.                              \\
Davies was the first to show \cite{Dav} 
(see also \cite{Car}, \cite{BMK}) that the unnormalized state of the 
$2$-level atom at time $t$ with initial state $\rho$, and 
conditioned on the outcome of the experiment being in a set $E\in\Sigma^t$ is given by: 
  \begin{equation*}
  \MC^t[E](\rho) = \int_E W_{t}(\omega)(\rho) d\mu(\omega),
  \end{equation*} 
where for $\omega = \{t_1, \ldots, t_k\} \in \Omega^t$ with 
$0 \le t_1 \le \ldots \le t_k < t$ we have
  \begin{equation*}
  W_{t}(\omega)(\rho) := \exp\big((t-t_k)\LC\big)\JC \ldots \JC\exp\big((t_2-t_1)\LC\big)\JC
  \exp\big(t_1\LC\big)(\rho).
  \end{equation*}
Furthermore, $\BB{P}_\rho^t[E] :=\mbox{Tr}(\MC^t[E](\rho))$ is the probability that the event 
$E$ occurs if the initial state is $\rho$. 
The family of prabability measures $\{\BB{P}_\rho^t\}_{t \ge 0}$ is consistent, 
i.e.\ $\BB{P}_\rho^{t+s}[E] = \BB{P}_\rho^t[E]$ for all $E \in \Sigma^t, s\ge 0$, 
see \cite{BMK}, hence by Kolmogorov's extension theorem it extends to a single 
probability measure $\BB{P}_\rho$ on
the $\sigma$-algebra $\Sigma^\infty$, of the set $\Omega^\infty$. \\
On the measure space $(\Omega^\infty, \Sigma^\infty, \BB{P}_\rho)$ we define the following random 
variables:
  \begin{equation*}
  N_t:\ \Omega^\infty \to \BB{N}:\ \omega \mapsto |\omega \cap [0,t)|,
  \end{equation*}    
counting the number of photons detected in the side channel up to time $t$. 
The counting  process $\{N_t\}_{t\geq 0}$ has differential $dN_t := N_{t+dt}-N_t$ satisfying 
$dN_t(\omega) = 1$ if $t\in \omega$ and $dN_t(\omega) = 0$ otherwise. 
Therefore we have the following It\^o rules: $dN_tdN_t = dN_t$ and $dN_tdt = 0$, 
(cf.\ \cite{BaB}). \\
To emphasise the fact that the evolution of the $2$-level atom is stochastic, we will regard the 
normalized density matrix as a random variable $\{\rho^t_\bullet\}_{t \ge 0}$ 
with values in the $2 \times 2$-density matrices defined as follows:
  \begin{equation}\label{rhoomega}
  \rho^t_\bullet:\ \Omega^\infty \to M_2:\ \omega \mapsto \rho^t_\omega := 
  \frac{W_{t}\big(\omega\cap [0,t)\big)(\rho)}
  {\mbox{Tr}\Big(W_{t}\big(\omega\cap [0,t)\big)(\rho)\Big)}.
  \end{equation}
The processes $N_t$ and $\rho^t_\bullet$ are related through the stochastic differential 
equation $d\rho^t_\bullet = \alpha_t dt + \beta_t dN_t$. Following \cite{BaB} we will 
now determine the processes $\alpha_t$ and $\beta_t$ by differentiating \eqref{rhoomega}. 
If $t \in \omega$ then $dN_t(\omega) = 1$, i.e.\ the differential
$dt$ is negligible compared to $dN_t=1$, therefore:
  \begin{equation}\label{beta}
  \beta_t(\omega) = \rho^{t+dt}_\omega - \rho^t_\omega = \frac{\JC(\rho^t_\omega)}
  {\mbox{Tr}\big(\JC(\rho^t_\omega)\big)} - \rho^t_\omega.
  \end{equation}
On the other hand, if $t \not \in \omega$ then $dN_t(\omega) = 0$, i.e.\ $dN_t$ is 
negligible compared to $dt$. Therefore it is only the $dt$ term that contributes:
  \begin{equation}\label{alpha}\begin{split}
  &\alpha_t(\omega) = \frac{d}{ds}\Big|_{s=t} \frac{\exp\big((s-t)\LC\big)(\rho^t_\omega)}
  {\mbox{Tr}\Big(\exp\big((s-t)\LC\big)(\rho^t_\omega)\Big)} =  \\
  &\LC(\rho^t_\omega) - \frac{\rho^t_\omega}{\mbox{Tr}(\rho^t_\omega)^2}
\mbox{Tr}\big(\LC(\rho^t_\omega)\big) 
  = \LC(\rho^t_\omega) + \mbox{Tr}\big(\JC(\rho^t_\omega)\big)\rho^t_\omega,
  \end{split}\end{equation}    
where we used that $\mbox{Tr}\big(\LC(\rho^t_\omega)\big) = - 
\mbox{Tr}\big(\JC(\rho^t_\omega)\big)$, 
as a consequence of the 
fact that \ $\mbox{Tr}\big(L(\sigma)\big) = 0$ for all density matrices $\sigma$.
Substituting \eqref{beta} and \eqref{alpha} into 
$d\rho^t_\bullet = \alpha_t dt  + \beta_t dN_t$ we 
get the following \emph{stochastic Schr\"odinger equation} for the
state evolution of the $2$-level atom if we are counting photons in the side channel 
(cf.\ \cite{BaB}, \cite{Car0}, \cite{BeM}):
  \begin{equation}\label{Belcount}
  d\rho^t_\bullet = L(\rho^t_\bullet)dt + \Big(\frac{\JC(\rho^t_\bullet)}
 {\mbox{Tr}\big(\JC(\rho^t_\bullet)\big)}- \rho^t_\bullet\Big)\Big
 (dN_t - \mbox{Tr}\big(\JC(\rho^t_\bullet)\big)dt\Big).
  \end{equation}
The differential $dM_t := dN_t - \mbox{Tr}\big(\JC(\rho^t_\bullet)\big)dt$ and the initial 
condition $M_0 = 0$ define an important process $M_t$ called the \emph{innovating martingale}, 
discussed in more detail in Section \ref{Belavkinslemma}.

\section{Homodyne detection}\label{Homodynedetection}

We change the experimental setup described in the previous section by introducing a 
\emph{local oscillator}, i.e.\ a one mode oscillator in a coherent state given
by the normalised vector in $l^2(\BB{N})$  
 \begin{equation}\label{eq.coherent}
 \psi(\alpha_t) :=\exp\big(\frac{-|\alpha_t|^2}{2}\big)(1, \alpha_t, \frac{\alpha_t^2}{\sqrt{2}}, 
 \frac{\alpha_t^3}{\sqrt{6}}, \ldots),
 \end{equation} 
for a certain $\alpha_t \in \BB{C}$. We take $\alpha_t = \frac{w_t}{\varepsilon}$, 
where $w_t$ is a complex number with modulus $|w_t| = 1$, and  $\varepsilon > 0$. 
The number $\varepsilon$ is inversely proportional to the intensity 
of the oscillator. Later on we will let the intensity go to infinity, i.e.\ 
$\varepsilon\to 0$. The phase $\phi_t$ of the oscillator is represented by 
$w_t = \exp(i\phi_t)$, with $\phi_t = \phi_0 + \omega_{lo} t$, where $\omega_{lo}$ 
is the frequency of the oscillator.                 \\
The local oscillator is coupled to a channel in the electromagnetic field, 
the local oscillator beam. The field is initially in the vacuum state. The 
local oscillator and the field are coupled in such a way that every time a photon is detected 
in the beam, a jump on the local oscillator occurs, given by the operation  
  \begin{equation}
  \JC_{lo}(\rho) = A_{lo}\rho A^*_{lo},
  \end{equation} 
where $A_{lo}$ is the annihilation operator corresponding to the mode of the local 
oscillator. The coherent state $\psi(\alpha_t)$ is an eigenstate of the jump operator $A_{lo}$ at 
eigenvalue $\alpha_t$.  \\
Now we are ready to discuss the homodyne detection scheme. Instead of directly counting photons
in the side channel we first mix the side channel with the local oscillator beam 
with the help of a fifty-fifty beam splitter. In one of the emerging beams a 
photon counting measurement is performed. A detected photon can come from the atom through the side 
channel or from the local oscillator via the local oscillator beam. 
Therefore the jump operator on states $\sigma$ of the atom and the oscillator together, 
is the sum of the respective jump operators:
  \begin{equation*}
  \JC_{a\ten lo}(\sigma) =   
  (\kappa_s V \ten I + I \ten A_{lo})\sigma
  (\overline{\kappa}_s V^* \ten I + I \ten A^*_{lo}).  
  \end{equation*}   
An initial product state $\rho \ten |\psi(\alpha_t)\rangle\langle\psi(\alpha_t)|$ of 
the $2$-level atom and the local oscillator will remain a product after the jump 
since $\psi(\alpha_t)$  is an eigenvector of the annihilation operator. Tracing out 
the local oscillator yields the following jump operation for the atom in the homodyne 
setup: 
  \begin{equation*}\begin{split}
  \JC_a(\rho) = 
  \mbox{Tr}_{lo} \Big(\JC_{a\ten lo}\big(\rho\ten \big|\psi(\alpha_t)\big\rangle
  \big\langle\psi(\alpha_t)\big|\big)\Big) 
  = \big(\kappa_s V + \frac{w_t}{\varepsilon}\big)
  \rho\big(\overline{\kappa}_s V^* + \frac{\overline{w}_t}{\varepsilon}\big).
  \end{split}\end{equation*}
In the same way as in Section \ref{Daviesprocess}, we can derive the following 
stochastic Schr\"odinger equation for the state evolution of the two-level atom when
counting photons after mixing the side channel and the local oscillator beam \cite{BaB} \cite{Car0}:
  \begin{equation}\label{Belcounthomodyne}
  d\rho^t_\bullet = L(\rho^t_\bullet)dt + \frac{1}{\varepsilon}\Big(\frac{\JC_a(\rho^t_\bullet)}
 {\mbox{Tr}\big(\JC_a(\rho^t_\bullet)\big)}- \rho^t_\bullet\Big)
 \varepsilon\Big(dN_t - \mbox{Tr}\big(\JC_a(\rho^t_\bullet)\big)dt\Big),
  \end{equation}   
where the extra $\varepsilon$'s  are introduced for future convenience. 
We will again use the abbreviation: $dM^a_t = dN_t - \mbox{Tr}\big(\JC_a(\rho^t_\bullet)\big)dt$ 
for the innovating martingale (see Section \ref{Belavkinslemma}).
In the homodyne detection scheme the intensity of the local oscillator beam is 
taken extremely large, i.e.\ we are interested in the limit $\varepsilon \to 0$ \cite{BaB}, 
\cite{Car}, \cite{WiM}. Then the number of detected photons becomes very large and it makes sense 
to scale and center $N_t$, obtaining in this way the process with differential 
$dW_t^\varepsilon := \varepsilon dN_t -dt/\varepsilon$ and $W_0^\varepsilon = 0$. 
We find the following It\^o rules for $dW_t^\varepsilon$:
  \begin{equation*}\begin{split}
  &dW^\varepsilon_tdW^\varepsilon_t = \big(\varepsilon dN_t - \frac{1}{\varepsilon}dt\big)
    \big(\varepsilon dN_t - \frac{1}{\varepsilon}dt\big) = \varepsilon^2 dN_t = 
    \varepsilon dW^\varepsilon_t + dt,                                                                 \\
  &dW^\varepsilon_t dt = 0.                                                    
  \end{split}\end{equation*}
In the limit $\varepsilon \to 0$ this becomes $dW_tdW_t = dt$ and $dW_tdt = 0$, i.e.\
the process $W_t := \lim_{\varepsilon \to 0} W_t^\varepsilon$ is a diffusion. 
It is actually this scaled and centered process that is being observed and not 
the individual photon counts $N_t$, see \cite{Car}. 
We pass now to the evaluation of the limit of \eqref{Belcounthomodyne}:
  \begin{equation*}
  \lim_{\varepsilon \to 0}\frac{1}{\varepsilon}\Big(\frac{\JC_a(\rho^t_\bullet)}
 {\mbox{Tr}\big(\JC_a(\rho^t_\bullet)\big)}- \rho^t_\bullet\Big) = w_t \overline{\kappa}_s
  \rho^t_\bullet V^* + \overline{w}_t\kappa_s V \rho^t_\bullet - \mbox{Tr}(w_t \overline{\kappa}_s
  \rho^t_\bullet V^* + \overline{w}_t\kappa_s V \rho^t_\bullet)\rho^t_\bullet.
  \end{equation*}
This leads to the following stochastic Schr\"odinger equation for the
homodyne detection scheme \cite{BaB}, \cite{Car0}, \cite{WiM}, \cite{BeM}
  \begin{equation}\label{Belhomodyne}
  d\rho^t_\bullet = L(\rho^t_\bullet)dt + \left(w_t \overline{\kappa}_s
  \rho^t_\bullet V^* + \overline{w}_t\kappa_s V \rho^t_\bullet - \mbox{Tr}(w_t \overline{\kappa}_s
  \rho^t_\bullet V^* + \overline{w}_t\kappa_s V \rho^t_\bullet)\rho^t_\bullet~\right)dM^{hd}_t, 
  \end{equation}
 for all states $\rho \in M_2$, where 
  \begin{equation}
  dM_t^{hd} := dW_t - \mbox{Tr}(w_t \overline{\kappa}_s
  \rho^t_\bullet V^* + \overline{w}_t\kappa_s V \rho^t_\bullet)dt. 
  \end{equation} 

Let $a_s(t)$ and $a_b(t)$ denote the annihilation operators for the side channel
and the local oscillator beam, respectively. They satisfy the canonical commutation relations
  \begin{equation*}
  [a_i(t), a^*_j(r)] = \delta_{i,j} \delta(t-r), \ \ \ i,j \in \{s,b\}. 
  \end{equation*}
Smearing with a quadratically integrable function $f$ gives 
  \begin{equation*}
  A_i(f) = \int f(t)a_i(t)dt, \ \ \ i \in \{s,b\}.
  \end{equation*}  
By definition, the stochastic process $\{N_t\}_{t \ge 0}$ 
counting the number of detected photons has the same law as the the number operator $\Lambda(t)$ 
up to time $t$ for the beam on which the measurement is performed. Formally we can write
   \begin{equation*}\begin{split}
  \Lambda(t) = 
  \int_0^t\big(a^*_s(r) \ten I + I \ten a^*_b(r)\big)
    \big(a_s(r) \ten I + I \ten a_b(r)\big)dr.   
  \end{split}\end{equation*}
The oscillator beam is at time $t$ in the coherent state 
$\psi\left(\frac{f_t}{\varepsilon}\right)$, where $f_t \in L^2(\BB{R})$ is the function 
$r \mapsto w_r\chi_{[0,t]}(r)$.  
Since the state of the local oscillator beam is an eigenvector of the annihilation operator 
$a_b(r)$ 
  \begin{equation*}
  a_b(r)\psi\left(\frac{f_t}{\varepsilon}\right)=\frac{w_r}{\varepsilon}
  \psi\left(\frac{f_t}{\varepsilon}\right),
  \end{equation*}
we find 
 \begin{equation*}\begin{split}
 \varepsilon \Lambda(t)-\frac{t}{\varepsilon} & =
   \varepsilon \Lambda_s(t)\ten I + 
   \varepsilon \int_0^t \big(\frac{w_r}{\varepsilon}a^*_s(r) + 
  \frac{\overline{w}_r}{\varepsilon}
  a_s(r)\big)\ten I + \frac{|w_r|^2}{\varepsilon^2}dr - \frac{t}{\varepsilon} \\
 & = \varepsilon \Lambda_s(t)\ten I + \big(A^*_s(f_t) + A_s(f_t)\big)\ten I.
 \end{split}\end{equation*}
The operator $X_\phi(t) :=  A^*_s(f_t) + A_s(f_t)$ is called a {\it field quadrature}.
We conclude that in the limit $\varepsilon\to 0$ the homodyne detection is a setup for 
continuous time measurement of the field quadratures $X_\phi(t)$ of the side channel.
(cf.\ \cite{Car}).

\section{Conditional expectations}\label{conditional expectation}

In the remainder of this article we will derive the equations \eqref{Belcount} and 
\eqref{Belhomodyne} in a different way. We will develop a general way to derive Belavkin 
equations (or stochastic Schr\"odinger equations). 
The counting experiment and the homodyne detection experiment, described in the 
previous sections, serve as examples in this general framework. 
The method we describe here closely
follows Belavkin's original paper on quantum filtering theory \cite{Bel}. 
The construction below, however, uses explicitly the decomposition
of operators over the measurement results. In the next section
it will turn out that this is done most naturally in the 
interaction picture.               

Let us remind the concept of conditional expectation from probability theory. 
Let $(\Omega, \Sigma, \mathbb{P})$ be a probability space describing the ``world'' 
and $\Sigma'\subset\Sigma$ a $\sigma$-algebra of events to which ``we have access''. 
A random variable $f$ on $(\Omega, \Sigma, \mathbb{P})$ with $\mathbb{E}(|f|)<\infty$ can 
be projected to its conditional expectation 
$\mathcal{E}(f)$ which is measurable with respect to $\Sigma'$ and satisfies 
 \begin{equation*} 
 \int_{E}f\mbox{d}\mathbb{P}= \int_{E}\mathcal{E}(f)\mbox{d}\mathbb{P} 
 \end{equation*}
for all events $E$ in $\Sigma'$. Our information about the state of that 
part of the world to which we have access, can be summarized in a probability distribution 
$\mathbb{Q}$ on $\Sigma'$. Then the predicted expectation of $f$ given this information is 
$\int_{\Omega}\mathcal{E}(f)\mbox{d}\mathbb{Q}$. 
We will extend this now to quantum systems and measurements.

The guiding example is that of an $n$ level atom described by the algebra $\BC:=M_n$ undergoing 
a transformation given by a completely positive unit preserving map $T:\BC\to \BC$ 
with the following Kraus decomposition $T(X)=\sum_{i\in\Omega}V_i^*XV_i$. 
The elements of $\Omega$ can be seen as the possible measurement outcomes. 
For any initial state $\rho$ of $\BC$ and measurement result $i\in\Omega$, 
the state after the measurement is given by 
 \begin{equation*}
 \rho_i=V_i\rho V_i^*/\mbox{Tr}(V_i\rho V_i^*),
 \end{equation*}
and the probability distribution of the outcomes is $p=\sum_{i\in\Omega} p_i\delta_i$ where 
$\delta_i$ is the atomic measure at $i$, and $p_i=\mbox{Tr}(V_i\rho V_i^*)$, which 
without loss of generality can be assumed to be strictly positive. 
We represent the measurement by an instrument, that is the completely positive map with the 
following action on states
 \begin{equation}\label{eq.instrument} 
 \MC:\ M_n^*\to M_n^* \tens\ell^1(\Omega):~\rho\mapsto \sum_{i\in\Omega}\rho_i\tens p_i \delta_i.
 \end{equation} 
Let $X\in\BC$ be an observable of the system. Its expectation after the measurement, given that 
the result $i\in\Omega$ has been obtained is $\mbox{Tr}(\rho_i X)$. The function
 \begin{equation*}  
 \EC(X):\ \Omega\to \BB{C}:~  i\mapsto \mbox{Tr}(\rho_i X) 
 \end{equation*}
is the {\it conditional expectation} of $X$ onto $\ell^\infty(\Omega)$. 
If $q=\sum q_i\delta_i$ is a probability distribution on $\Omega$ then $\sum q_i\EC(X)(i)$ 
represents the expectation of $X$ on a statistical ensemble for which the distribution 
of the measurement outcomes is $q$.
We extend the conditional expectation to the linear map 
 \begin{equation*}  
 \EC:\ \BC\tens\ell^\infty(\Omega)\to\ell^\infty(\Omega)\subset \BC\tens\ell^\infty(\Omega) 
 \end{equation*} 
such that for any element $A:i\mapsto A_i$ in 
$\BC\tens\ell^\infty(\Omega)\cong \ell^\infty(\Omega\to\BC)$ we have
 \begin{equation*}  
 \EC(A):\ i\mapsto \mbox{Tr}(\rho_i A_i). 
 \end{equation*}
This map has the following obvious properties: it is idempotent and has norm one. 
Moreover, it is the unique linear map with these properties preserving the state 
$\MC(\rho)$ on $\BC\tens\ell^\infty(\Omega)$. For this reason we will call $\EC$, 
the conditional expectation with respect to $\MC(\rho)$. 
Its dual can be seen as an extension of probability distribitions 
$q\in \ell^1(\Omega)$ to states on $\BC\tens\ell^\infty(\Omega)$
 \begin{equation*} 
 \EC^*:\ q\mapsto \sum_{i\in\Omega}\rho_i\tens q_i\delta_i.
 \end{equation*}
Thus while the measurement \eqref{eq.instrument} provides a state  $\MC(\rho)$ on  
$\BC\tens\ell^\infty(\Omega)$, the conditional expectation with respect to $\MC(\rho)$ 
extends probability distributions $q\in\ell^1(\Omega)$ of outcomes, to states on 
$\BC\tens\ell^\infty(\Omega)$, and in particular on $\BC$ which represents the state after the 
measurement given the outcomes distribution $q$.

With this example in mind we pass to a more general setup which will be needed in deriving the 
stochastic Schr\"odinger equations. 
Let $\AC$ be a unital $^*$-algebra of bounded operators on a Hilbert space $\BB{H}$ whose 
selfadjoint elements represent the observables of a quantum system. 
It is natural from the physical point of view to assume that $\AC$ is strongly closed, i.e.\ 
if $\{A_n\}_{n\geq 0}$ is a sequence of operators in $\AC$ such that 
$\|A_n\psi\|\to\|A\psi\|$ for any vector $\psi$ in $\BB{H}$ and a fixed bounded operator $A$, 
then $A\in\AC$. From the mathematical point of view this leads to the rich theory of 
von Neumann algebras inspired initially by quantum mechanics, but can as well be seen 
as the generalization of probability theory to the non-commutative world of quantum mechanics. 
Indeed, the building blocks of quantum systems are matrix algebras, while probability spaces 
can be encoded into their {\it commutative} algebra of bounded random variables 
$L^\infty(\Omega, \Sigma, \BB{P})$ which appeared already in the example above. 
A state is described by a density matrix in the first case or a probability distribution 
in the second, in general it is a positive normalized linear functional $\psi:\AC\to\mathbb{C}$ 
which is continuous with respect to the weak*-topology, the natural topology on a von Neumann 
algebra seen as the dual of a Banach space \cite{KaR}.   
\begin{de}
Let $\BC$ be a von Neumann subalgebra of a von Neumann algebra $\AC$ of operators on a (separable)
Hilbert space $\BB{H}$. 
A \emph{conditional expectation} of $\AC$ onto $\BC$
is a linear surjective map $\EC:\ \AC \to \BC$, such that:
  \begin{enumerate}
  \item $\EC^2 = \EC$\ \ ($\EC$ is idempotent), 
  \item $\forall_{A \in \AC}:\ \| \EC(A)\| \le \| A\|$\ \ ($\EC$ is normcontractive).
  \end{enumerate}
\end{de}
In \cite{Tom} it has been shown that the conditions 
$1$ and $2$ are equivalent to $\EC$ being an identity preserving, completely positive map, 
and satisfying the \emph{module property}
 \begin{equation}\label{eq.moduleproperty}
 \EC(B_1AB_2) = 
  B_1\EC(A)B_2, \qquad\mbox{for all} ~B_1, B_2 \in \BC, ~\mbox{and}~ A \in \AC, 
 \end{equation}
generalizing a similar property of conditional expectations in 
classical probability theory (cf.\ \cite{Wil}).                                                                                     

In analogy to the classical case we are particularly 
interested in the conditional expectation which leaves a given state $\rho$ on 
$\AC$ invariant, i.e.\ $\rho\circ\EC=\rho$. 
However such a map does not always exist, but if it exists then it is unique \cite{Tak} and 
will be denoted $\EC_\rho$. 
Using $\EC_\rho$ we can extend states $\sigma$ on $\BC$ to states 
$\sigma \circ \EC_\rho$ of $\AC$ which should be interpreted as the updated state of $\AC$ 
after receiving the information (for instance through a measurement) that the subsystem 
$\BC$ is in the state $\sigma$ (cf.\ \cite{Kum3}).                     \\
In the remainder of this section we will construct the conditional expectation $\EC_\rho$
from a von Neumann algebra $\AC$ onto its \emph{center} 
$\CC := \{C \in \AC;\ AC = CA \mbox{\ for all\ } A \in \AC\}$ 
leaving a given state $\rho$ on $\AC$ invariant. The center $\CC$ is a commutative
von Neumann algebra and is therefore isomorphic to some $L^\infty(\Omega, \Sigma, \mathbb{P})$. 
In our guiding example the center of $\BC\tens\ell^\infty(\Omega)$ is $\ell^\infty(\Omega)$. 
Later on (see section \ref{QSC}) this role will be played by the 
commutative algebra of the observed process with $\Omega$ the space of all paths of 
measurement records.                               \\
\begin{stel}\label{condexp}
There exists a unique conditional expectation $\EC_\rho: \AC \to \CC$ which 
leaves the state $\rho$ on $\AC$ invariant.
\end{stel}
\begin{proof}
The proof is based on the central decomposition of $\AC$ \cite{KaR}. In our guiding example, 
$\BC\tens\ell^\infty(\Omega)$ is isomorphic to $\oplus_{i\in\Omega}\BC_i$ where 
the $\BC_i$'s are copies of $\BC$. In general we can identify the center $\CC$ with some 
$L^\infty(\Omega, \Sigma, \mathbb{P})$ where $\mathbb{P}$ corresponds to the restriction of 
$\rho$ to $\CC$. We will ignore for simplicity all issues related with measurability in the 
following constructions. 
The Hilbert space $\BB{H}$ has a direct integral representation 
$\BB{H}=\int^\oplus_\Omega \BB{H}_\omega\mathbb{P}(\mbox{d}\omega)$ in the sense that there exists 
a family of Hilbert spaces $\{\BB{H}_\omega\}_{\omega\in\Omega}$ and for any 
$\psi\in\BB{H}$ there exists a map $\omega \mapsto \psi_\omega \in \BB{H}_\omega$ such that 
 \begin{equation*}
 \langle \psi,\phi\rangle = \int_\Omega \langle \psi_\omega, \phi_\omega\rangle 
 \BB{P}(d\omega).
 \end{equation*}
The von Neumann algebra $\AC$ has a 
{\it central decomposition} $\AC= \int^\oplus_\Omega \AC_\omega \BB{P}(d\omega)$ in the sense 
that there exists a family $\{\AC_\omega\}_{\omega\in\Omega}$ of von Neumann algebras 
with trivial center, or factors, and for any $A\in\AC$ there is a map 
$\omega \mapsto A_\omega \in \AC_\omega$ such that $(A\psi)_\omega=A_\omega\psi_\omega$ 
for all $\psi\in\BB{H}$ and $\mathbb{P}$-almost all $\omega\in\Omega$. 
The state $\rho$ on $\AC$ has a decomposition in states $\rho_\omega$ on $\AC_\omega$ 
such that for any $A\in\AC$ its expectation is obtained by integrating with 
respect to $\mathbb{P}$ the expectations of its components $A_\omega$:
 \begin{equation}\label{eq.statepreserving}
 \rho(A) =  \int_\Omega \rho_\omega(A_\omega)\BB{P}(d\omega).
 \end{equation}
The map $\EC_\rho:\ \AC \to \CC$ defined by 
 \begin{equation*}
 \EC_\rho(A):\ \omega\mapsto\rho_\omega(A_\omega) 
 \end{equation*}
for all $A\in\AC$ is the desired conditional expectation. One can easily verify that this 
map is linear, identity preserving, 
completely positive (as a positive map onto a commutative von Neumann algebra), and has the 
module property. Thus, $\EC_\rho$ is a conditional expectation and leaves the state $\rho$ 
invariant by \ref{eq.statepreserving}. Uniqueness follows from \cite{Tak}.
\end{proof}

It is helpful to think of the state $\rho$ and an arbitrary operator $A$ as maps 
$\rho_\bullet:~\omega\mapsto \rho_\omega$, and respectively $A_\bullet:~\omega\mapsto A_\omega$. 
The conditional expectation $\EC_\rho(A)$ is the function 
$ \rho_\bullet(A_\bullet):~\omega\mapsto \rho_\omega(A_\omega)$.

\section{The dilation}\label{dilation}

Let $\BC$ be the observable algebra of a given quantum system on the Hilbert space $\BB{H}$. 
In the case of resonance fluorescence $\BC$ will be all $2 \times 2$ matrices $M_2$, 
the algebra of observables for the $2$-level atom. 
The irreversible evolution of the system in the Heisenberg picture is given by the 
norm continuous semigroup $\{T_t\}_{t\ge 0}$ of completely positive maps 
$T_t:\ \BC \to \BC$. By Lindblad's theorem 
\cite{Lin} we have $T_t = \exp(tL)$ where the generator $L:\ \BC \to \BC$ has the following action 
  \begin{equation}\label{eq.lindblad}
  L(X) = i[H,X] + \sum_{j=1}^k V_j^*XV_j - \frac{1}{2}\{V_j^*V_j, X\},
  \end{equation} 
where $H$ and the $V_j$'s are fixed elements of $\BC$, $H$ being selfadjoint.                          \\
We can see the irreversible evolution as stemming from a {\it reversible} 
evolution of the system $\BC$ coupled to an environment, 
which will be the electromagnetic field. 
We model a channel in the field by the bosonic or symmetric Fock space over the 
Hilbert space $L^2(\BB{R})$ of square integrable wave functions on the real line, i.e.\ 
 \begin{equation*}
 \FC := \BB{C} \oplus \bigoplus_{n=1}^\infty L^2(\BB{R})^{\ten_sn}.
 \end{equation*} 
The algebra generated by the field observables on $\FC$  contains all bounded operators and 
we denote it by $\WC$. For the dilation we will need $k$ independent copies of this algebra 
$\WC^{\ten k}$.                         \\
The free evolution of the field is given by the unitary group $S_t$, 
the second quantization of the {\it left} shift $s(t)$ on $L^2(\BB{R})$ , 
i.e. $s(t):f\mapsto f(\cdot +t)$. 
In the Heisenberg picture the evolution on $\WC$ is 
 \begin{equation*}
 W\mapsto S_t^*WS_t:=\mbox{Ad}[S_t](W).
 \end{equation*} 
The atom and field together form a closed quantum system, thus their joint evolution is 
given by a one-parameter group $\{\hat{T}_t\}_{t \in \BB{R}}$ 
of $*$-automorphisms on $\BC \ten \WC^{\ten k}$:
 \begin{equation*}
 X\mapsto \hat{U}_t^*X\hat{U}_t:=\mbox{Ad}[\hat{U}_t](X).
 \end{equation*}  
The group $\hat{U}_t$ is a perturbation of the free evolution without interaction. 
We describe this perturbation by the family of unitaries  
$U_t:=S^{\ten k}_{-t}\hat{U}_t$ for all $t\in\BB{R}$ satisfying the {\it cocycle} identity 
 \begin{equation*}
 U_{t+s} = S_{-s}^{\ten k} U_t S_{s}^{\ten k} U_s,\qquad \mbox{for all}~t,s\in\BB{R}. 
 \end{equation*}  
The direct connection between the reduced evolution of the atom given by \eqref{eq.lindblad} and 
the cocycle $U_t$ is one of the important results of quantum stochastic calculus \cite{HuP} 
which makes the object of Section \ref{QSC}. 
For the moment we only mention that in the Markov limit, 
$U_t$ is the solution of the stochastic differential equation \cite{HuP}, \cite{Par}, \cite{Mey}
  \begin{equation}\label{HuP1}
  dU_t = \{V_j dA_{j}^*(t) - V_{j}^*dA_{j}(t) -(iH + \frac{1}{2} V^*_jV_j)dt\}U_t, \qquad  U_0 
  = \mathbf{1},
  \end{equation}
where the repeated index $j$ is meant to be summed over.  
The quantum Markov dilation can be summarized by the following diagram 
(see \cite{Kum1}, \cite{Kum2}):
  \begin{equation}\label{dildiag}\begin{CD}
     \BC @>T_t>> \BC              \\
     @V{\Id \ten \mathbf{1}^{\ten k}}VV        @AA{\Id \ten \phi^{\ten k}}A      \\
     \BC\ten\WC^{\ten k} @>\hat{T}_t>> \BC\ten \WC^{\ten k}           \\
  \end{CD}\end{equation}
i.e.\ for all $X \in \BC:\ T_t(X) = \big(\Id \ten \phi^{\ten k}\big)
\big(\hat{T}_t(X\ten \mathbf{1}^{\ten k})\big)$, where $\phi$ is the vacuum  state on $\WC$, 
and $\mathbf{1}$ is the identity operator in $\WC$. Any dilation of the semigroup $T_t$ with 
Bose fields is unitarily equivalent with the above one under certain minimality requirements.  
The diagram can also be read in the Schr\"odinger picture if we reverse the arrows: 
start with a state $\rho$ of the system $\BC$ in the upper right hand corner, 
then this state undergoes the following sequence of maps:
  \begin{equation*}
  \rho \mapsto 
  \rho \ten \phi^{\ten k} \mapsto 
  (\rho \ten \phi^{\ten k})\circ\hat{T}_t =\hat{T}_{t*}(\rho \ten \phi^{\ten k})
  \mapsto 
  \mbox{Tr}_{\FC^{\ten k}}\big(\hat{T}_{t*}(\rho \ten \phi^{\ten k})\big).
  \end{equation*} 
This means that at $t=0$, the atom in state $\rho$ is coupled to the $k$ channels in the 
vacuum state, and after $t$ seconds of unitary evolution we take the 
partial trace taken over the $k$ channels.                                                                           

We would now like to introduce the measurement process. It turns out that this can be best 
described in the {\it interaction picture}, where we let the shift part of 
$\hat{U}_t=S^{\tens k}_tU_t$ act on the observables while the cocycle part acts on the states:
 \begin{equation}\label{eq.stateinteraction}
 \rho^t(X) := \rho\ten\phi^{\ten k}(U_t^*XU_t)
 \end{equation} 
for all $X \in \BC\ten \WC^{\ten k}$. It is well known that for the Bose field for arbitrary 
time $t$ we can split the noise algebra as a tensor product
 \begin{equation*}
 \WC = \WC_{0)} \ten \WC_{[0,t)} \ten \WC_{[t}
 \end{equation*} 
with each term being the algebra generated by those fields over test functions with support in 
the corresponding subspace of $L^2(\BB{R})$:
 \begin{equation*}
L^2(\BB{R}) = 
L^2\big((-\infty, 0)\big)\oplus L^2\big([0, t)\big) \oplus L^2\big([t, \infty)\big).
\end{equation*} 
Such a continuous tensor product structure is called a {\it filtration} and it is essential in the 
development of quantum stochastic calculus reviewed in Section \ref{QSC}. 
The observables which we measure in an arbitrary time interval $[0,t)$ form a commuting 
family of selfadjoint operators $\{Y_s\}_{0\leq s\leq t}$ 
whose spectral projections belong to the  middle part of the tensor product $\WC_{[0,t)}$. 
In the Davies process $Y_s=\Lambda(s)$, i.e.\ the number operator up to time $s$, while  in 
the homodyne case $Y_s=X_\phi(s)$. 
Notice that the part $\WC_{0)}$ will not play any significant role as it corresponds to ``what 
happened before we started our experiment''.

Let $\CC_t$ be the commutative von Neumann generated by 
the observed process up to time $t$, $\{Y_s\}_{0 \le s \le t}$ $(t \ge 0)$, 
seen as a subalgebra of $\BC\ten \WC^{\ten k}$. By a theorem on von Neumann algebras, 
$\CC_t$ is equal to the double commutant of the observed process up 
to time $t$: $\CC_t =\{Y_s;\ 0\le s \le t\}''$, with the \emph{commutant} $\SC'$ of a 
subset $\SC$ of $\BC\ten \WC^{\ten k}$ being defined by 
$\SC' := \{X \in \BC\ten \WC^{\ten k};\ XS = SX\ \forall S\in \SC \}$. 
The algebras $\{\CC_t\}_{t\geq 0}$ form a growing family, that is 
$\CC_s\subset\CC_t$ for all $s\leq t$. Thus we can define the 
inductive limit $\CC_{\infty}:=\lim_{t\to\infty}\CC_t$, which is the smallest von Neumann 
algebra containing all $\CC_{t}$. On the other hand for each $t\geq 0$ we have a 
state on $\CC_t$ given by the restriction of the state $\rho^t$ of the whole system 
defined by \eqref{eq.stateinteraction}. We will show now that the states 
$\rho^t$ for different times ``agree with each other''.
\begin{stel} 
On the commutative algebra $\CC_{\infty}$ there exists a unique 
state $\rho^\infty$ which coincides with $\rho^t$ when restricted to $\CC_t\subset\CC_{\infty}$, 
for all $t\geq 0$. In particular there exists a measure space 
$(\Omega,\Sigma,\mathbb{P}_\rho)$ such that $(\CC_\infty,\rho^\infty)$ is isomorphic with 
$L^\infty(\Omega,\Sigma,\mathbb{P}_\rho)$ and a growing family $\{\Sigma_t\}_{t\geq 0}$ of 
$\sigma$-subalgebras of $\Sigma$ such that  
$(\CC_t,\rho^t)\cong L^\infty(\Omega,\Sigma_t,\mathbb{P}_\rho)$.
\end{stel}

\begin{proof} 
In the following we will drop the extensive notation of tensoring identity operators 
when representing operators in $\WC_{[s,t)}$ for all $s,t\in\BB{R}$. 
Let $X\in \CC_s$, in particular $X\in\WC_{[0,s)}^{\ten k}$. 
By \eqref{HuP1}, $U_t\in \BC\ten\WC_{[0,t)}^{\ten k}$, because the coefficients of the stochastic 
differential equation lie in $\BC \ten \WC_{[0,t)}^{\ten k}$. This implies that 
$S_{-s}^{\ten k}U_tS_s^{\ten k}\in \BC\ten \WC_{[s,t+s)}^{\ten k}$. Using the tensor product structure 
of $\WC^{\ten k}$, we see that $\WC_{[0,s)}^{\ten k}$ and $\BC\ten\WC_{[s,t+s)}^{\ten k}$ commute, 
and in particular $X$ commutes with $S_{-s}^{\ten k}U_tS_s^{\ten k}$. Then 
\begin{eqnarray}\label{consistent}
  \rho^{t+s}(X) &=& \rho^{0}(U_{t+s}^* X U_{t+s}) = 
  \rho^0(U_s^*(S_{-s}^{\ten k}U_tS_s^{\ten k})^*  X S_{-s}^{\ten k}U_tS_s^{\ten k}U_s)  
  \nonumber\\
 &=&\rho^0\big(U_s^* X U_s\big) =  \rho^s(X). 
  \end{eqnarray}
This implies that the limit state $\rho^\infty$ on $\CC_\infty$ with the desired 
properties exists, in analogy to the Kolmogorov extension theorem for probability measures. 
As seen in the previous section, $(\CC_\infty,\rho^\infty)$ is isomorphic to 
$L^\infty(\Omega, \Sigma, \BB{P}_\rho)$ for some 
probability space $(\Omega, \Sigma, \BB{P}_\rho)$. 
The subalgebras $(\CC_t,\rho^t)$ are isomorphic to 
$L^\infty(\Omega,\Sigma_t,\mathbb{P}_\rho)$ for some growing family $\{\Sigma_t\}_{t\geq 0}$ of 
$\sigma$-subalgebras of $\Sigma$.               
\end{proof} 

\noindent\textbf{Remark.}                                     
From spectral theory it follows that the measure space $(\Omega^t,\Sigma_t)$ coincides with the 
joint spectrum of $\{Y_s\}_{s\leq t}$, i.e.\ $\Omega^t$ is the set of all paths of the 
process up to time $t$. For the example of the counting process this means that $\Omega^t$ is 
the Guichardet space of the interval $[0,t)$, which is the set of all sets of instants 
representing a "click" of the photon counter, i.e.\ it is the set of all paths of the 
counting process.              

We define now $\ \AC_t := \CC_t'$ for all $t\geq 0$ , 
i.e.\ $\AC_t$ is the commutant of $\CC_t$, then 
$\CC_t$ is the center of the von Neumann algebra $\AC_t$. Notice that the observable algebra 
of the atom $\BC$ is contained in $\AC_t$. By Theorem \ref{condexp} we can construct a family 
of conditional expectations $\{\EC_{\rho^t}^t:\ \AC_t \to \CC_t\}_{t\ge 0}$. 
For each $t$, $\EC_{\rho^t}^t$ depends on the state of the ``world'' at that moment 
$\rho^t$, keeping this in mind we will simply denote it by $\EC^t$. An important property
of $\EC^t$ is that $\rho^\infty \circ \EC^t = \rho^t \circ \EC^t = \rho^t$, since the range
of $\EC^t$ is $\CC_t$ and $\EC^t$ leaves $\rho^t$ invariant. 

For an element $X \in \AC_t$, $\EC^t(X)$ is an element in $\CC_t$, i.e.\ a 
function on $\Omega_t$. Its value in a point $\omega\in\Omega_t$, i.e.\ an outcomes 
record up to time $t$, is the expectation value of $X$ given the observed path 
$\omega$ after $t$ time units. 
We will use the notation $\EC^t(X):=\rho^t_\bullet(X_\bullet)$ defined in the end of Section 
\ref{conditional expectation} to emphasise the fact that this is a function on $\Omega_t$. 
When restricted to $\BC\tens \CC_t$ the conditional expectation is precisely of the type 
discussed in our guiding example in Section \ref{conditional expectation}.\\
There exists no conditional expectation from $\BC \ten \WC$ onto $\CC_t$ since 
performing the measurement has \emph{demolished} the information about observables that 
do not commute with the observed process \cite{Bel}. 
We call $\AC_t$ the algebra of observables that are 
\emph{not demolished} \cite{Bel} by observing the process $\{Y_s\}_{0\le s\le t}$. 
This means that performing the experiment and ignoring the outcomes gives the same time 
evolution on $\AC_t$ as when no measurement was done. 

From classical probability it follows that 
for all $t\ge 0$ there exists a unique conditional expectation $\BB{E}^t_\rho:\ \CC_\infty \to \CC_t$
that leaves the state $\rho^\infty$ invariant, i.e.\ $\rho^\infty \circ \BB{E}^t_\rho = \rho^\infty$.
These conditional expectations have the {\it tower property}, 
i.e.\ $\BB{E}^s_\rho\circ\BB{E}^t_\rho = \BB{E}^s_\rho$ for all $t \ge s \ge 0$, which is often very
useful in calculations. $\BB{E}^0_\rho$ is the expectation with respect to $\BB{P}_\rho$, and will
simply be denoted $\BB{E}_\rho$. Note that the tower property for $s=0$ is exactly the invariance of 
the state $\rho^\infty (= \BB{E}_\rho)$.

\section{Quantum stochastic calculus}\label{QSC}

In this section we briefly discuss the quantum stochastic calculus developed by Hudson and 
Parthasarathy \cite{HuP}. For a detailed treatment of the subject we refer to 
\cite{Par} and \cite{Mey}. 
Let $\FC(\HC)$ denote the symmetric (or bosonic) Fock space over the one particle space 
$\HC :=\mathbb{C}^k\ten L^2(\BB{R}_+) = L^2(\{1,2,\ldots k\}\times \BB{R}_+)$. 
The space $\mathbb{C}^k$ describes the $k$ channels we identified in the 
electromagnetic field. As in the previous section we denote the algebra of bounded operators 
on the one channel Fock space $\FC (\BB{R}_+)$ by $\WC$, and on the $k$ channels 
$\FC (\HC)$ by $\WC^{\tens k}$. \\
For every $f \in \HC$ we define the \emph{exponential vector} $e(f) \in \FC(\HC)$ 
in the following way:
\begin{equation*}
e(f) := 1 \oplus \bigoplus_{n=1}^\infty \frac{1}{\sqrt{n!}}f^{\ten n}, 
\end{equation*}
which differs from the coherent vector by a normalization factor. The inner products of
two exponential vectors $e(f)$ and $e(g)$ is 
$\langle e(f),e(g)\rangle=\mbox{exp}(\langle f,g\rangle)$. 
Note that the span of all exponential vectors, denoted $\DC$, forms a dense subspace of 
$\FC(\HC)$. 
Let $f_j$ be the $j$'th component of $f \in \HC$ for $j=1,2,\ldots,k$. 
The annihilation operator $A_j(t)$, creation operator $A^*_j (t)$ and number operator 
$\Lambda_{ij} (t)$ are defined on the domain $\DC$ by
  \begin{equation*}\begin{split}
  &A_j(t)e(f) = \langle \chi_{[0,t]},\, f_j \rangle e(f) = \int_0^t f_j(s)ds~ e(f)\\
  &\big\langle e(g),\, A^*_j(t) e(f) \big\rangle = 
   \langle g_j,\, \chi_{[0,t]} \rangle \big\langle e(g),\, e(f) \big\rangle = 
   \int_0^t \overline{g}_j(s)ds ~ \mbox{exp}(\langle f,g\rangle)           \\
  &\big\langle e(g),\, \Lambda_{ij}(t) e(f) \big\rangle =
  \langle g_i,\, \chi_{[0,t]}f_j\rangle\big\langle e(g),\, e(f) \big\rangle = 
  \int_0^t \overline{g}_i(s) f_j(s)ds ~\mbox{exp}(\langle f,g\rangle) .
  \end{split}\end{equation*} 
The operator $\Lambda_{ii}(t)$ is the usual counting operator for the $i$'th channel. 
Let us write $L^2(\BB{R^+})$ as direct sum $L^2([0,t]) \oplus L^2([t,\infty])$, 
then $\FC(L^2(\BB{R}_+))$ is unitarily equivalent with 
$\FC(L^2([0,t])\ten \FC(L^2[t,\infty))$ through the identification 
$e(f) \cong e(f_{t]}) \ten e(f_{[t})$, with $f_{t]} = f\chi_{[0,t]}$ and 
$f_{[t} = f\chi_{[t,\infty)}$. We will also use the notation $f_{[s,t]}$ for $f\chi_{[s,t]}$ 
and omit the tensor product signs between exponential vectors. 
The same procedure can be carried out for all the $k$ channels.   
                      
Let $M_t$ be one of the processes $A_j(t), A^*_j(t)$ or $\Lambda_{ij}(t)$.
The following factorisation property \cite{HuP}, \cite{Par} makes  
the definition of stochastic integration against $M_t$ possible
  \begin{equation*}
  (M_t - M_s)e(f) = e(f_{s]})\big\{(M_t-M_s) e(f_{[s,t]})\big\}e(f_{[t}), 
  \end{equation*} 
with $(M_t-M_s) e(f_{[s,t]}) \in \FC\big(\mathbb{C}^k\tens L^2([s,t])\big)$. 
We first define the stochastic integral for the so called  
{\it simple} operator processes  with values in the 
atom and noise algebra $\BC \ten \WC^{\tens k}$ where $\BC:=M_n$.

\begin{de} 
Let $\{L_s\}_{0 \le s \le t}$ be an adapted 
(i.e.\ $L_s \in \BC \ten \WC_{s]}$ for all $0 \le s \le t$) simple 
process with respect to the partition $\{s_0=0, s_1,\dots, s_p= t\}$ in the sense that 
$L_s = L_{s_j}$ whenever $s_j \le s < s_{j+1}$. Then the stochastic integral of 
$L$ with respect to $M$ on $\BB{C}^n \ten \DC$ is given by \cite{HuP}, \cite{Par}:  
  \begin{equation*}
  \int_0^t L_s dM_s  ~f e(u) := 
\sum_{j=0}^{p-1} \big(L_{s_j}fe(u_{s_j]})\big)\big((M_{s_{j+1}}- M_{s_{j}})
   e(u_{[s_j, s_{j+1}]})\big)e(u_{[s_{j+1}}).
  \end{equation*}
\end{de}
By the usual approximation by simple processes we can extend the definition of the 
stochastic integral 
to a large class of {stochastically integrable processes} \cite{HuP}, \cite{Par}. 
We simplify our notation by writing $dX_t = L_tdM_t$ for 
$X_t = X_0 + \int_0^t L_sdM_s$. Note that the definition of the stochastic integral 
implies that the increments 
$dM_s$ lie in the future, i.e.\ $dM_s \in \WC_{[s}$. Another consequence of the definition 
of the stochastic 
integral is that its expectation with respect to the vacuum state $\phi$ is always $0$ due to 
the fact that the increments $dA_j, dA^*_j,d\Lambda_{ij}$ have zero expectation values in the 
vacuum. This will often simplify calculations of expectations, our strategy being that of 
trying to bring these increments to act on the 
vacuum state thus eliminating a large number of differentials. 

The following theorem of Hudson and Parthasarathy extends the It\^o rule 
of classical probability theory.\\
\begin{stel}\label{Itorule}\textbf{(Quantum It\^o rule \cite{HuP}, \cite{Par})}
Let $M_1$ and $M_2$ each be one of the processes $A_j, A^*_j$ or $\Lambda_{ij}$. Then 
$M_1M_2$ is an adapted process satisfying the relation:
  \begin{equation*}
  d(M_1M_2) = M_1dM_2 + M_2dM_1 + dM_1dM_2,
  \end{equation*}
where $dM_1dM_2$ is given by the quantum It\^o table:
\begin{center}
{\large \begin{tabular} {l|lll}
$dM_1\backslash dM_2$ & $dA^*_i$ & $d\Lambda_{ij}$ & $dA_i$ \\
\hline 
$dA^*_k$ & $0$ & $0$ & $0$ \\
$d\Lambda_{kl}$ & $\delta_{li}dA^*_k$ & $\delta_{li}d\Lambda_{kj}$ & $0$  \\
$dA_k$ & $\delta_{ki}dt$ & $\delta_{ki}dA_j$ & $0$ 
\end{tabular} }
\end{center}
\end{stel}
\noindent\textbf{Notation.} The quantum It\^o rule will be used for 
calculating differentials of products of It\^o integrals. 
Let $\{Z_i\}_{i=1,\dots, p}$ be It\^o integrals, then 
 \begin{equation*} 
 d (Z_1Z_2\dots Z_p)= \sum_{\substack{\nu\subset\{1,\dots, p\} \\ \nu \neq \emptyset}}[\nu]
 \end{equation*}
where the sum runs over all {\it non-empty} subsets of $\{1,\dots, p\}$ and for any 
$\nu=\{i_1,\dots i_k\}$, the term $[\nu]$ is the contribution 
to $d (Z_1Z_2\dots Z_p)$ coming from differentiating only the terms with indices 
in the set $\{i_1,\dots i_k\}$ and preserving the order of the factors in the product. 
For example the differential $d(Z_1Z_2Z_3)$ contains terms of the type 
$[2]=Z_1(dZ_2)Z_3$, $[13]=(dZ_1)Z_2(dZ_3)$, and $[123]=(dZ_1)(dZ_2)(dZ_3)$.

Let $V_j$ for $j = 1,2,\ldots,k$, and $H$ be operators in $\BC$ with $H$ is selfadjoint. 
Let $S$ be a unitary operator on $\BB{C}^n \ten l^2(\{1,2,\ldots, k\})$ with 
$S_{ij}=\langle i, S j \rangle\in\BC$ the ``matrix elements'' in the basis 
$\{|i>:i=1,\dots, k\}$ of $\mathbb{C}^k$. Then there exists a unique unitary 
solution for the following quantum stochastic differential equation \cite{HuP}, \cite{Par}
  \begin{equation}\label{HuP2}
  dU_t = \big\{V_jdA_j^*(t) + (S_{ij}- \delta_{ij})d\Lambda_{ij}(t) - 
  V^*_iS_{ij}dA_j(t) -(iH + \frac{1}{2} V_j^*V_j)dt\big\}U_t, 
  \end{equation}
with initial condition $U_0 = \I$, where again repeated indices have been summed. 
Equation \eqref{HuP1}, providing the cocycle of unitaries perturbing the free evolution of the 
electromagnetic field is an example of such an equation. 
The terms $d\Lambda_{ij}$ in equation $\eqref{HuP2}$ describe 
direct scattering between the channels in the electromagnetic field \cite{BaL}.
We have omitted this effect for the sake of simplicity, i.e.\ we always take 
$S_{ij} = \delta_{ij}$.  \\ 
We can now check the claim made in Section \ref{dilation} that the dilation diagram 
\ref{dildiag} commutes. It is easy to see that following the lower part of the diagram 
defines a semigroup on $\BC$. We have to show it is generated by $L$. 
For all $X \in \BC$ we have 
 \begin{equation*}
 d~\Id\ten\phi^k\big(\hat{T}_t(X\ten \I^{\ten k})\big)
 = \Id\ten\phi^k(d~U_t^* X\ten \I^{\ten k} U_t). 
 \end{equation*}
Using the It\^o rules we obtain
 \begin{equation*}
 d~U_t^* X\ten \I^{\ten k} U_t = (dU_t^*)X\ten \I^{\ten k} U_t + U_t^*X\ten \I^{\ten k}dU_t + 
 (dU_t^*)X\ten \I^{\ten k} dU_t. 
 \end{equation*}
With the aid of the It\^o table we can evaluate these terms. 
We are only interested in the $dt$-terms since the expectation with respect to the 
vacuum kills the other terms. 
Then we obtain: $d~\Id\ten\phi^k(U_t^* X\ten \I^{\ten k} U_t) = 
\Id\ten\phi^k(U_t^* L(X)\ten \I^{\ten k} U_t)dt$, proving the claim. 

Now we return to the example of resonance fluorescence. Suppose the laser is off,
then we have spontaneous decay of the $2$-level atom into the field which is in the vacuum state. 
For future convenience we already distinguish
a \emph{forward} and a \emph{side} channel in the field, the Liouvillian  is then given by
  \begin{equation*}
  L(X) = i[H, X] + \sum_{\sigma = f,s} V_\sigma^* X V_\sigma - 
  \frac{1}{2}\{V^*_\sigma V_\sigma, X\},  
  \end{equation*}  
where 
  \begin{equation*}
  V = \begin{pmatrix} 0 & 0 \\ 1 & 0 \end{pmatrix},\ \ \ V_f = \kappa_fV,\ \ \ V_s = 
  \kappa_sV,\ \ \ 
  |\kappa_f|^2 + |\kappa_s|^2 = 1, 
  \end{equation*} 
with $|\kappa_f|^2$ and $|\kappa_s|^2$ the decay rates into the forward and side channel 
respectively.   \\
The dilation of the quantum dynamical system $(M_2, \{T_t =\exp(tL)\}_{t \ge 0})$, 
is now given by the closed system 
$(M_2\ten\WC_f\ten\WC_s, \{\hat{T}_t\}_{t\in \BB{R}})$ with unitary 
cocycle given by
  \begin{equation*}
  dU^{sd}_t = \{V_fdA^*_f(t)-V^*_fdA_f(t) + V_sdA^*_s(t) - V_s^*dA_s(t) - 
  (iH+\frac{1}{2} V^*V)dt\}U^{sd}_t, \ \ 
  U^{sd}_0 = \I,
  \end{equation*}  
where the superscript $sd$ reminds us of the fact that the laser is off, 
i.e.\ we are considering spontaneous decay. We can summarize this in the following 
dilation diagram
  \begin{equation*}\begin{CD}
     \BC @>T_t = \exp(tL)>> \BC              \\
     @V{\Id \ten \I \ten \I}VV        @AA{\Id \ten \phi\ten\phi}A      \\
     \BC\ten\WC_f\ten\WC_s @>\hat{T}^{sd}_t = \mbox{Ad}[\hat{U}^{sd}_t]>> \BC\ten \WC_f\ten\WC_s           \\
  \end{CD}\end{equation*}
where $\hat{U}^{sd}_t$ is given by $S_t\ten S_t U^{sd}_t$ for $t \ge 0$.      

We change this setting by introducing a laser on the forward channel, i.e.\ 
the forward channel is now in a coherent state (see \ref{eq.coherent}) 
$\gamma_h := \langle\psi(h),\,\cdot\,\psi(h)\rangle$ for some $h \in L^2(\BB{R}_+)$. 
This leads to the following dilation diagram
  \begin{equation}\label{laserdiagram}\begin{CD}
     \BC @>T^h_t>> \BC              \\
     @V{\Id \ten \I \ten \I}VV        @AA{\Id \ten \gamma_h\ten\phi}A      \\
     \BC\ten\WC_f\ten\WC_s @>\hat{T}^{sd}_t = \mbox{Ad}[\hat{U}^{sd}_t]>> \BC\ten \WC_f\ten\WC_s           \\
  \end{CD}\end{equation}
i.e.\ the evolution on $\BC$ has changed and it is in general {\it not} a semigroup. 
Denote by $W(h)$ the unitary \emph{Weyl} or 
\emph{displacement operator} defined on $\DC$ by: 
$W(h)\psi(f) = \exp(-2i\mbox{Im}\langle h, f\rangle)$ $\psi(f+h)$. 
Note that $W(h)\phi = W(h)\psi(0) = \psi(h)$, so that we can write
  \begin{equation*}\begin{split}
  &T^h_t(X) = \mbox{Id} \ten \gamma_h \ten \phi({U^{sd}_t}^*X\ten \I\ten \I U^{sd}_t) = \\
  &\mbox{Id} \ten \phi \ten \phi\big(W_f(h)^* {U^{sd}_t}^*X\ten \I\ten \I U^{sd}_t W_f(h)\big) =  \\
  &\mbox{Id} \ten \phi \ten \phi\big(W_f(h_{t]})^* {U^{sd}_t}^*X\ten \I\ten \I U^{sd}_t W_f(h_{t]})\big),
  \end{split}\end{equation*}                    
where $h_{t]} := h\chi_{[0,t]}$ and $W_f(h):=\I\tens W(h)\tens\I$. 
Defining $U_t := U^{sd}_tW_f(h_{t]})$, together with the stochastic differential 
equation for $W_f(h_{t]})$ \cite{Par}
  \begin{equation*}
  dW_f(h_{t]}) = \{h(t) dA^*_f(t) -\overline{h}(t)dA_f(t) - 
  \frac{1}{2}|h(t)|^2dt\}W_f(h_{t]}),\ \ \ \ W_f(h_0) = \I, 
  \end{equation*}
and the It\^o rules leads to the following quantum stochastic differential equation for $U_t$:
  \begin{eqnarray*}
  dU_t=& \big\{(V_f+h(t))dA^*_f(t) -(V_f^* + \overline{h}(t))dA_f(t)\ + 
  V_sdA^*_s(t) - V^*_sdA_s(t)\ -  \\
  &-\big(iH+\frac{1}{2}(|h(t)|^2 + V^*V + 2h(t)V_f^*)\big)dt~\big\} U_t,\ \ \ U_0 = \I.
  \end{eqnarray*}
Define $\tilde{V}_f := V_f + h(t)$, $\tilde{V}_s := V_s$ and 
$\tilde{H} := H + i\frac{1}{2}(\overline{h}(t)V_f - h(t)V_f^*)$ then this reads
  \begin{equation}\label{cocyclelaser}
dU_t =\ \sum_{\sigma = f,s}\big\{ \tilde{V}_\sigma dA^*_\sigma(t) - 
  \tilde{V}^*_\sigma dA_\sigma 
  -(i \tilde{H} + \frac{1}{2}\tilde{V}^*_\sigma\tilde{V}_\sigma)dt\big\}U_t, \ \ \ U_0 = \I.
  \end{equation}
The time dependent generator of the dissipative evolution in the presence of 
the laser on the forward channel is 
  \begin{equation}\label{timedepgenerator}
  L(X) = i[\tilde{H}, X] + \sum_{\sigma = f,s}\tilde{V}^*_\sigma X\tilde{V}_\sigma - 
       \frac{1}{2}\{\tilde{V}^*_\sigma\tilde{V}_\sigma, X\}.
  \end{equation}
Therefore the diagram for resonance fluorescence \eqref{laserdiagram} is equivalent to
  \begin{equation*}\label{laserII}\begin{CD}
     \BC @>T^h_t>> \BC              \\
     @V{\Id \ten \I \ten \I}VV        @AA{\Id \ten \phi\ten\phi}A      \\
     \BC\ten\WC_f\ten\WC_s @>\hat{T}_t = \mbox{Ad}[\hat{U}_t] >> \BC\ten W_f\ten\WC_s           \\
  \end{CD}\end{equation*}  
where $\hat{U}_t$ is given by $S_t\ten S_t U_t$ for $t \ge 0$.
For $h(t) = -i \frac{\Omega}{2\kappa_f}$, we find the master equation for resonance 
fluorescence \eqref{Master}. From now on we will no longer suppress the oscillations of 
the laser, i.e.\ we take $h(t) = -i \exp(i\omega t)\frac{\Omega}{2\kappa_f}$. Then we find
  \begin{equation*}
  L(X) = i[H, X] - 
  i\frac{\Omega}{2}[e^{-i\omega t}V+e^{i\omega t}V^*, X] - \frac{1}{2}\{V^*V,X\} + V^* X V,
  \end{equation*}
note that the laser is resonant when $\omega = \omega_0$.

\section{Belavkin's stochastic Schr\"odinger equations}\label{Belavkinslemma}

Now we are ready to derive a stochastic differential equation for the process $\EC^t(X)$. 
In the next section we will see that this equation leads to the 
stochastic Schr\"odinger equations \eqref{Belcount} and \eqref{Belhomodyne}, that we 
already encountered in Sections \ref{Daviesprocess} and \ref{Homodynedetection}. 

\begin{de}\label{defMt}
Let $X$ be an element of $\BC:=M_n$. Define the process $\{M^X_t\}_{t\ge0}$ in the algebra 
$\CC_\infty \cong L^\infty(\Omega, \Sigma, \BB{P}_\rho)$, generated by the observed process 
$\{Y_t\}_{t\ge 0}$ (see Section \ref{dilation}) by
  \begin{equation*}
  M^X_t := \EC^t(X) - \EC^0(X) - \int_0^t \EC^r\big(L(X)\big)dr, 
  \end{equation*}
where $L:\ \BC \to \BC$ is the Liouvillian. 
In the following we suppress the superscript $X$ in $M^X_t$ 
to simplify our notation.  
\end{de}

Note that from the above definition it is clear that $M_t$ is an element of $\CC_t$ for all 
$t\ge 0$. The following theorem first appeared (in a more general form and with a different proof) 
in \cite{Bel} and is at the heart of quantum filtering theory. We prove it using the properties 
of conditional expectations. For simplicity we have restricted to observing a process in the 
field $\WC^{\ten k}$. The theory can be extended to processes that are in $\BC\ten \WC^{\ten k}$,
transforming it into a more interesting filtering theory. For the stochastic Schr\"odinger equations
arising in quantum optics our approach is general enough. 

\begin{stel}\label{BelavkinI}
The process $\{M_t\}_{t\ge 0}$ of definition \ref{defMt} is a martingale with respect to 
the filtration $\{\Sigma_t\}_{t\ge 0}$ of $\Omega$ and the measure $\BB{P}_\rho$, i.e.\ 
for all $t \ge s \ge 0$ we have: $\BB{E}_\rho^s(M_t) = M_s$. 
\end{stel} 
\begin{proof}
From the module property of the conditional expectation it follows that $\BB{E}_\rho^s(M_t)$ $= M_s$ 
for $t \ge s \ge 0$ is equivalent to $\BB{E}_\rho^s(M_t -M_s) = 0$ for $t \ge s \ge 0$. This means 
we have to prove for all $t \ge s \ge 0$ and $E \in \Sigma_s$: 
  \begin{equation*}
  \int_E\BB{E}_\rho^s(M_t-M_s)(\omega)\BB{P_\rho}(d\omega) = 0,
  \end{equation*}
which, by the tower property, is equivalent to
  \begin{equation}\label{martprop}
  \int_E (M_t-M_s)(\omega)\BB{P_\rho}(d\omega) = 0,
  \end{equation}
i.e.\  $\BB{E}_\rho\big(\chi_E(M_t-M_s)\big) = 0$. 
Now using Definition \ref{defMt} and again the module property of the conditional 
expectation we find,
writing $E$ also for the projection corresponding to $\chi_E$
  \begin{equation*}\begin{split}
  \BB{E}_\rho\big(\chi_E(M_t-M_s)\big) &= 
  \rho^\infty\Big(\EC^t(X\ten E) - \EC^s(X\ten E) - 
   \int_s^t\EC^r\big(L(X)\ten E\big)dr\Big)                                 \\
  &= \rho^t(X \ten E) - \rho^s(X \ten E) - \int_s^t \rho^r\big(L(X)\ten E\big)dr. 
  \end{split}\end{equation*}
This means we have to prove: $d\rho^t(X \ten E) - \rho^t\big(L(X) \ten E\big)dt = 0$,
for all $t \ge s$. Note that $\rho^t(X\ten E) = \rho^0(U^*_t X \ten E U_t) = 
\rho \ten \phi^{\ten k}(U^*_t X \ten E U_t)$. Therefore $d\rho^t(X\ten E) = 
\rho \ten \phi^{\ten k}\big(d(U^*_t X\ten E U_t)\big)$. We will use the notation
below Theorem \ref{Itorule} with $Z_1 = U^*_t$ and $Z_2 = X \ten E U_t$. Using
the quantum It\^o table and the fact that only the $dt$ terms 
survive after taking a vacuum expectation, we find:
  \begin{equation*}\begin{split}
  d\rho^0(U^*_tX \ten EU_t) &= \rho^0\big([1]\big) + \rho^0\big([2]\big) + 
     \rho^0\big([12]\big), \ \ \mbox{where} \\
  \rho^0\big([1]\big) + \rho^0\big([2]\big) & = 
     \rho^0\big(U^*_t(i[H,X]\ten E - \frac{1}{2}\{V_j^*V_j, X\}\ten E) U_t\big)dt \\
  \rho^0\big([12]\big) & = \rho^0\big(U^*_t(V^*_jXV_j)\ten EU_t\big)dt. 
  \end{split}\end{equation*}
This means $d\rho^t(X\ten E) = \rho^t\big(L(X)\ten E\big)dt$, for all $t \ge s$, 
proving the theorem.
\end{proof}
Note that in the proof of the above theorem we have used that the projection $E \in \CC_s$ commutes 
with the increments $dA_{j}(s),\, dA^*_{j}(s),\, ds$ and with the processes in front of the 
increments in equation \eqref{HuP1}, i.e.\ $V_j,\, V_j^*,\, V_j^*V_j$ and $H$. 
If the theory is extended to a more 
general filtering theory \cite{Bel}, then these requirements become real restrictions on 
the process $\{Y_t\}_{t \ge 0}$. If they are satisfied the observed process 
$\{Y_t\}_{t \ge 0}$ is said to be \emph{self non demolition} \cite{Bel}.          \\
Definition \ref{defMt} implies the following stochastic differential equation 
for the process $\EC^t(X)$
  \begin{equation}\label{Belav1}
  d\EC^t(X) = \EC^t\big(L(X)\big)dt + dM_t, 
  \end{equation}
called the \emph{Belavkin equation}. The only thing that remains to be done is linking the 
increment $dM_t$ to the increment of the observed process $dY_t$.                        \\
Let us assume that the observed process $\{Y_t\}_{t \ge 0}$ satisfies a quantum stochastic 
differential equation
  \begin{equation*}
  dY_t =  \alpha_j(t) dA^*_{j}(t) + \beta_{ij}(t) d\Lambda_{ij}(t) +  \alpha_j^*(t)dA_{j}(t)  + 
         \delta(t) dt,
  \end{equation*}
for some adapted stochastically integrable processes 
$\alpha_j, \beta_{ij}$, and $\delta$, such that 
$\alpha_j(t),$ $\beta_{ij}(t),$ $\delta(t)\in\WC^{\ten k}_{t]}$ for all $t\geq 0$, and 
$\beta_{ij}^*=\beta_{ji},\, \delta=\delta^*$ since $Y_t$ is selfadjoint. Furthermore, since 
the observed process $\{Y_t\}_{t\ge 0}$ is commutative, 
we have $[dY_t, Y_s] = 0$ for all $s \le t$, which 
leads to
  \begin{equation*}\begin{split}
  &[\alpha_j(t), Y_s]dA^*_j(t) + [\beta_{ij}(t), Y_s]d\Lambda_{ij}(t) + [\alpha^*_j(t), Y_s]dA_j(t)
  + [\delta(t), Y_s]dt = 0  \ \  \Rightarrow \\
  & [\alpha_j(t), Y_s] = 0, \ \ [\beta_{ij}(t), Y_s] = 0, \ \ [\alpha^*_j(t), Y_s] = 0, \ \ 
  [\delta(t), Y_s] = 0,
  \end{split}\end{equation*}
i.e.\ $\alpha_j(t), \beta_{ij}(t), \alpha^*_j(t), \delta(t) \in \AC_t$.
This enables us to define a process $\tilde{Y}_t$ by 
  \begin{equation}\label{dYtilde}\begin{split}
  d\tilde{Y}_t =&\ \Big( \alpha_j(t)dA^*_j(t) - \EC^t\big(V_j^*\alpha_j(t)\big)dt \Big)\ + 
            \Big(\beta_{ij}(t) d\Lambda_{ij}(t) - \EC^t\big(V_i^*\beta_{ij}(t)V_j\big)dt\Big)\ + \\
                &\Big(\alpha_j^*(t)dA_j(t) - \EC^t\big(\alpha_j^*(t)V_j\big)dt\Big) , 
           \ \ \ \ \ \tilde{Y}_0 = 0,  
  \end{split}\end{equation} 
i.e.\ we have the following splitting of $Y_t$:
  \begin{equation}\label{DoobMeyer}
  Y_t = Y_0 + \tilde{Y}_t + \int_0^t \Big(\EC^s\big(V_j^*\alpha_j(s)\big) + 
  \EC^s\big(V^*_i\beta_{ij}(s)V_j\big) 
   + \EC^s\big(\alpha_j^*(s)V_j\big) + \delta(s)\Big)ds, 
  \end{equation}
which in view of the following theorem is the semi-martingale splitting of $Y_t$. The process 
$\tilde{Y}_t$ is called the \emph{innovating martingale} of the observed process $Y_t$.

\begin{stel}\label{BelavkinII}
The process $\{\tilde{Y}_t\}_{t \ge 0}$ is a martingale with respect to 
the filtration $\{\Sigma_t\}_{t\ge 0}$ of $\Omega$ and the measure $\BB{P}_\rho$, i.e.\ 
for all $t \ge s \ge 0$ we have: $\BB{E}_\rho^s(\tilde{Y}_t) = \tilde{Y}_s$.
\end{stel}
\begin{proof}
We need to prove that for all $t \ge s \ge 0:\ \BB{E}^s_\rho(\tilde{Y}_t -\tilde{Y}_s) = 0$. 
This means we have to prove for all $t \ge s \ge 0$ and $E \in \Sigma_s$:
  \begin{equation*}\begin{split}
  &\int_E \BB{E}^s_\rho(\tilde{Y}_t -\tilde{Y}_s)(\omega) \BB{P}_\rho(d\omega) = 0 \iff 
   \int_E (\tilde{Y}_t -\tilde{Y}_s)(\omega) \BB{P}_\rho(d\omega) = 0 \iff 
  \BB{E}_\rho \Bigg(Y_tE - Y_sE \\ 
  &- \int_s^t \Big(\EC^r\big(V_j^*\alpha_j(r)\big)E + 
  \EC^r\big(V^*_i\beta_{ij}(r)V_j\big)E 
   + \EC^r\big(\alpha_j^*(r)V_j\big)E + \delta(r)E\Big)dr\Bigg) = 0 \iff \\
  &\rho^t(Y_tE) - \rho^s(Y_sE) =  \\ 
  &\int_s^t \rho^r\Big(\EC^r\big(V_j^*\alpha_j(r)\big)E + 
  \EC^r\big(V^*_i\beta_{ij}(r)V_j\big)E 
  + \EC^r\big(\alpha_j^*(r)V_j\big)E + \delta(r)E\Big)dr.
  \end{split}\end{equation*}
For $t=s$ this is okay, so it remains to be shown that for all 
$t \ge s \ge 0$ and $E \in \Sigma_s$:
  \begin{equation*}\begin{split}
  &d\rho^t(Y_tE) = \rho^t\Big(\EC^t\big(V_j^*\alpha_j(t)\big)E +
  \EC^t\big(V^*_i\beta_{ij}(t)V_j\big)E + \EC^t\big(\alpha_j^*(t)V_j\big)E + 
  \delta(t)E\Big)dt\ \iff \\
  &d\rho^0(U_t^*Y_tEU_t) = \rho^t\Big(\EC^t\big(V_j^*\alpha_j(t)\big)E +
  \EC^t\big(V^*_i\beta_{ij}(t)V_j\big)E + \EC^t\big(\alpha_j^*(t)V_j\big)E + \delta(t)E\Big)dt.
  \end{split}\end{equation*}
We define: $Z_1(t) := U^*_t,\ Z_2(t) := Y_tE$ and $Z_3(t) := U_t$ then we find, using the
notation below Theorem \ref{Itorule}: 
$d\rho^0(U_t^*Y_tEU_t) = \rho^0([1] + [2] + [3] + [12] + [13] + 
[23] + [123])$. 
Remember $\rho^0 = \rho \ten \phi^{\ten k}$, i.e.\ we are only interested in the $dt$ terms, 
since the vacuum kills all other terms. The terms $[1], [3]$ and $[13]$ together make up 
the usual Lindblad term and since $L(\I) = 0$ we do not have to consider them.                  \\
Furthermore, term [2] contributes $U^*_t\delta(t)EU_tdt$, term [12] contributes 
$U^*_tV_j^*\alpha_j(t)E$ $U_tdt$,
term $[23]$ contributes $U^*_t\alpha^*_j(t)V_jEU_tdt$ and term $[123]$ contributes 
$U^*_tV_i^*\beta_{ij}(t)V_j$ $U_tdt$, therefore we get
  \begin{equation*}\begin{split}
  &d\rho^0(U_t^*Y_tEU_t) = \\
  &\rho^0\big(U^*_t\alpha^*_j(t)V_jEU_t + U^*_tV_i^*\beta_{ij}(t)V_jU_t +
  U^*_tV_j^*\alpha_j(t)EU_t + U^*_t\delta(t)EU_t\big)dt = \\
  &\rho^t\big(\alpha^*_j(t)V_jE + V_i^*\beta_{ij}(t)V_j + V_j^*\alpha_j(t)E + 
     \delta(t)E\big)dt = \\
  &\rho^t\Big(\EC^t\big(V_j^*\alpha_j(t)\big)E +
  \EC^t\big(V^*_i\beta_{ij}(t)V_j\big)E + \EC^t\big(\alpha_j^*(t)V_j\big)E + \delta(t)E\Big)dt,
  \end{split}\end{equation*}
proving the theorem.      
\end{proof} 

\noindent\textbf{Remark.} In the probability literature an adapted process which can be written 
as the sum of a martingale and a finite variation process is called a semimartingale \cite{RoW}. 
The Theorems \ref{BelavkinI} and \ref{BelavkinII} show that $M_t$ and $Y_t$ are semimartingales.

We now represent the martingale $M_t$ from Definition \ref{defMt} as an integral over the 
innovating martingale (cf.\ \cite{Kal}) by
\begin{equation}\label{martingalerep}
dM_t = \eta_t d\tilde{Y}_t
\end{equation} 
for some stochastically integrable process $\eta_t$, which 
together with equation \eqref{DoobMeyer} provides the link between $dM_t$ and $dY_t$. 
We are left with the problem of determining 
$\eta_t$, which we will carry out in the next section for the examples 
of Section \ref{Daviesprocess} and \ref{Homodynedetection}. 
Here we just give the recipe for finding $\eta_t$. 

\textbf{Recipe.} Define for all integrable adapted 
processes $b_t$ and $c_t$ a process $B_t$ in $\CC_\infty$ by
  \begin{equation}\label{eq.Bt}
  dB_t = b_t d\tilde{Y_t} + c_t dt.
  \end{equation}
These processes form a dense subalgebra of $\CC_\infty$. 
Now determine $\eta_t$ from the fact that 
$\EC^t$ leaves $\rho^t$ invariant \cite{Bel}, i.e.\ for all $B_t$
  \begin{equation*}
  \rho^t\big(\EC^t(B_tX)\big) = \rho^t(B_tX). 
  \end{equation*}
From this it follows that for all $B_t$ 
 \begin{equation}\label{eq.for.k}
  d\rho^0\big(U_t^*B_t(\EC^t(X) - X)U_t\big) = 0. 
  \end{equation}
We evaluate the differential $d\big(U_t^*B_t(\EC^t(X) - X)U_t\big)$ using the 
quantum It\^o rules. Since $\rho^0 = \rho \ten \phi^{\ten k}$ we can restrict to the 
$dt$ terms, since the others die on the vacuum. We will use the notation below 
Theorem \ref{Itorule} with $Z_1(t) = U_t^*,\ Z_2(t) = B_t, 
Z_3(t) = \EC^t(X)-X$ and $Z_4(t) = U_t$. The following lemma simplifies the 
calculation considerably.  

\begin{lem} The sum of all terms in which $Z_2$ is not differentiated has zero expectation:  
$\rho^0 ([1]+[3]+[4]+[13]+[14]+[34]+[134])=0$.
\end{lem}

\begin{proof}
The $dt$ terms of $[3]$ are $U^*_tB_t\EC^t\big(L(X)\big)U_tdt$ and 
$-U^*_tB_t\eta_t\big(\EC^t(V^*_j\alpha_j) + \EC^t(V^*_i\beta_{ij}$ $V_j) + 
\EC^t(\alpha^*_jV_j)\big)U_tdt$. Using the fact that $\EC^t$ leaves $\rho^t$ invariant
we see that the term $U^*_tB_t\EC^t\big(L(X)\big)U_tdt$ cancels 
against the $dt$ terms of  $[1], [4]$ and $[14]$, which make up the Lindblad generator $L$
with a minus sign. The other term of $[3]$ is cancelled in expectation against the $dt$ 
terms of $[13], [34]$ and $[134]$, since 
  \begin{equation*}\begin{split}
  &\rho^0([13]) = \rho^t(B_t\eta_t V^*_j\alpha_j)dt =  
    \rho^t\big(\EC^t(B_t\eta_t V^*_j\alpha_j)\big)dt = 
    \rho^t\big(B_t\eta_t \EC^t(V^*_j\alpha_j)\big)dt\\
  &\rho^0([34]) = \rho^t(B_t\eta_t \alpha^*_jV_j)dt =  
    \rho^t\big(\EC^t(B_t\eta_t \alpha^*_jV_j)\big)dt =
    \rho^t\big(B_t\eta_t\EC^t(\alpha^*_jV_j)\big)dt \\
  &\rho^0([134]) = \rho^t(B_t\eta_t V^*_i\beta_{ij}V_j)dt =
     \rho^t\big(\EC^t(B_t\eta_t V^*_i\beta_{ij}V_j)\big)dt = 
     \rho^t\big(B_t\eta_t \EC^t(V^*_i\beta_{ij}V_j)\big)dt.
  \end{split}\end{equation*}  
\end{proof}

Using equation \eqref{dYtilde}, the fact that $\EC^t$ leaves $\rho^t$ invariant and the 
module property, we find that the term $[2]$ has expectation 
zero as well 
 \begin{equation*}\begin{split} 
  &\rho^0([2]) = \rho^t\left(b_td\tilde{Y}_t(\EC^t(X)-X)\right) =  \\
  &-\rho^t\left(b_t\EC^t(V_j^*\alpha_j(t)+\alpha_j^*(t)V_j+V_i^*\beta_{ij}V_j)(\EC^t(X)-X)\right) dt = \\
  &-\rho^t\left(b_t\EC^t(V_j^*\alpha_j(t)+\alpha_j^*(t)V_j+V_i^*\beta_{ij}V_j)\EC^t(\EC^t(X)-X)\right)dt = 
  0.
 \end{split}\end{equation*}
Thus, only the terms containing no $B_t$ nor $c_t$ can contribute non-trivially. This leads to an  
equation allowing us to obtain an expression for $\eta_t$ by solving
  \begin{equation}
  \rho^0([12]+[23]+[24]+[123]+[124]+[234]+[1234])=0.
  \end{equation}
Although this can be carried out in full generality, we will provide the solution only for our 
main examples, the photon counting and homodyne detection experiments for  
a resonance fluorescence setup, in the next section.

\section{Examples}

We now return to the example considered in Section \ref{Daviesprocess}. 
We were considering a $2$-level
atom in interaction with the electromagnetic field. The interaction was given by a cocycle $U_t$ 
satisfying equation \eqref{cocyclelaser}. The observed process is
the number operator in the side channel, i.e.\ $Y_t = \Lambda_{ss}(t)$. 
Therefore $d\tilde{Y}_t = d\Lambda_{ss}(t) - \EC^t(V_s^*V_s)dt$. 
Recall now the notation $Z_1(t) = U_t^*,\ Z_2(t) = B_t, Z_3(t) = \EC^t(X)-X$ and $Z_4(t) = U_t$,
their differentials are given by
  \begin{equation*}\begin{split}
  dU^*_t =&\ U^*_t \sum_{\sigma = f,s} \big\{ \tilde{V}^*_\sigma dA_\sigma(t) - 
\tilde{V}_\sigma dA^*_\sigma(t) 
  - (-i \tilde{H} + \frac{1}{2}\tilde{V}^*_\sigma\tilde{V}_\sigma)dt\big\}                         \\
  dB_t =&\ b_td\Lambda_{ss}(t) + \big(c_t - b_t\EC^t(V_s^*V_s)\big)dt                        \\
  d(\EC^t(X)-X) =&\ \eta_t d\Lambda_{ss}(t) + \Big(\EC^t\big(L(X)\big) - 
\eta_t\EC^t(V_s^*V_s)\Big)dt \\
  dU_t =&\ \sum_{\sigma = f,s}\big\{ \tilde{V}_\sigma dA^*_\sigma(t) - 
\tilde{V}^*_\sigma dA_\sigma(t) 
  - (i \tilde{H} + \frac{1}{2}\tilde{V}^*_\sigma\tilde{V}_\sigma)dt\big\}U_t.
  \end{split}\end{equation*}
Following the recipe of the previous section we now only have to determine the $dt$ terms of
$[12],[23],[24],$ $[124], [123], [124]$ and $[1234]$. All of these terms are zero in expectation 
with respect to $\rho^0$, except for $[124]$ and $[1234]$ 
  \begin{equation*}\begin{split}
  \rho^0\big([124]\big) &= \rho^0\Big(U_t^* b_tV^*_s\big(\EC^t(X) -X\big)V_s U_t\Big)dt  \\
  \rho^0\big([1234]\big) &= \rho^0\big(U_t^*b_t\eta_tV_s^*V_s U_t\big)dt.
  \end{split}\end{equation*}
For all $b_t$ the sum of these terms has to be  $0$ in expectation, i.e.
  \begin{equation*}\begin{split}
  &\forall b_t:\ \rho^t\Bigg(b_t\Big(V^*_s\big(\EC^t(X) -X\big)V_s +     
  \eta_tV_s^*V_s\Big) \Bigg)dt = 0 \iff \\
  &\forall b_t:\ \rho^t\Bigg(\EC^t\Bigg(b_t\Big(V^*_s\big(\EC^t(X) -X\big)V_s +     
  \eta_tV_s^*V_s\Big) \Bigg)\Bigg)dt = 0 \iff \\
  &\forall b_t:\ \rho^t\Bigg(b_t\Big(\EC^t(X)\EC^t(V^*_sV_s) -\EC^t(V^*_sXV_s) +     
  \eta_t\EC^t(V_s^*V_s)\Big) \Bigg)dt = 0 \iff \\
  &\eta_t = \frac{\EC^t(V^*_sXV_s)}{\EC^t(V_s^*V_s)}- \EC^t(X).
  \end{split}\end{equation*} 
Substituting the expressions for $\eta_t$ and $\tilde{Y}_t$ into equation \eqref{Belav1} 
we obtain the 
Belavkin equation for photon counting in the side channel
  \begin{equation}\label{BelavcountHeis}
  d\EC^t(X) = \EC^t\big(L(X)\big)dt + 
  \Big(\frac{\EC^t(V^*_sXV_s)}{\EC^t(V_s^*V_s)}
  - \EC^t(X)\Big)\big(d\Lambda_{ss}(t) - \EC^t(V_s^*V_s)dt\big).
  \end{equation}
Now recall that $\EC^t(X) = \rho^t_\bullet (X_\bullet)$, i.e.\ it is the function 
$\Omega_t \to \BB{C}:\ 
\omega \mapsto \rho^t_\omega(X_\omega)$. For all $X \in \BC = M_2$, the $M_2$ valued function 
$X_\bullet$
is the constant function $\omega \mapsto X$. Therefore for all $X$ in $\BC$, 
the Belavkin equation \eqref{BelavcountHeis} is equivalent to
  \begin{equation*}
  d\rho_\bullet^t(X) = \rho_\bullet^t\big(L(X)\big)dt + 
  \Big(\frac{\rho_\bullet^t(V^*_sXV_s)}{\rho_\bullet^t(V_s^*V_s)}
  - \rho_\bullet^t(X)\Big)\big(d\Lambda_{ss}(t) - \rho_\bullet^t(V_s^*V_s)dt\big),
  \end{equation*}
which is equivalent to the Belavkin equation of Section \ref{Daviesprocess}, equation 
\eqref{Belcount}. In simulating the above equation we can take for $Y_t = \Lambda_{ss}(t)$
the unique jump process with independent jumps and rate $\rho^t_\bullet(V_s^*V_s)$, since
$\Lambda_{ss}(t) - \int_0^t \rho^r_\bullet(V_s^*V_s)dr$ has to be a martingale.

Let us now turn to the homodyne detection scheme which we already discussed in Section 
\ref{Homodynedetection}.
The observed process is now $Y_t = X_\phi(t) = A^*_s(f_t) + A_s(f_t)$ (see Section 
\ref{Homodynedetection} for the definition of $f_t$). 
This means the 
innovating martingale $\tilde{Y}_t$ satisfies 
$d\tilde{Y}_t = e^{i\phi_t}dA^*_s(t) + e^{-i\phi_t}dA_s(t) - \EC^t(e^{i\phi_t}V^*_s + e^{-i\phi_t}V_s)dt$,
where $\phi_t = \phi_0 + \omega_{lo}t$ with $\omega_{lo}$ the frequency of the local oscillator.
Therefore we find different differentials for $B_t$ and $\EC^t(X)-X$ than we had in the photon counting
case
  \begin{equation*}\begin{split}
  dB_t =&\ b_t\big(e^{i\phi_t}dA^*_s(t) + e^{-i\phi_t}dA_s(t)\big) + 
     \big(c_t - b_t\EC^t(e^{i\phi_t}V^*_s + e^{-i\phi_t}V_s)\big)dt                        \\
  d(\EC^t(X)-X) =&\ \eta_t\big(e^{i\phi_t}dA^*_s(t) + e^{-i\phi_t}dA_s(t)\big)\  +         \\
      &\Big(\EC^t\big(L(X)\big) - \eta_t\EC^t(e^{i\phi_t}V^*_s + e^{-i\phi_t}V_s)\Big)dt \\
  \end{split}\end{equation*}
Following the recipe of the previous section we now only have to determine the $dt$ terms of
$[12],[23],$ $[24], [124], [123], [124]$ and $[1234]$. All of these terms are zero in expectation 
with respect to $\rho^0$, except for $[12], [23]$ and $[24]$ 
  \begin{equation*}\begin{split}
  \rho^0\big([12]\big) &= \rho^0\Big(U_t^* e^{i\phi_t}V^*_sb_t\big(\EC^t(X)-X\big)U_t\Big)dt\\
  \rho^0\big([23]\big) &= \rho^0(U_t^*b_t\eta_t U_t)dt\\
  \rho^0\big([24]\big) &= \rho^0\Big(U_t^*b_t\big(\EC^t(X)-X\big)e^{-i\phi_t}V_s U_t\Big)dt.
  \end{split}\end{equation*}
For all $b_t$ the sum of these terms has to be  $0$ in expectation, i.e.
  \begin{equation*}\begin{split}
  &\forall b_t:\ \rho^t\Bigg(b_t\Big(e^{i\phi_t}V^*_s\big(\EC^t(X)-X\big)+ 
  \big(\EC^t(X)-X\big)e^{-i\phi_t}V_s + \eta_t\Big)\Bigg)dt = 0 \iff \\
  &\forall b_t:\ \rho^t\Bigg(\EC^t\Bigg(b_t\Big(e^{i\phi_t}V^*_s\big(\EC^t(X)-X\big)+ 
  \big(\EC^t(X)-X\big)e^{-i\phi_t}V_s + \eta_t\Big)\Bigg)\Bigg)dt = 0 \iff \\
  &\forall b_t:\ \rho^t\Big(b_t\big( -\EC^t(e^{i\phi_t}V^*_sX  + e^{-i\phi_t}XV_s)\ + \\
   &\ \ \ \ \ \ \ \EC^t(e^{i\phi_t}V^*_s  + e^{-i\phi_t}V_s)\EC^t(X)+ \eta_t \big)\Big)dt = 0 \iff \\
  &\eta_t = \EC^t(e^{i\phi_t}V^*_sX  + e^{-i\phi_t}XV_s) - \EC^t(e^{i\phi_t}V^*_s  + 
e^{-i\phi_t}V_s)\EC^t(X).
  \end{split}\end{equation*} 
Substituting the expressions for $\eta_t$ and $\tilde{Y}_t$ into equation \eqref{Belav1} we obtain the 
Belavkin equation for the homodyne detection scheme
  \begin{equation}\label{BelavhomoHeis}\begin{split}
  &d\EC^t(X) = \EC^t\big(L(X)\big)dt + 
  \big(\EC^t(e^{i\phi_t}V^*_sX  + e^{-i\phi_t}XV_s) - \EC^t(e^{i\phi_t}V^*_s  + 
e^{-i\phi_t}V_s)\EC^t(X)\big)\times\\
  &\times \big(e^{i\phi_t}dA^*_s(t) + e^{-i\phi_t}dA_s(t) - \EC^t(e^{i\phi_t}V^*_s + 
e^{-i\phi_t}V_s)dt\big).
  \end{split}\end{equation}
Now recall that $\EC^t(X) = \rho^t_\bullet (X_\bullet)$, i.e.\ it is the function 
$\Omega_t \to \BB{C}:\ \omega \mapsto \rho^t_\omega(X_\omega)$. For all $X \in \BC = M_2$, the $M_2$ 
valued function $X_\bullet$
is the constant function $\omega \mapsto X$. 
Therefore for all $X$ in $\BC$, the Belavkin equation \eqref{BelavhomoHeis} is equivalent to
  \begin{equation*}\begin{split}
  &d\rho_\bullet^t(X) = \rho_\bullet^t\big(L(X)\big)dt + 
  \big(\rho_\bullet^t(e^{i\phi_t}V^*_sX  + e^{-i\phi_t}XV_s) - \rho_\bullet^t(e^{i\phi_t}V^*_s  + 
  e^{-i\phi_t}V_s)\rho_\bullet^t(X)\big)\times  \\
  &\times \big(e^{i\phi_t}dA^*_s(t) + e^{-i\phi_t}dA_s(t) - 
  \rho_\bullet^t(e^{i\phi_t}V^*_s + e^{-i\phi_t}V_s)dt\big),
  \end{split}\end{equation*}
which is equivalent to the Belavkin equation of Section \ref{Homodynedetection}, equation 
\eqref{Belhomodyne}. 
Since $A^*_s(f_t) + A_s(f_t) - \int_0^t\rho_\bullet^r(e^{i\phi_r}V^*_s + e^{-i\phi_r}V_s)dr$ 
is a martingale with variance $t$ on the space of the Wiener process, it must be the Wiener process itself.

\newpage
\thispagestyle{empty}

%% file: sec.tex
\chapter{Squeezing enhanced control}\label{ch sec}

\begin{center}
{\large Luc Bouten}\\
\vspace{1cm}
\emph{Mathematisch Instituut, Katholieke Universiteit Nijmegen \\
Toernooiveld 1, 6526 ED Nijmegen, The Netherlands}
\vspace{0.4cm}
\end{center}

{\small
\begin{center}
\textbf{Abstract\footnote{This 
chapter is based on \cite{Bou}.}}
\end{center}
We study an open system in contact with its 
environment, the electromagnetic field. The 
information gained by measuring a quadrature 
of the field is used to send control pulses 
to the system. Goal is to fix the unknown state 
of the system in time. We show that in the special case of 
an essentially commutative interaction this goal 
can be achieved. In dealing with spontaneous decay 
we approximate the essentially commutative situation 
by bringing the field in a squeezed state. We show that 
when squeezing goes to infinity, the state can again 
be kept fixed.}

\section{Introduction}

The last decade there have been rapid developments 
in quantum information theory, initiated mainly by some 
fundamental papers \cite{Sho}, \cite{Fey1} showing
the increased possibilities when quantum features are 
exploited in computations.
However, implementation of the proposed  
algorithms on real physical qubits still poses
a great challenge. One of the problems is 
the interaction with the environment, i.e.\ the electromagnetic 
field, and the decoherence that 
goes along with it. Dealing with this problem motivates the 
development of theory and methods for coherently 
manipulating, or controlling, quantum systems. 

Decoherence is a result of ignoring information  
lost from an open quantum system to its environment 
via their interaction. However, the lost information can be 
retrieved, at least partially, by observing the environment, 
i.e.\ by performing measurements on it. The decoherence 
can be combatted by using the retrieved information 
in a scheme for controlling the quantum system, see also \cite{GrW}.      

Since the electromagnetic field and the open system 
are in interaction, information on the system itself is 
gained when measuring some observables of the 
field. Hence conditioning on the obtained measurement results
provides a back-action of the measurement in the field 
on the open system. One of the pioneers in this area is
Belavkin who extended many ideas in classical 
filtering theory, cf.\ \cite{Str}, to the quantum 
regime \cite{Bel2}, \cite{Bel}. Quantum filtering 
theory \cite{Bel88}, \cite{Bel} explains how the state, conditioned 
on the result of a continuous time measurement in the environment,
evolves in time. Note that since the results of the continuous time 
measurement are random, the conditioned state is also a 
random state. Quantum filtering theory provides a 
stochastic differential equation, the \emph{Belavkin equation}, 
for the state evolution in which the measurement process 
is one of its driving terms \cite{Bel88}, \cite{Bel}. 

Another approach to the back-action due to conditioning,
is via quantum trajectory theory as developed 
in quantum optics in the late 1980's and
early 1990's \cite{Car}, but already envisioned
by Davies \cite{SrD}, \cite{Dav} in the 1970's. 
In this approach photon counting measurements are analysed to obtain a 
continuous time evolution of the open system interrupted 
by jumps the moments at which photons are detected. 
Differentiation of the trajectory evolution leads to 
a \emph{stochastic Schr\"odinger equation} \cite{Car0}, 
which is a stochastic differential equation for the 
evolution of the state conditioned on the outcomes 
of the counting experiment. A diffusive limit of photon 
counting in which the jumps in the state space decrease 
in size but become increasingly frequent, makes it possible 
to incorporate homodyne and heterodyne detection 
schemes into quantum trajectory theory \cite{BaB}, 
\cite{Car}, \cite{WiM}.  The stochastic Schr\"odinger 
equations encountered in quantum optics are equivalent 
to the Belavkin equations from quantum 
filtering theory \cite{Bel}, \cite{BGM}.    

The result of the continuous time measurement in the field
can be used to exert control over the system. The solution
to the quantum filtering problem \cite{Bel2}, \cite{Bel88} makes it possible 
to directly carry over many ideas in classical 
control theory \cite{Str}, \cite{Kus}, \cite{Kus1}
to the quantum regime, \cite{Bel00}, \cite{Bel000}, 
\cite{Bel88}, \cite{Bel3}, \cite{DHJMT}, \cite{HSM}. Coming from the quantum 
trajectory approach, other pionering work in quantum
control was done by Wiseman and Milburn in the 
first half of the 1990's, see \cite{Wis}, \cite{Wis2}, \cite{WiM2}.
Two different objectives in control problems can 
be distinguished, one where the state is controlled
in order to let it follow a certain path in time \cite{Bel3}, 
\cite{DHJMT}, and one
where the semigroup describing the dissipative evolution
of the open system, i.e.\ the channel itself, 
is being controlled \cite{MaZ}, \cite{GrW}, \cite{ADL}, \cite{AWM}.  

In this paper a problem of the second type is 
considered. The question addressed here is how to keep an 
unknown state of an open system fixed in time, 
i.e.\ how to keep its dynamical semigroup as close 
to identity as possible. In this article we 
will not be concerned with optimality results. 
The main issue is to find or engineer situations where 
the control is perfect, in which case the control scheme 
is said to \emph{restore quantum information} \cite{GrW}.
Furthermore, we will not be concerned here with 
encoding our system into the code space of a larger system 
and then protecting just this code space \cite{ABCDGM}, \cite{AWM}. 
 
The control scheme consists of two parts. The first part 
is an evolution over a period of $\tau$ time units in which 
a quadrature of the field is observed. This evolution 
is governed by the Belavkin equation corresponding to 
this measurement. In the second part the result of  
the measurement is used to construct a laser pulse designed,
if at all possible, to take the system through a Rabi cycle that  
corrects the evolution of the past $\tau$ 
time units. This scheme is studied in the limit for 
very small $\tau$, i.e.\ the control pulses are sent
at very high frequency. 
 
In general the above control scheme will not be able to 
restore quantum information. Since the interaction of 
the field and the system is studied in the weak coupling 
limit, the field acts as two classical noises. However 
these two noises, represented by two different quadratures
of the field, do not commute with each other. Therefore
only one of these noises can be observed and its disturbing 
effect on the system corrected. An idealised interaction 
of system and field in which there is only one instead of two classical 
noises present is called \emph{essentially commutative} \cite{KuM0}. 
In the essentially commutative case it will turn out that the 
above control scheme restores quantum information.  

For the more realistic situation where both noises are present 
our strategy will be to manipulate the state of the field 
in order to approximate the essentially commutative case. 
This is done by putting the field in 
a squeezed state, i.e.\ one quadrature's variance increases
while the other quadrature's variance decreases, 
\cite{Gar1}, \cite{GaP}, \cite{KoC}. The 
idea is to measure the noise with the large variance and 
correct its disturbing effect on the system. It will turn out
that when squeezing goes to infinity the control
scheme described above will restore quantum information.

This paper is organised as follows. Section \ref{sec dilation}
describes the dissipative evolution of the open system within
the Markov approximation. The joint evolution of system and 
field is given by unitaries satisfying a quantum stochastic differential 
equation in the sense of \cite{HuP}. In the next section a brief
exposition of quantum stochastic calculus \cite{HuP} is given.
This enables us to make sense of the quantum 
stochastic differential equation providing the unitaries 
of section \ref{sec dilation}. Sections \ref{sec dilation}
and \ref{sec qsc} describe dilation theory and quantum 
stochastic calculus in a nutshell. Section \ref{sec condexp}
is a brief exposition of quantum filtering theory. 
It contains a derivation of the Belavkin equation 
for field quadrature measurement.

Sections \ref{sec escom} and \ref{sec con-squeez} 
deal with controlling the state of an open system 
in the essentially commutative case and the more 
realistic situation of spontaneous decay of a two-level
system, respectively. Here we show that for the essentially 
commutative case it is possible to restore quantum 
information. For spontaneous decay, however,
problems are encountered motivating the investigation in the 
remainder of the paper.

Section \ref{sec squeez} shows how to describe 
the interaction of system and field when the field 
is in a squeezed state. To do this we have to 
do quantum stochastic calculus in the GNS-representation 
space of the squeezed state. In the last section 
the Belavkin equation for measuring a quadrature 
of a squeezed field is given and a control scheme 
based on this measurement is presented. It turns out 
that when squeezing goes to infinity, the scheme 
restores quantum information.

\section{The dilation}\label{sec dilation}

Let $\BC := M_n$ stand for the algebra of 
observables of an $n$-dimensional quantum system. 
On this algebra $\{T_t\}_{t \ge 0}$ is a semigroup 
of completely positive identity preserving operators. 
It represents the irreversible time 
evolution of the system in the Heisenberg 
picture. Lindblad's theorem \cite{Lin} asserts that 
$T_t = \exp(tL)$ where the generator $L$ 
is given by 
  \begin{equation*}
  L(X) = i[H,X] + \sum_{j=1}^k V_j^*XV_j - \frac{1}{2}\{V_j^*V_j, X\},\ \ \ \ X \in \BC,
  \end{equation*} 
with $H$ and the $V_j's$ fixed elements of $\BC$, 
$H$ being selfadjoint. The notation $\{X,Y\}$ stands 
for the anticommutator $XY+YX$. For simplicity, 
we take $k =1$ and $H =0$, i.e.\
  \begin{equation}
  L(X) =  V^*XV - \frac{1}{2}\{V^*V, X\}.
  \end{equation}
This paper deals mainly with two special cases 
of the above situation. In the first special case 
we have either $V=V^*$  or $V = -V^*$. 
This case is called \emph{essentially commutative} 
\cite{KuM0}, see section \ref{sec escom}.
In the second special case we have
  \begin{equation}\label{eq V}
  V=\begin{pmatrix}0 & 0 \\ 1 & 0\end{pmatrix}.
  \end{equation}
Then the semigroup $T_t$ describes spontaneous 
decay to the ground state of a two-level system,  
see sections \ref{sec con-squeez} and \ref{sec con+squeez}.

The system $\BC$ and its environment, the electromagnetic 
field, evolve reversibly in time. The irreversible evolution 
$T_t$ of $\BC$ is the result after tracing out the field.  
Up to section \ref{sec squeez} the electromagnetic 
field to which the system $\BC$ is coupled, will be 
taken in the vacuum state or a coherent state. 
Then, see section \ref{sec squeez} for more details, 
a decay channel in the field can be modelled by the 
\emph{bosonic} or \emph{symmetric Fock space} over the 
Hilbert space $L^2(\BB{R})$ of square integrable wave 
functions on the real line, i.e.\
  \begin{equation*}
  \FC := \BB{C} \oplus \bigoplus_{n=1}^\infty L^2(\BB{R})^{\ten_s n}.
  \end{equation*}   
The algebra generated by the field observables 
on $\FC$ contains all bounded operators and it is 
denoted by $\WC$.

For future convenience we already distinguish 
two decay channels in the field, i.e.\ we rewrite
$L$ as 
  \begin{equation}\label{eq Lindblad}
  L(X) = V^*_fXV_f - \frac{1}{2}\{V^*_fV_f,X\} + 
   V^*_sXV_s - \frac{1}{2}\{V^*_sV_s, X\},\ \ \ \ X \in \BC,
  \end{equation}
where $V_f = \kappa_fV,\ V_s = \kappa_sV$ and 
$\kappa_f,\kappa_s \in \BB{R}$ such that 
$|\kappa_f|^2 + |\kappa_s|^2=1$. The subscripts $f$ and 
$s$ stand for \emph{forward} and \emph{side channel}, respectively. 
On the forward channel in the field we will put a laser 
with which we want to control the system, while in the 
side channel of the field we are going to perform a 
measurement. The decay 
rates into the forward and side channel are 
given by $|\kappa_f|^2$ and $|\kappa_s|^2$, respectively.
Since the field is modelled by these two decay 
channels, we need two copies of the algebra $\WC$,
denoted $\WC^f\ten\WC^s$.

The free evolution of a channel in the field is 
given by the unitary group $S_t$, the second 
quantization of the left shift $s(t)$ on $L^2(\BB{R})$, 
i.e.\ $s(t):\ f \mapsto f(\cdot + t)$. In the 
Heisenberg picture the evolution on $\WC^f\ten\WC^s$ is 
  \begin{equation*}
  W \mapsto (S_t\ten S_t)^*W(S_t\ten S_t) 
  := \mbox{Ad}[S_t\ten S_t](W), \ \ \ \ W\in \WC^f\ten\WC^s.
  \end{equation*}
The system $\BC$ and field together form a closed 
system, thus their joint evolution is given by 
a one-parameter group $\{\hat{T}_t\}_{t\in \BB{R}}$
of *-automorphisms on $\BC \ten \WC^f\ten\WC^s$
  \begin{equation*}
  X \mapsto \hat{U}_t^*X\hat{U}_t 
  := \mbox{Ad}[\hat{U}_t](X), \ \ \ \ X\in \BC\ten\WC^f\ten\WC^s.
  \end{equation*}
The group $\hat{U}_t$ is a perturbation of the free 
evolution without interaction. We describe this 
perturbation by the family of unitaries 
$U_t := (S_{-t}\ten S_{-t})\hat{U}_t$ for all 
$t \in \BB{R}$ satisfying the \emph{cocycle} identity
  \begin{equation*}
  U_{t+s} = (S_{-s}\ten S_{-s})U_t(S_s\ten S_s)U_s,
  \ \ \ \ \mbox{for all \ } t,s \in \BB{R}.
  \end{equation*} 
The direct connection between the reduced evolution 
of $\BC$ given by \eqref{eq Lindblad} 
and the cocycle $U_t$ is one of the important results 
of quantum stochastic calculus \cite{HuP} 
which is the object of the next section. 
For the moment we only mention that in the 
weak coupling limit \cite{AFLu}, $U_t$ is the solution 
of the stochastic differential equation 
\cite{HuP}, \cite{Par}, \cite{Mey}
  \begin{equation}\label{eq HuP}
  dU_t = \{V_fdA_f^*(t) - V_f^*dA_f(t) + V_sdA_s^*(t) - V_s^*dA_s(t)  
  - \frac{1}{2}V^*Vdt\}U_t, \ \ \ \ U_0 = \I.
  \end{equation} 
We will see in the next section that if $U_t$ 
satisfies \eqref{eq HuP} the following
\emph{dilation diagram} \cite{Kum1}, \cite{Kum2} 
commutes:
  \begin{equation}\label{diag dildiag}\begin{CD}
     \BC @>T_t>> \BC              \\
     @V{\Id \ten \I \ten \I}VV        @AA{\Id \ten \phi \ten\phi}A      \\
     \BC\ten\WC^f\ten\WC^s @>\hat{T}_t=\mbox{Ad}[\tilde{U}_t]>> \BC\ten\WC^f\ten\WC^s   \\
  \end{CD}\end{equation}
  
i.e.\ for all $X \in \BC:\ T_t(X) = 
\big(\mbox{Id}\ten\phi\ten\phi\big)\big(\hat{T}_t(X\ten\I\ten\I)\big)$,
where $\phi$ is the vacuum state on $\WC$, 
and $\I$ is the identity operator in $\WC$.
Any dilation of the semigroup $T_t$ with Bose 
fields is unitarily equivalent with the above 
one under certain minimality requirements.

The dilation diagram can also be read in the 
Schr\"odinger picture if we reverse the arrows: 
start with a state $\rho$ of the system $\BC$ 
in the upper right hand corner, then this 
state undergoes the following sequence of maps
  \begin{equation*}
  \rho \mapsto \rho\ten\phi\ten\phi \mapsto 
  \rho\ten\phi\ten\phi \circ \hat{T}_t = \hat{T}_{t*}(\rho\ten\phi\ten\phi)
  \mapsto \mbox{Tr}_{\FC^f\ten\FC^s}(\hat{T}_{t*}(\rho\ten\phi\ten\phi)).
  \end{equation*}
This means that at $t=0$, the atom in the 
state $\rho$ is coupled to the electromagnetic
field in the vacuum state, after $t$ seconds 
of unitary evolution the partial trace 
over the field is taken.

\section{Quantum stochastic calculus}\label{sec qsc}

Here, we briefly discuss the quantum stochastic 
calculus developed by Hudson and Parthasarathy \cite{HuP}.
For a detailed treatment of the subject we refer to 
\cite{Par} and \cite{Mey}. The exposition here is a bit
broader than strictly necessary for the construction of the 
cocycle of the previous section. However, the general 
description \cite{Par} presented here is needed in section 
\ref{sec squeez}.

Let $\HC$ be a Hilbert space.
We define the \emph{bosonic} or \emph{symmetric Fock 
space} over $\HC$ by
  \begin{equation*}
  \FC(\HC) := \BB{C} \oplus \bigoplus_{k = 1}^\infty \HC^{\ten_s k}.
  \end{equation*}
In the previous section we had $\HC = L^2(\BB{R})$.
For every $f \in \HC$ we define the \emph{exponential vector}
$e(f) \in \FC(\HC)$ in the following way
  \begin{equation*}
  e(f) := 1 \oplus \bigoplus_{k = 1}^\infty \frac{1}{\sqrt{k!}}f^{\ten k}.
  \end{equation*}
The inner product of two exponential vectors 
$e(f)$ and $e(g)$ is $\langle e(f),e(g)\rangle 
= \exp(\langle f,g\rangle)$. We denote the 
\emph{vacuum vector} $e(0)= 1\oplus 0\oplus0\oplus\ldots$ 
also by $\Phi$. The span of all exponential vectors, 
denoted $\DC$, forms a dense subspace of $\FC(\HC)$. 

Let $\xi$ be a projection (on the Hilbert space $H$) 
valued measure on $\BB{R}$ with no jump points, 
i.e.\ $\xi(\{t\}) = 0$ for all $t \in \BB{R}$. 
Denote by $\HC_{t]}, \HC_{[s,t]}$ and $\HC_{[t}$ the ranges of 
the projections $\xi((-\infty,t]), \xi([s,t])$ and 
$\xi([t, \infty))$, respectively. For a vector $f\in \HC$ we denote
$f_{t]} := \xi((-\infty,t])f, f_{[s,t]} := \xi([s,t])f$ 
and $f_{[t} := \xi([t,\infty))f$. Let us write $\HC$ as 
the direct sum $\HC_{t]}\oplus\HC_{[t}$, then $\FC(\HC)$
is unitarily equivalent with $\FC(\HC_{t]})\ten \FC(\HC_{[t})$
through the identification $e(f) \cong e(f_{t]})\ten e(f_{[t})$.
For notational convenience the tensor product 
signs between exponential vectors are often omitted. 
The algebra $\WC := \BC\big(\FC(\HC)\big)$ also splits 
as a tensor product $\WC_{t]}\ten\WC_{[t}$ where 
$\WC_{t]} := \BC\big(\FC(\HC_{t]})\big)$ and 
$\WC_{[t} := \BC\big(\FC(\HC_{[t})\big)$.
 
A map $m:\ \BB{R}_+ \to \HC:\ t \mapsto m_t$ is 
called a \emph{$\xi$-martingale} if
$m_t \in \HC_{t]}$ for all $t$ and 
$\xi([0,s])m_t = m_s$ for all $s < t$. For $m$ and 
$m'$ $\xi$-martingales, there exists a complex 
valued measure (of finite variation on every 
bounded interval), 
denoted $\langle\langle m, m'\rangle\rangle$ on $\BB{R}_+$,
satisfying
  \begin{equation}\label{eq defmeasure}
  \langle\langle m, m'\rangle\rangle\big([0,t]\big) = \langle m_t,m'_t\rangle,
  \end{equation}
for all $t \ge 0$. Let $m$ be a $\xi$-martingale. The annihilation
operator $A(m_t)$ and creation operator $A^*(m_t)$ are defined on 
the domain $\DC$ by
  \begin{equation}\label{eq defAandAster}\begin{split}
  &A(m_t)e(g) = \langle m_t, g\rangle e(g),\ \ \ \ g\in \HC, \\  
  &\big\langle e(h), A^*(m_t)e(g) \big\rangle_{\FC(\HC)} = 
  \langle h, m_t\rangle\big\langle e(h),e(g)\big\rangle_{\FC(\HC)}, \ \ \ h,g \in \HC.
  \end{split}\end{equation}

Let $M_t$ be one of the processes $A(m_t)$ or 
$A^*(m_t)$ for some $\xi$-martingale $m$. 
The following factorisation property 
\cite{HuP}, \cite{Par} makes the definition of 
stochastic integration against $M_t$ possible
  \begin{equation*}
  (M_t - M_s)e(f) = e(f_{s]})\big\{(M_t-M_s)e(f_{[s,t]})\big\}e(f_{[t}),
  \end{equation*}
with $(M_t-M_s)e(f_{[s,t]}) \in \FC(\HC_{[s,t]})$.
We first define the stochastic integral for the 
so-called {\it simple} operator processes  
with values in the atom and noise algebra 
$\BC \ten \WC$ where $\BC:=M_n$ and $\WC$
is the algebra of all bounded operators on 
the Fock space $\FC(\HC)$.

\begin{de} 
Let $\{L_s\}_{0 \le s \le t}$ be an adapted 
(i.e.\ $L_s \in \BC \ten \BC(\FC(H_{s]}))$ for all $0 \le s \le t$) simple 
process with respect to the partition $\{s_0=0, s_1,\dots, s_p= t\}$ in the sense that 
$L_s = L_{s_j}$ whenever $s_j \le s < s_{j+1}$. Then the stochastic integral of 
$L$ with respect to $M$ on $\BB{C}^n \ten \DC$ is given by \cite{HuP}, \cite{Par}:  
  \begin{equation*}
  \int_0^t L_s dM_s  ~f e(u) := 
\sum_{j=0}^{p-1} \big(L_{s_j}fe(u_{s_j]})\big)\big((M_{s_{j+1}}- M_{s_{j}})
   e(u_{[s_j, s_{j+1}]})\big)e(u_{[s_{j+1}}).
  \end{equation*}
\end{de}

By the usual approximation by simple processes we can 
extend the definition of the stochastic integral 
to a large class of {stochastically integrable 
processes} \cite{HuP}, \cite{Par}. We simplify 
our notation by writing $dX_t = L_tdM_t$ for 
$X_t = X_0 + \int_0^t L_sdM_s$. Note that the 
definition of the stochastic integral implies 
that the increments $dM_s$ lie in the future, 
i.e.\ $dM_s \in \WC_{[s}$. Another consequence 
of the definition of the stochastic 
integral is that its expectation with respect to 
the vacuum state $\langle\Phi,\,\cdot\,\Phi\rangle$ 
is always $0$ due to the fact that the increments 
$dA(m_t)$ and $dA^*(m_t)$ have zero expectation 
values in the vacuum. This will often simplify 
calculations of expectations, our strategy being 
that of trying to bring these increments to act 
on the vacuum state thus eliminating a large number 
of differentials. 

The following theorem of Hudson and Parthasarathy 
extends the It\^o rule of classical probability theory.
\begin{stel}\label{Itorules}\textbf{(Quantum It\^o rule \cite{HuP}, \cite{Par})}
Let $M_1$ and $M_2$ each be one of the processes $A(m_t)$ or $A^*(m_t')$. Then 
$M_1M_2$ is an adapted process satisfying the relation:
  \begin{equation*}
  d(M_1M_2) = M_1dM_2 + M_2dM_1 + dM_1dM_2,
  \end{equation*}
where $dM_1dM_2$ is given by the quantum It\^o table:
\begin{center}
{\large \begin{tabular} {l|lll}
$dM_1\backslash dM_2$ & $dA^*(m'_t)$ & $dA(m'_t)$ \\
\hline 
$dA^*(m_t)$ & $0$ & $0$ \\
$dA(m_t)$ & $d\langle\langle m,m'\rangle\rangle$ & $0$ 
\end{tabular} }
\end{center}
\end{stel}
\noindent\textbf{Notation.} The quantum It\^o rule will 
be used for calculating differentials of products 
of It\^o integrals. Let $\{Z_i\}_{i=1,\dots, p}$ 
be It\^o integrals, then 
 \begin{equation*} 
 d (Z_1Z_2\dots Z_p)= \sum_{\substack{\nu\subset\{1,\dots, p\} \\ \nu \neq \emptyset}}[\nu]
 \end{equation*}
where the sum runs over all {\it non-empty} subsets 
of $\{1,\dots, p\}$ and for any $\nu=\{i_1,\dots i_k\}$, 
the term $[\nu]$ is the contribution to 
$d (Z_1Z_2\dots Z_p)$ coming from differentiating 
only the terms with indices in the set 
$\{i_1,\dots i_k\}$ and preserving the order 
of the factors in the product. For example 
the differential $d(Z_1Z_2Z_3)$ contains 
terms of the type $[2]=Z_1(dZ_2)Z_3$, 
$[13]=(dZ_1)Z_2(dZ_3)$, and 
$[123]=(dZ_1)(dZ_2)(dZ_3)$.

Let us return to the setup of section \ref{sec dilation}. 
We now make sense of equation \eqref{eq HuP}. Note that 
the Hilbert space $\HC$ is $L^2(\BB{R})\oplus L^2(\BB{R})$. 
The forward and side channel both have their 
own copy of $L^2(\BB{R})$. The projection valued 
measure $\xi$ is given by
  \begin{equation*}
  \xi(I) (f_f \oplus f_s) = (\chi_If_f) \oplus (\chi_If_s),
  \ \ \ \ \ f_f,f_s\in L^2(\BB{R}),
  \end{equation*}
for all Borel subsets $I$ of $\BB{R}$. Here $\chi_I$ denotes 
the \emph{indicator function} of $I$, i.e.\ the function that
takes the value $1$ on $I$ and is $0$ elsewhere. 

The maps 
$m^f:\ \BB{R}_+ \to \HC:\ t \mapsto \chi_{[0,t]}\oplus 0$ 
and $m^s:\ \BB{R}_+ \to \HC:\ t \mapsto 0\oplus \chi_{[0,t]}$
are  $\xi$-martingales. We denote
the annihilation $A(m^f_t), A(m^s_t)$  and creation 
operators $A^*(m^f_t), A^*(m^s_t)$, defined on 
$\DC$ by \eqref{eq defAandAster}, more compactly 
by $A_f(t), A_s(t), A_f^*(t)$ and $A^*_s(t)$, respectively. 
The calculus for stochastic integrals with respect to $A_\sigma(t)$
and $A_\nu^*(t),\ \sigma, \nu \in \{f,s\}$ is 
then given by the \emph{Hudson-Parthasarathy It\^o
table} \cite{HuP}, \cite{Par}:
\begin{center}
{\large \begin{tabular} {l|lll}
$dM_1\backslash dM_2$ & $dA_\nu^*(t)$ & $dA_\nu(t)$ \\
\hline 
$dA_\sigma^*(t)$ & $0$ & $0$ \\
$dA_\sigma(t)$ & $\delta_{\sigma\nu}dt$ & $0$ 
\end{tabular} }
\end{center}

Let us introduce the selfadjoint quantum noise $\beta_t$ 
describing the interaction between the quantum system 
$\BC = M_n(\BB{C})$ and the electromagnetic field
  \begin{equation}\label{eq quantumnoise}
  d\beta_t := -i\big(V_fdA_f^*(t) -V_f^*dA_f(t)+
  V_sdA_s^*(t) -V_s^*dA_s(t)\big), \ \ \ \ \ \ \ \beta_0 = 0.
  \end{equation}
It is clear in our example of spontaneous decay of a 
two-level system that this noise represents
an interaction consisting of creations of excitations
of the two-level system accompanied by annihilations
of photons in the decay channels and vice versa. 
It describes the interaction of the electromagnetic 
field, in which we distinguished two decay channels, 
and the two-level system in the weak coupling limit \cite{AFLu}. 
We let the cocycle $U_t$ of section \ref{sec dilation},
providing the evolution in the weak coupling limit 
of the two-level system and field together, be given by the quantum 
stochastic differential equation
  \begin{equation*}\begin{split}
  &dU_t = \{id\beta_t - \frac{1}{2}(d\beta_t)^2\}U_t = \\
  &\{V_fdA_f^*(t) -V_f^*dA_f(t)+
  V_sdA_s^*(t) -V_s^*dA_s(t) 
  -\frac{1}{2}V^*Vdt\}U_t, \\
  &U_0 = \I.  
  \end{split}\end{equation*}

We can now check that the dilation diagram \eqref{diag dildiag}
commutes. Using the continuous tensor product structure of 
the Fock space $\FC(\HC)$, it is easy to see that 
following the lower part of diagram \eqref{diag dildiag} 
defines a semigroup on $\BC$, i.e.\ we only have to show 
that it is generated by the Lindblad operator $L$ 
of equation \eqref{eq Lindblad}. 
For all $X \in \BC$
  \begin{equation*}
  d\mbox{Id}\ten\phi\ten\phi\big(\hat{T}_t(X \ten \I\ten\I)\big) = 
  \mbox{Id}\ten\phi\ten\phi\big(d(U^*_t(X \ten \I \ten\I)U_t)\big).
  \end{equation*}
Using the notation below Theorem \ref{Itorules} with $Z_1 = U_t^*$
and $Z_2 = (X\ten \I\ten\I) U_t$, we find
  \begin{equation*}
  d\mbox{Id}\ten\phi\ten\phi\big(\hat{T}_t(X \ten \I\ten\I)\big) 
  = \mbox{Id}\ten\phi\ten\phi\big([1]+[2]+[12]\big).
  \end{equation*}
With the aid of the Hudson-Parthasarathy 
It\^o table we can evaluate these terms. 
We are only interested in the $dt$-terms since 
the expectation with respect to the vacuum kills 
the other terms. The terms $[1]$ and 
$[2]$ provide the anticommutators 
$-\frac{1}{2}\{V_f^*V_f, X\}dt$ and  
$-\frac{1}{2}\{V_s^*V_s, X\}dt$ and $[12]$ provides 
the terms $V_f^*XV_fdt$ and $V_s^*XV_sdt$, proving our claim.

We now change the situation in diagram \eqref{diag dildiag} 
by introducing a laser on the forward channel, i.e.\ 
the forward channel is now in a \emph{coherent state} 
$\gamma_h := \langle\psi(h), \cdot\,\psi(h)\rangle$ where
$\psi(h):= \exp(-\frac{1}{2}\p h\p^2)e(h)$, the exponential 
vector $e(h)$ for some $h \in L^2(\BB{R_+})$ normalised 
to unity. The laser will be used to send 
control-pulses to the system $\BC$. This leads 
to the following dilation diagram
  \begin{equation}\label{diag dildiaglas}\begin{CD}
     \BC @>T^h_t>> \BC              \\
     @V{\Id \ten \I \ten \I}VV        @AA{\Id \ten \gamma_h \ten\phi}A      \\
     \BC\ten\WC^f\ten\WC^s @>\hat{T}_t=\mbox{Ad}[\tilde{U}_t]>> \BC\ten\WC^f\ten\WC^s \\
  \end{CD}\end{equation} 

i.e. the evolution on $\BC$ has changed and it is 
in general \emph{not} a semigroup. Denote by $W(h)$ 
the unitary \emph{Weyl} or \emph{displacement operator} 
defined on $\DC$ by: 
$W(h)\psi(g) = \exp(-2i\mbox{Im}\langle h,g\rangle)\psi(g+h)$. 
Note that $W(h)\Phi = W(h)\psi(0) = \psi(h)$, so 
that we can write for all $X\in \BC$
  \begin{equation*}\begin{split}
  &T^h_t(X) = \mbox{Id}\ten\gamma_h\ten\phi(U_t^*X\ten\I\ten\I U_t) = \\
  &\mbox{Id}\ten\phi\ten\phi\big(W_f(h)^*U_t^*X\ten\I\ten\I U_tW_f(h)\big) =\\
  &\mbox{Id}\ten\phi\ten\phi\big(W_f(h_{t]})^*U_t^*X\ten\I\ten\I U_tW_f(h_{t]})\big),
  \end{split}\end{equation*}
where $h_{t]} = h\chi_{(0,t]}$ and 
$W_f(h) := \I\ten W(h)\ten\I$. Defining 
$U_t^h := U_tW_f(h_{t]})$, together with the 
quantum stochastic differential equation for 
$W_f(h_{t]})$ \cite{Par}
  \begin{equation*}
  dW_f(h_{t]}) = \{h(t)dA_f^*(t) -\overline{h}(t)dA_f(t)-
  \frac{1}{2}|h(t)|^2dt\}W_f(h_{t]}), \ \ \ W_f(h_{0]}) = \I,
  \end{equation*}
and the It\^o rules leads to the following quantum 
stochastic differential equation for $U^h_t$
  \begin{equation}\label{eq U+cont}\begin{split}
  &dU^h_t = \Big\{\big(V_f+h(t)\big)dA^*_f(t) - \big(V^*_f+\overline{h}(t)\big)dA_f(t))
    + V_sdA^*_s(t) - V^*_sdA_s(t)\ - \\
  &\ \ \ \ \ \ \ \ \ \ \ \ \ \
  \frac{1}{2}\big(|h(t)|^2 + V^*V + 2h(t)V^*_f\big)dt\Big\}U^h_t, \ \ \ \ U^h_0 = \I.
  \end{split}\end{equation}
Therefore, the dilation diagram \eqref{diag dildiaglas}
is equivalent to
  \begin{equation}\label{diag dildiaglaser}\begin{CD}
     \BC @>T^h_t>> \BC              \\
     @V{\Id \ten \I \ten \I}VV        @AA{\Id \ten \phi \ten\phi}A      \\
     \BC\ten\WC^f\ten\WC^s @>\hat{T}^h_t=\mbox{Ad}[\tilde{U}^h_t]>> \BC\ten\WC^f\ten\WC^s \\
  \end{CD}\end{equation}  

In the following, we will often omit the superscript $h$ to 
simplify the notation. Define  a Hamiltonian by 
$H:= i\big(\overline{h}(t)V_f - h(t)V^*_f\big)$, then 
following the lower part of diagram \eqref{diag dildiaglaser}
and using It\^o's rules, see Theorem \ref{Itorules}, shows
that the time dependent generator of the dissipative 
evolution $T^h_t$ in the presence of the laser on the 
control channel is given by
  \begin{equation}\label{eq gen}
  L(X) = i [H,X]+ V^* X V
  -\frac{1}{2}\{V^*V, X\}.
  \end{equation} 
Later on we will choose $h$ in a suitable way in order to exert 
control on the system $\BC$.

\section{The Belavkin equation}\label{sec condexp}

Let us now turn our attention to the side channel. 
In this channel an observable is  measured continuously 
in time. Goal is to briefly show how to 
derive a stochastic differential equation for 
the stochastic state evolution of the system $\BC$ 
conditioned on the outcome of the measurement process. 
The method described below is known as quantum filtering, 
see \cite{Bel} and \cite{BGM} for a more detailed 
treatment.

In this paper the observable $Y^s_t$ of the field that 
is measured continuously in time will always be a 
field quadrature, i.e.
  \begin{equation}\label{eq defY}
  Y^s_t := \I_{\BC}\ten\I_{W^f}\ten\Big(
  \big(e^{-i\phi}A_s(t) + e^{i\phi}A_s^*(t)\big)\ten \I_{\WC^s_{[t}}\Big)\
  \in\ \BC\ten\WC^f\ten\Big(\WC^s_{t]}\ten\WC^s_{[t}\Big),
  \end{equation}
for some phase $\phi \in [0,2\pi)$. Such a field quadrature
measurement can be performed by a homodyne detection experiment.
See \cite{Bel}, \cite{BGM} for measurement of other observables. 
Let $\rho$ be the initial state of the quantum system
$\BC$. We describe the measurement process in the 
interaction picture, i.e.\ the shift part of
$\hat{U}_t := (S_t\ten S_t)U_t$ 
acts on the observables while 
the cocycle part $U_t$, given by equation \eqref{eq U+cont} 
with the superscript $h$ suppressed,
acts on the states
  \begin{equation*}
  \rho^t(X) := \rho\ten\phi(U_t^*XU_t),\ \ \ \ X \in \BC\ten\WC^f\ten\WC^s.
  \end{equation*} 

Let $\CC_t$
be the von Neumann algebra generated by the 
family of observables $\{Y^s_r;\ 0 \le r \le t\}$. 
Since $Y^s_r$ and $Y^s_t$ commute for all $r,t \ge 0$
the algebra $\CC_t$ is commutative. The algebras 
$\{\CC_t\}_{t\ge 0}$ form a growing family, that 
is $\CC_{s} \subset \CC_t$ for all $s \le t$. 
Thus we can define the inductive limit 
$\CC_\infty := \lim_{t\to\infty}\CC_t$, which is
the smallest von Neumann algebra containing all 
$\CC_t$. It follows via Kolmogorov's extension 
theorem, see \cite{BGM} Theorem 5.1, that
there exists a unique state $\rho^\infty$ on $\CC_\infty$ 
which coincides with $\rho^t$ when restricted to 
$\CC_t \subset \CC_\infty$ for all $t \ge 0$. 
From spectral theory it follows that there 
exists a measure space 
$(\Omega, \Sigma, \BB{P}_\rho)$ and a growing 
family $\{\Sigma_t\}_{t\ge 0}$ of 
$\sigma$-subalgebras of $\Sigma$, such that 
$(\CC_\infty,\rho^\infty)$ and $(\CC_t, \rho^t)$
are isomorphic to $L^\infty(\Omega,\Sigma, \BB{P}_\rho)$
and $L^\infty(\Omega,\Sigma_t, \BB{P}_\rho)$, respectively.
The space $\Omega$ should be interpreted as the 
paths of the observed process $Y^s_r$ when the 
measurement is continued infinitely long. The 
$\sigma$-algebras $\Sigma_t$ contain the events up to 
time $t$. 

In the Heisenberg picture,
when a measurement of an observable $Y$ with discrete spectrum 
$Sp(Y)$ has been performed, all observables 
in $\BC\ten\WC^f\ten\WC^s$ have to be sandwiched with the projection
corresponding to the observed measurement result. If 
the result of the measurement is unknown, but the measurement
has taken place, an observable takes the form of a direct sum 
over all possible outcomes of the original observable sandwiched
with the projections corresponding to the outcomes, i.e. 
  \begin{equation*}
  X_{\mbox{after meas.}} = \bigoplus_{y \in Sp(Y)} P_y XP_y 
  \ \ \ \ X \in \BC\ten\WC^f\ten\WC^s.
  \end{equation*}
Note that this procedure
destroys all coherences between different measurement 
results. Moreover, it maps all observables in $\BC \ten \WC^f\ten\WC^s$ to 
the commutant of the algebra generated by the measured 
observable. 

Therefore, in analogy with the above, when a process $\{Y^s_r\}_{0\le r\le t}$ 
has been measured continuosly in time, we can restrict to the 
algebra $\AC_t \subset \BC\ten\WC^f\ten\WC^s$ which is
the \emph{commutant} of the observed process 
  \begin{equation*}
  \AC_t := \CC_t' := \{X \in \BC\ten\WC^f\ten\WC^s; \ XC = CX,\ \forall C \in \CC_t\}.
  \end{equation*}
We call $\AC_t$ the algebra of observables that are \emph{not 
demolished} \cite{Bel} by observing the process 
$\{Y^s_r\}_{0\le r\le t}$. Note that  
from the double commutant theorem 
it follows that $\CC_t$ is the 
\emph{center} of $\AC_t$, i.e.\ 
$\CC_t = \{C \in \AC_t;\ AC = CA,\ \forall A \in \AC_t\}$.
  
We investigate the situation of the previous paragraph 
more abstractly for a moment, i.e.\ let $\AC$ be a von Neumann algebra 
of operators on some Hilbert space $\BB{H}$ and let 
$\CC$ be its center. Let $\rho$ denote a state on the 
algebra $\AC$. We will now explain the 
decomposition of $\AC$ over its center $\CC$, 
see \cite{KaR} for all details and proofs.
We can identify the center $\CC$ with some 
$L^\infty(\Omega,\Sigma,\BB{P})$ where $\BB{P}$ corresponds
to the restriction of $\rho$ to $\CC$. The Hilbert space 
$\BB{H}$ has a direct integral representation 
$\BB{H} = \int^{\oplus}\BB{H}_\omega\BB{P}(d\omega)$ in 
the sense that there exists a family of Hilbert spaces
$\{\BB{H}_\omega\}_{\omega \in \Omega}$ and for any 
$\psi \in \BB{H}$ there exists a map 
$\omega \mapsto \psi_\omega \in \BB{H}_\omega$ such that
  \begin{equation*}
  \langle \psi,\phi\rangle = 
  \int_\Omega \langle\psi_\omega,\phi_\omega\rangle\BB{P}(d\omega)
  \ \ \ \ \psi,\phi \in \BB{H}.
  \end{equation*}
The von Neumann algebra $\AC$ has a \emph{central decomposition}
$\AC = \int^\oplus \AC_\omega\BB{P}(d\omega)$ in the sense that
there exists a family $\{\AC_\omega\}_{\omega \in \Omega}$ of 
von Neumann algebras with trivial center, or factors, and for any
$A \in \AC$ there is a map $\omega \mapsto A_\omega \in \AC_\omega$
such that $(A\psi)_\omega = A_\omega\psi_\omega$ for 
all $\psi \in \BB{H}$ and $\BB{P}$-almost all $\omega \in \Omega$. 
The state $\rho$ on $\AC$ has a decomposition in states $\rho_\omega$ on $\AC_\omega$ 
such that for any $A\in\AC$ its expectation is obtained by integrating with 
respect to $\mathbb{P}$ the expectations of its components $A_\omega$:
 \begin{equation*}
 \rho(A) =  \int_\Omega \rho_\omega(A_\omega)\BB{P}(d\omega).
 \end{equation*}
 
Loosely speaking the component $A_\omega \in \AC_\omega$ is the 
operator $A \in \AC$ sandwiched with the projection 
corresponding to a measurement result $\omega$. Moreover, 
the state $\rho_\omega$ is the state $\rho$  conditioned on the 
measurement result $\omega$. 
For all $X \in \AC$ we denote by $\rho_\bullet(X_\bullet)$ 
the function $\omega \mapsto \rho_\omega(X_\omega)$.
The complex number $\rho_\omega(X_\omega)$ is the expectation 
of the observable $X$ in the state $\rho$ conditioned on 
measurement result $\omega$.

Define a map $\EC_\rho:\ \AC \to 
\CC \cong L^\infty(\Omega,\Sigma,\BB{P})$ by
$\EC_\rho(X) := \rho_\bullet(X_\bullet)$ for all $X\in \AC$. 
It is easily verified, see also \cite{BGM}, that this map 
is linear, surjective, identity preserving, 
completely positive, it satisfies the \emph{module property}
  \begin{equation*}
  \EC_\rho(C_1XC_2) = C_1\EC_\rho(X)C_2, \ \ \ \ \ C_1,C_2\in\CC,\ X\in A,
  \end{equation*} 
and it leaves the state $\rho$ invariant, i.e.\ 
$\rho\big(\EC_\rho(X)\big) = \rho(X)$ for all $X \in \AC$. 
These properties uniquely determine the map $\EC_\rho$, see \cite{Tak}.
It is called the \emph{conditional expectation} of $\AC$
onto $\CC$ with respect to $\rho$. 
Returning to the original problem, i.e.\ a whole family of 
algebras $\AC_t$ with center $\CC_t$, we get a  
family of conditional expectations $\EC_{\rho^t}:\ \AC_t \to \CC_t$.
We denote $\EC_{\rho^t}$ more compactly by $\EC^t$. 

Apart from the family of quantum mechanical conditional expectations
$\EC^t$, there is also a family of conditional expectations
in the classical sense that plays an important role in the 
following. Denote by $\BB{E}_\rho^t$ the unique classical
conditional expectation from $\CC_\infty \cong 
L^\infty(\Omega,\Sigma, \BB{P}_\rho)$ onto 
$\CC_t \cong L^\infty(\Omega, \Sigma_t, \BB{P}_\rho)$
that leaves the state $\rho^\infty$, or equivalently, the 
expectation with respect to $\BB{P}_\rho$ invariant, i.e.\
$\rho^\infty \circ \BB{E}^t = \rho^\infty$. These conditional
expectations satisfy the \emph{tower property}, that is
$\BB{E}^s_\rho\big(\BB{E}^t_\rho(C)\big) = \BB{E}^s_\rho(C)$
for all $C \in \CC_\infty$ and $t \ge s \ge 0$. $\BB{E}^0_\rho$
is the expectation with respect to $\BB{P}_\rho$ and will 
simply be denoted $\BB{E}_\rho$. Note that the tower property
for $s=0$ is just the invariance of the state 
$\rho^\infty (= \BB{E}_\rho)$.

For all $t \ge 0$ and $X \in \BC$ the operator $X\ten\I\ten\I 
\in \BC\ten\WC^f\ten\WC^s$
commutes with the observed process $\{Y^s_r\}_{0 \le r \le t}$ up to 
time $t$, i.e.\ $\BC \subset \AC_t$. Therefore we can define for 
all $X \in \BC$ a process $\{M_t^X\}_{t\ge 0}$ in the algebra 
$\CC_\infty \cong L^\infty(\Omega, \Sigma, \BB{P}_\sigma)$
by
  \begin{equation}\label{eq defM}
  M_t^X := \EC^t(X) - \EC^0(X) - \int_0^r\EC^r\big(L(X)\big)dr,
  \end{equation}
where $L:\ \BC \to \BC$ is the Liouvillian of equation \eqref{eq gen}. 
From the definition it is clear that $M^X_t$ is an
element of $\CC_t$ for all $t\ge 0$. The process $\{M^X_t\}_{t\ge 0}$
is a martingale, i.e.\ for all $0 \le s \le t$ we have
$\BB{E}^s_\rho(M^X_t) = M^X_s$,
see \cite{Bel}, \cite{BGM} for details and a proof.
In differential form equation \eqref{eq defM} reads
  \begin{equation*}
  d\rho^t_\bullet(X) = \rho^t_\bullet\big(L(X)\big)dt +dM^X_t,
  \end{equation*}
where we have used that $X_\bullet$ is the constant function 
$\omega \mapsto X$. This equation is the 
\emph{Belavkin equation} \cite{Bel88}, \cite{Bel}, \cite{BGM}. 

Denote by $\tilde{Y}^s_t$ the process given by the following 
stochastic differential equation
  \begin{equation*}
  d\tilde{Y}^s_t = dY^s_t - \EC^t(e^{i\phi}V_s^*+e^{-i\phi}V_s)dt,\ \ \ \tilde{Y}^s_0 =0.
  \end{equation*}
The process $\tilde{Y}^s_t$ is a martingale, i.e.\ for all $0 \le r \le t$ 
we have $\BB{E}^r_\rho(\tilde{Y}^s_t) = \tilde{Y}^s_r$, 
see \cite{Bel}, \cite{BGM} for details and a proof.
We call $\tilde{Y}^s_t$ the \emph{innovating martingale}
of the observed process $Y^s_t$. The link between the martingale
$M_t^X$ and the observed process $Y^s_t$ is provided by 
the martingale representation theorem which states that there 
exists a stochastically integrable process $\eta^X_t$ such 
that 
  \begin{equation*}
  dM_t^X = \eta^X_td\tilde{Y}^s_t = 
  \eta^X_t\big(dY^s_t-\EC^t(e^{i\phi}V_s^*+e^{-i\phi}V_s)dt\big). 
  \end{equation*}
The process $\eta^X_t$ can be calculated by using that $\EC^t$ leaves
$\rho^t$ invariant \cite{Bel}. We refer to \cite{BGM} for the details, 
the result is
  \begin{equation*}
  \eta^X_t = \EC^t(e^{i\phi}V_s^*X  + e^{-i\phi}XV_s) - \EC^t(e^{i\phi}V_s^*  + 
  e^{-i\phi}V_s)\EC^t(X). 
  \end{equation*}
This leads to the Belavkin equation \cite{Bel}, \cite{BGM}
    \begin{equation}\label{eq Belavkin}\begin{split}
  d\rho_\bullet^t(X) = \rho_\bullet^t\big(&L(X)\big)dt + 
  \Big(\rho_\bullet^t(e^{i\phi}V_s^*X  + e^{-i\phi}XV_s) - \rho_\bullet^t(e^{i\phi}V_s^*  + 
  e^{-i\phi}V_s)\rho_\bullet^t(X)\Big)\times  \\
  &\times \Big(dY^s_t - 
  \rho_\bullet^t(e^{i\phi}V_s^* + e^{-i\phi_t}V_s)dt\Big) \ \ \ \ X\in \BC.
  \end{split}\end{equation}
This equation tells us how the state of the system $\BC$ evolves
over an infinitesimal time $dt$ depending on what we observe 
for the measurement process $dY^s_t$. Since $\tilde{Y}^s_t$
is a martingale with variance $t$ on the space of the 
Wiener process, it must be the Wiener process itself.

\section{Control: the essentially commutative case}\label{sec escom}

In this section we focus on dilations that are essentially 
commutative \cite{KuM0}. We will use the results of the 
measurement of $Y^s_t$ to control the time evolution $T_t$
of the system $\BC$ in order to bring it as close to 
the identity map as possible. For essentially commutative
dilations this can be done (nearly) perfectly. This section
serves as a guiding example for the more realistic situations 
described in sections \ref{sec con-squeez} and 
\ref{sec con+squeez}.

Let $V$ be selfadjoint, i.e.\ $V=V^*$. The discussion below can 
easily be adapted to fit the situation where $V = -V^*$. 
Define for $\sigma = f,s$ field observables 
$Y^\sigma_t := i(A_\sigma^*(t) - A_\sigma(t))\in\WC^\sigma_{t]}$. 
Using $V=V^*$, equation \eqref{eq HuP}, i.e.\ the laser on the forward 
channel is off, simplifies to 
  \begin{equation}\label{eq escomHuP}
  dU_t = \Big\{-iV_fdY^f_t -iV_sdY^s_t- \frac{1}{2}V^2dt\Big\}U_t, 
  \ \ \ \ \ U_0 = \I.
  \end{equation}
This means that for $t\ge 0$ the solution $U_t$ is an element 
of $\BC \ten \CC_t$, with $\CC_t$ the commutative von Neumann 
algebra generated by the process 
$\{Y^f_r\ten Y^s_r\}_{0 \le r\le t}$. 
(We have dropped the extensive notation with the identities
tensored to the $Y_r$'s.)
This means that we can restrict the dilation of diagram 
\eqref{diag dildiaglaser} to $\BC\ten\CC_\infty$, i.e.
  \begin{equation}\label{diag escomdildiag}\begin{CD}
     \BC @>T_t>> \BC              \\
     @V{\Id \ten \I \ten\I}VV        @AA{\Id \ten \phi\ten\phi}A      \\
     \BC\ten\CC_\infty @>\hat{T}_t>> \BC\ten \CC_\infty           \\
  \end{CD}\end{equation}
A dilation for which the relative commutant of the embedding 
of the algebra $\BC$ into the subalgebra of $\BC\ten\WC^f\ten\WC^s$
generated by $\{U_t^*X\ten\I\ten\I U_t;\ X\in \BC,\, t\ge 0\}$
is commutative, is called \emph{essentially
commutative} \cite{KuM0}. Although we restrict the discussion to the 
essentially commutative dilation determined by equation 
\eqref{eq escomHuP}, the results of this section can be 
extended to all essentially commutative dilations \cite{KuM0}.

If the dilation is essentially commutative the derivation 
of the Belavkin equation is extremely simple. Since $U_t$
is not demolished by observing $\{Y^s_r\}_{0\le r\le t}$,
i.e.\ it is an element of the commutant of $\CC_t$, we can 
just calculate $d(U^*_tX\ten\I\ten\I U_t)$ using the quantum It\^o
rules and decompose it over the paths of the measurement 
process. It is clear that this leads exactly 
to the Belavkin equation of the previous section 
\eqref{eq Belavkin} with $\phi = \frac{\pi}{2}$ 
and $V=V^*$  
  \begin{equation}\label{eq escomBelavkin}
  d\rho^t_\bullet(X) = 
  \rho^t_\bullet\big(L(X)\big)dt + i\rho^t_\bullet\big([V_s,X]\big)dY^s_t, 
  \end{equation}  
where $L$ is as in equation \eqref{eq gen} with $H$ is $0$, i.e.\
there has been no control yet.
In general, however, we do not have the decomposition of $U_t$ over the 
center and we have to resort to the methods of the previous section.
Note that for $\phi = \frac{\pi}{2}$ and $V=V^*$ we have 
$d\tilde{Y}^s_t = dY^s_t$, i.e.\ $Y^s_t$ is the Wiener process. 
This means that the measurement 
process is \emph{non-informative} \cite{GrW}, i.e.\ since here 
there is no state dependent drift term, we do not 
gain information about the state $\rho^t_\bullet$ by observing 
$Y^s_t$.

Let $\rho^0$ be the density matrix of the initial state 
of the system $\BC$. We observe $Y^s_t$ from time $0$ 
to time $\tau$. Suppose that the laser is off in that 
time interval, i.e. $h(t) = 0$ for $0 \le t < \tau$. 
Then the stochastic density matrix at time 
$\tau$ is given by \eqref{eq escomBelavkin}
  \begin{equation*}
  \tilde{\rho}_\bullet^\tau = \rho^0 + \int_0^\tau 
  V\tilde{\rho}^t_\bullet V^*dt -\frac{1}{2}\{V^*V, \tilde{\rho}^t_\bullet\}dt
  + i[\tilde{\rho}^t_\bullet, V_s]dY^s_t,  
  \end{equation*} 
where the tilde has been introduced to remind us that this 
is the state \emph{before} control has taken place.  
In the time interval from $0$ to $\tau$ we have observed 
$Y^s_t$ and therefore we can determine the difference 
$\Delta(\tau) := Y^s_\tau - Y^s_0$ at time $\tau$. Then we want 
to control the state $\tilde{\rho}^\tau_\bullet$ with a 
unitary
  \begin{equation*}
  U^\tau_{c} := \exp\big(i\Delta(\tau)V_s\big),
  \end{equation*}
i.e.\ the density matrix after control is given by
$\rho^\tau_\bullet = U^\tau_c\tilde{\rho}^\tau_\bullet U^{\tau*}_{c}$.
This can be done by supplying a very sharply peaked 
laser pulse to the system, i.e.\ take
  \begin{equation*}
  h(t) = -i\frac{\kappa_s\Delta(\tau)}{2\kappa_f}\delta_\tau(t),\ \ \ \ 0 \le t < 2\tau, 
  \end{equation*}
where $\delta_\tau$ is the delta function at time $\tau$.
Then $H = -\Delta(\tau)\delta_\tau V_s$ in equation \eqref{eq gen}, 
i.e.\ at time $\tau$ all terms in equation \eqref{eq gen} 
are negligable with respect to the commutator with $H$. At time $\tau$
this commutator performs a Rabi oscillation exactly of size $U^\tau_c$.
After having applied the control unitaries the 
state $\rho^\tau_\bullet$ is taken 
as the new initial state $\rho^0$ and  
the control scheme is repeated after every 
$\tau$ time units. 

Note that the control unitary $U^\tau_c$ satisfies the 
following stochastic differential equation
  \begin{equation*}
  dU^\tau_c = \{iV_sdY^s_\tau - \frac{1}{2}V_s^2d\tau\}U^\tau_c
  = U^\tau_c\{iV_sdY^s_\tau - \frac{1}{2}V_s^2d\tau\}, \ \ \ \ U^0_c=\I.
  \end{equation*}
Recall that we have the It\^o rules $dY^s_tdY^s_t = dt,\ 
dY^s_tdt = dtdY^s_t =0,\ dY^f_tdt = dtdY^f_t =0$ and
$dY^s_tdY^f_t = dY^f_tdY^s_t = 0$. 
Using the notation below  Theorem \ref{Itorules} with 
$Z_1 = U^\tau_c,\, Z_2 = \tilde{\rho}^\tau_\bullet$ and $Z_3 = U^{\tau*}_c$  
we find infinitesimally at $\tau = 0$, i.e.\ $\tau$ should be 
very small or equivalently we should correct with extremely 
high frequency
  \begin{equation}\label{eq controlrho}
  d\rho^\tau_\bullet\Big|_{\tau = 0} = 
  \big([1] + [2] + [3] + [12] + [13] + [23] + [123]\big)\Big|_{\tau=0}.
  \end{equation}
Note that it immediately follows from It\^o's rules that 
$[123] =0$.  

For $W \in \BC$ we denote by $L_W$ the Lindblad operator
corresponding to $W$ acting on density matrices 
$\rho$, i.e.\ 
  \begin{equation*} 
  L_W(\rho) := W\rho W^* - \frac{1}{2}\{W^*W,\rho\}. 
  \end{equation*} 
Then we can write $([1]+[3]+[13])|_{\tau=0} = 
L_{V_s}(\rho^0)d\tau - 
i[\rho^0, V_s]dY^s_0$ and 
$([2]+[12]+[23])|_{\tau=0} = L_V(\rho^0)d\tau + 
i[\rho^0, V_s]dY^s_0 - 
2L_{V_s}(\rho^0)d\tau$. Therefore we get
$d\rho^\tau_\bullet|_{\tau=0} =  L_{V_f}(\rho^0)d\tau$ and since 
we repeat the control every $\tau$ time units 
with $\tau$ very small, i.e.\ we take $\tau$ infinitesimal, 
this leads to the following deterministic state evolution
  \begin{equation*}
  d\rho^t = L_{V_f}(\rho^t)dt.
  \end{equation*}
This means we only have dissipation into the forward channel.
We can take $\kappa_f$ arbitrarily small which means we have succeeded
in freezing the state evolution nearly perfectly, i.e.\ the control 
scheme \emph{restores quantum information} in the sense of \cite{GrW}.

\section{Control without squeezing}\label{sec con-squeez}

We now return to the more realistic situation
of spontaneous decay of a two-level atom to its ground state. 
We are again interested in controlling the state of a system 
in order to get as close as possible to freezing its state 
evolution. However, in trying to do this, we encounter 
problems that motivate the investigation put forward in the 
sections to come.

Guided by the previous section we write 
$V$ of equation \eqref{eq V} as the sum 
$V= V_R + iV_I$ with $V_R$ and $V_I$ 
selfadjoint, i.e.\ 
  \begin{equation}\label{eq defVRI}
  V_R := \frac{V+V^*}{2} = \frac{1}{2}\begin{pmatrix}0 & 1 \\ 1 & 0\end{pmatrix}
  \ \ \ \ \mbox{and} \ \ \ \ 
  V_I := \frac{V-V^*}{2i} = \frac{1}{2}\begin{pmatrix}0 & i \\ -i & 0 \end{pmatrix}. 
  \end{equation}
Denote for $\sigma = f,s:\ V_R^\sigma := \kappa_\sigma V_R,\ 
V_I^\sigma := \kappa_\sigma V_I,\  
Y^\sigma_R(t) := i\big(A_\sigma^*(t)-A_\sigma(t)\big)$ 
and $Y^\sigma_I(t) := A_\sigma^*(t)+A_\sigma(t)$. Then  
equation \eqref{eq HuP}, i.e.\ the laser is off, can be written as
  \begin{equation}\label{eq quadHuP}\begin{split}
  dU_t := \Big\{\Big(\sum_{\sigma = f,s} iV^\sigma_IdY^\sigma_I(t)  
  -iV^\sigma_RdY^\sigma_R(t)\Big) -
  \frac{1}{2}V^*Vdt\Big\}U_t, \ \ \ \ U_0 = \I. 
  \end{split}\end{equation}
Since the noises $Y^s_R(t)$ and $Y^s_I(t)$ in 
the side channel do not commute we can 
not observe them both simultaneously. 

In the following we choose to observe $Y^s_R(t)$ and to 
keep notation simple we denote it by $Y_t$. 
The Belavkin equation for observation of $Y_t$ follows 
from equation \eqref{eq Belavkin}
  \begin{equation*}\begin{split}
  d\rho_\bullet^t = L&(\rho_\bullet^t)dt + 
  i\Big(\rho_\bullet^tV_s^*  - V_s\rho_\bullet^t - \mbox{Tr}(\rho_\bullet^tV_s^*  - 
  V_s\rho_\bullet^t)\rho_\bullet^t\Big)\times\\
  &\times\Big(dY_t - 
  i\mbox{Tr}(\rho_\bullet^tV_s^*  - 
  V_s\rho_\bullet^t)dt\Big).
  \end{split}\end{equation*}
where $L$ is given by equation \eqref{eq gen} with $H =0$.  
Using the relation 
$\rho_\bullet^tV_s^* - V_s\rho_\bullet^t = [\rho_\bullet^t,V^s_R] - 
i\{\rho_\bullet^t,V^s_I\}$, this equation simplifies to
  \begin{equation}\label{eq BelYRs}
  d\rho_\bullet^t = L(\rho_\bullet^t)dt + 
  \Big(i[\rho_\bullet^t,V^s_R] + 
  \{\rho_\bullet^t,V^s_I\} - 2\mbox{Tr}(\rho_\bullet^tV^s_I) \rho_\bullet^t\Big)
  \Big(dY_t - 
  2\mbox{Tr}(\rho_\bullet^tV^s_I)dt\Big).
  \end{equation} 
Note that $Y_t$ is a Wiener process plus a stochastic 
drift term that depends on the state of the two-level 
atom. By observing $Y_t$ we can estimate this drift 
term and in this way obtain information about the state
$\rho^t_\bullet$.
   
We run a control scheme similar to the one 
in the previous section, i.e.\ we choose 
$h(t) := -i\frac{\kappa_s\Delta(\tau)}{2\kappa_f}\delta_\tau(t)$ 
for $0\le t<2\tau$. Then we get a control unitary 
$U^\tau_c = \exp\big(i\Delta(\tau)V^s_R\big)$, 
satisfying the stochastic differential equation
  \begin{equation*}
  dU^\tau_c = \{iV^s_RdY_\tau - \frac{1}{2}{V^s_R}^2d\tau\}U^\tau_c
  = U^\tau_c\{iV^s_RdY_\tau - \frac{1}{2}{V^s_R}^2d\tau\}, \ \ \ \ U^0_c =\I.  
  \end{equation*} 
The state after control is again given by 
$\rho^\tau_\bullet := U^\tau_c\tilde{\rho}^\tau_\bullet {U^\tau_c}^*$ where 
$\tilde{\rho}^\tau_\bullet$ is given by the Belavkin 
equation \eqref{eq BelYRs}. We use the notation below 
Theorem \ref{Itorules} with 
$Z_1 = U^\tau_c,\, Z_2 = \tilde{\rho}^\tau_\bullet$ and $Z_3 = U^{\tau*}_c$. 
For infinitesimal $\tau$ evaluated at $\tau =0$, 
this leads to equation \eqref{eq controlrho}, i.e.\
  \begin{equation*}
  d\rho^\tau_\bullet\Big|_{\tau=0} = 
  \big([1]+[2]+[3]+[12]+[13]+[23]+[123]\big)\Big|_{\tau=0}.
  \end{equation*}  
Again $[123] =0$ and further $([1]+[3]+[13])|_{\tau=0} = 
L_{V_R^s}(\rho^0)d\tau - i[\rho^0, V_R^s]dY_0$.
Furthermore  we have
  \begin{equation*}\begin{split}
  &[2]\Big|_{\tau=0} = L_V(\rho^0)d\tau + 
  \Big(i[\rho^0,V^s_R] + 
  \{\rho^0,V^s_I\} - 2\mbox{Tr}(\rho^0 V^s_I) \rho^0\Big)
  \Big(dY_0 - 
  2\mbox{Tr}(\rho^0 V^s_I)d\tau\Big), \\
  &\big([12]+[23]\big)\Big|_{\tau = 0} = -2L_{V_R^s}(\rho^0)d\tau
  +i[V_R^s,\{\rho^0,V^s_I\}]d\tau 
  - 2\mbox{Tr}(\rho^0 V^s_I)i [V_R^s, \rho^0]d\tau.
  \end{split}\end{equation*}
A calculation shows that 
  \begin{equation*}
   L_V(\rho^0)-L_{V_R^s}(\rho^0)
  +i[V_R^s,\{\rho^0,V^s_I\}] =  
  L_{V_f}(\rho^0) + L_{V^s_I}(\rho^0) 
  + \frac{i}{2}[V^s_IV^s_R + V_R^sV_I^s, \rho^0].
  \end{equation*}
Since for spontaneous decay $V^s_IV^s_R + V_R^sV_I^s =0$, 
we get
  \begin{equation*}
  d\rho^\tau_\bullet\Big|_{\tau = 0} = L_{V_f}(\rho^0)d\tau +
  L_{V_I^s}(\rho^0)d\tau +
  \Big(\{\rho^0,V^s_I\} - 
  2\mbox{Tr}(\rho^0 V^s_I) \rho^0\Big)
  \Big(dY_0 - 
  2\mbox{Tr}(\rho^0 V^s_I)d\tau\Big).
  \end{equation*}
Since we repeat the control every $\tau$ time units with $\tau$ very small, i.e.\
we take $\tau$ infinitesimal, this leads to the following stochastic time evolution
for the density matrix of the two-level atom
  \begin{equation*}
  d\rho^t_\bullet = L_{V_f}(\rho^t_\bullet)dt +
  L_{V_I^s}(\rho^t_\bullet)dt +
  \Big(\{\rho^t_\bullet,V^s_I\} - 
  2\mbox{Tr}(\rho^t_\bullet V^s_I) \rho^t_\bullet\Big)
  \Big(dY_t - 
  2\mbox{Tr}(\rho^t_\bullet V^s_I)dt\Big).
  \end{equation*}
The first term on the right hand side is harmless, we already encountered it in 
the previous section, by taking $\kappa_f$ small enough 
we can make it as small as we want. The third term is 
also harmless. Since $Y_t-\int_0^t 2\mbox{Tr}(\rho^r_\bullet V^s_I)dr$ 
is a martingale it vanishes when we average over all possible 
outcomes for $Y_t$. However, the second term reflects the 
fact that we can not observe $Y_I^s$ and correct it 
simultaneously with $Y_R^s$. The next sections 
are devoted to finding a way around this problem.

\section{Squeezed states and their calculus}\label{sec squeez}

In this section we drop the assumption that the side channel 
of the field is initially in the vacuum state. We take a 
step back and rethink our model for (a channel in) the field. 
For the vacuum state we are going to end up with the description 
we have already used this far. Goal of the description below 
is to incorporate the situation where the initial state 
of the side channel is a so-called \emph{squeezed} state.
In a sqeezed state the variance of one of the quadratures
$Y_R^s$ and $Y^s_I$ decreases while the other one increases
as a result of Heisenberg's uncertainty relation. In the 
next section we will observe the increased quadrature 
and correct it. The disturbing effect of the other quadrature
has decreased as a result of the squeezing.

Let $H$ be the real space of quadratically 
integrable $\BB{R}^2$-valued functions on $\BB{R}$. On $H$ we define 
a symplectic form $\sigma:\ H\times H \to \BB{R}$ by
  \begin{equation*}
  \sigma(f,g) := -\int_{\BB{R}}\begin{pmatrix}f_1 f_2\end{pmatrix} 
  \begin{pmatrix}0 & -1 \\ 1 & 0\end{pmatrix}
  \begin{pmatrix}g_1 \\g_2 \end{pmatrix}d\lambda, \ \ \ \ \ 
    f = \begin{pmatrix}f_1 \\f_2 \end{pmatrix},\
     g= \begin{pmatrix}g_1 \\g_2 \end{pmatrix} \in H,  
  \end{equation*}
where $\lambda$ denotes the Lebesgue measure on $\BB{R}$. For 
notational convenience we define
  \begin{equation*}
  J_0 := \begin{pmatrix}0 & -1 \\ 1 & 0\end{pmatrix}.
  \end{equation*}
We will describe (a channel in) the electromagnetic 
field by the $C^*$-algebra of \emph{canonical 
commutation relations} $CCR(H,\sigma)$ over 
the symplectic space $(H,\sigma)$.

The algebra $CCR(H,\sigma)$ is defined as the $C^*$-algebra generated 
by abstract elements $\{W(f);\ f\in H\}$ satisfying relations
  \begin{equation}\begin{split}\label{eq defCCR} 
  &1.\ \ \ \ \  W(f)^* = W(-f),\ \ \ \ \ f \in H,                                      \\
  &2.\ \ \ \ \  W(f)W(g) = \exp\big(-i\sigma(f,g)\big)W(f+g),\ \ \ \ \ f,g \in H.
  \end{split}\end{equation} 
The second relation is called the \emph{Weyl relation}. 
It follows from \cite{Sla} that the $C^*$-algebra 
$CCR(H,\sigma)$ exists and moreover that it is unique 
up to isomorphism. Furthermore it immediately follows 
from \eqref{eq defCCR} that $W(f)$ is unitary for all
$f\in H$.

Let $\alpha:\ H\times H \to \BB{R}$ be a symmetric positive 
bilinear form satisfying
  \begin{equation}\label{eq existence}
  \sigma(f,g)^2 \le \alpha(f,f)\alpha(g,g),\ \ \ \ \ f,g \in H.
  \end{equation}
It is well known (cf.\ \cite{Pet}) that if $\alpha$ 
satisfies \eqref{eq existence} then there exists a 
unique state $\gamma$ on the $C^*$-algebra $CCR(H,\sigma)$ 
satisfying
  \begin{equation}\label{eq quasifree}
  \gamma\big(W(f)\big) = \exp\big(-\frac{1}{2}\alpha(f,f)\big), \ \ \ \ \ f \in H.
  \end{equation}
Such a state $\gamma$ on $CCR(H,\sigma)$ is called a 
\emph{quasifree} state.  
  
In this paper we focus on a particular class of quasifree 
states $\gamma_{nc}$ indexed by a parameter $n \in \BB{R}$ 
and a complex parameter $c = a+ib, a,b \in \BB{R}$. These 
states will turn out to be the \emph{squeezed white noise} 
states of the field as they are encountered in quantum optics 
after a Markov approximation is made (cf.\ \cite{GaZ}). 
They are defined through equation \eqref{eq quasifree} 
with a symmetric positive bilinear form $\alpha_{nc}$ given 
by \cite{HHKKR}
  \begin{equation*}
  \alpha_{nc}(f,g) = \int_{\BB{R}}\begin{pmatrix}f_1 f_2\end{pmatrix}
  \begin{pmatrix}2n+1+2a & 2b \\ 2b & 2n +1 -2a\end{pmatrix}
  \begin{pmatrix}g_1 \\ g_2 \end{pmatrix}d\lambda, \ \ \ \ \ f,g \in H.   
  \end{equation*}
For notational convenience we define
  \begin{equation*}
  Q_{nc} := \begin{pmatrix}2n+1+2a & 2b \\ 2b & 2n+1-2a\end{pmatrix}.
  \end{equation*}
Condition \eqref{eq existence} leads to the restrictions 
$n(n+1) \ge |c|^2$ and $n \ge 0$. For $n=c=0$ we get the 
usual \emph{vacuum state} and for $c =0$ we end up with 
a chaotic temperature state. More details on the interpretation
of this class of states will follow below.
  
A real linear map $J:\ H \to H$ is called \emph{multiplication  by $i$} 
if it satisfies $J^2 = -\mbox{id}$. Then $H$ is a complex vector 
space with the usual addition and the scalar multiplication given 
by $(x+iy)f = xf + yJf$ for all $x,y \in \BB{R}$.
  \begin{lem}\label{lem Fockstate}
  Let $n \ge 0$ and $n(n+1) \ge |c|^2$.
  $H$ can be considered as a complex vector space 
  equipped with an inner product given by 
    \begin{equation}\label{eq inner product}
    \langle f,g\rangle_{nc} = \alpha_{nc}(f,g) + i\sigma(f,g),\ \ \ \ \ f,g \in H,
    \end{equation}
    if and only if $n(n+1) = |c|^2$. In this case multiplication
    by $i$ is given by $J_{nc} = J_0 Q_{nc}$.
  \end{lem}
\begin{proof}
Since the inner product \eqref{eq inner product} is linear in its 
second argument $J_{nc}$ has to satisfy 
$\sigma(f,J_{nc}g) = \alpha_{nc}(f,g)$ for all $f,g\in H$.
It easily follows from $n \ge 0$ and $n(n+1) \ge |c|^2$ 
that $Q_{nc}$ is non degenerate. Therefore $J_{nc}$ 
has to satisfy $-J_0J_{nc} = Q_{nc}$ which is equivalent 
to $J_{nc} = J_0Q_{nc}$. $J_{nc}$ is multiplication 
by $i$ if and only if $J_{nc}^2=-\mbox{id}$, which
is equivalent to
     \begin{equation*}\begin{split}
  &\begin{pmatrix} 0 & -1 \\ 1 & 0 \end{pmatrix}
  \begin{pmatrix}2n + 1 + 2a & 2b \\ 2b & 2n + 1 - 2a \end{pmatrix}
  \begin{pmatrix} 0 & -1 \\ 1 & 0 \end{pmatrix}
  \begin{pmatrix}2n + 1 + 2a & 2b \\ 2b & 2n + 1 - 2a \end{pmatrix}   = \\
  &\ \ \ \ \ \ \ \ \ \ \ \ \ \ \ \ \ \ \ \ \ \ \ \ \ \ 
  \begin{pmatrix}-1 & 0 \\ 0 & -1 \end{pmatrix}\ \ \ \ \ \ \ \ \iff \\
  &\begin{pmatrix}4b^2 -(2n+1+2a)(2n+1-2a) & 0 \\ 
   0 & 4b^2 -(2n+1+2a)(2n+1-2a)\end{pmatrix} =                 \\
  &\ \ \ \ \ \ \ \ \ \ \ \ \ \ \ \ \ \ \ \ \ \ \ \ \ \ 
  \begin{pmatrix}-1 & 0 \\ 0 & -1\end{pmatrix}\ \ \ \ \ \ \ \ \iff \ \ \ \ \ \ \ \ \
  |c|^2 = n^2+ n.
  \end{split}\end{equation*}  
\end{proof}  

In the following we will always be in the situation of 
Lemma \ref{lem Fockstate}, i.e.\ $n \ge 0$ and $n(n+1) = |c|^2$.
The states of Lemma \ref{lem Fockstate}, i.e.\ states that allow 
for the definition of an inner product on $H$ through 
\eqref{eq inner product}, are called \emph{Fock states} (name will become
apparent in a minute).   
We denote the complex Hilbert space given by the pair $(H,J_{nc})$
equipped with the inner product of \eqref{eq inner product} by $H_{nc}$.   
Note that $H_{00}$ is just the space $L^2(\BB{R})$ of all 
quadratically integrable functions on the real line $\BB{R}$. 
The representation of $CCR(H,\sigma)$ discussed below is actually 
the GNS-representation with respect to a Fock state $\gamma_{nc}$, see
\cite{Pet} for the details.

Fix $n \ge 0$ and $c \in \BB{C}$ such that $n(n+1) = |c|^2$. Recall that
the bosonic Fock space over $H_{nc}$ was defined as
  \begin{equation*}
  \FC_{nc} := \BB{C} \oplus \bigoplus_{k = 1}^\infty H_{nc}^{\ten_s k}, 
  \end{equation*} 
and that for all $f$ in $H_{nc}$ the \emph{exponential vector} is 
given by $e(f) := 1 \oplus \bigoplus_{k = 1}^\infty 
\frac{1}{\sqrt{k!}}f^{\ten k}$.
The span of all exponential vectors was denoted $\DC$ and the 
\emph{vacuum vector} $e(0)= 1\oplus 0 \oplus 0 \oplus \ldots$ 
was also written as $\Phi$. On the dense domain $\DC$ we 
define for all $f\in H_{nc}$ operators $W_{nc}(f)$ by
  \begin{equation*}
  W_{nc}(f)e(g) := \exp\big(-\langle f,g\rangle_{nc} - 
  \frac{1}{2}\alpha_{nc}(f,f)\big)e(f+g), \ \ \ \ f,g \in H_{nc}.
  \end{equation*}
They are isometric and therefore uniquely extend to unitary 
operators on $\FC_{nc}$. The mapping $\Pi_{nc}:\ W(f)\mapsto W_{nc}(f)$ 
uniquely defines a linear map
$\Pi_{nc}$ from $CCR(H,\sigma)$ into the bounded operators on 
the bosonic Fock space. The map $\Pi_{nc}$ preserves the 
relations $1.$ and $2.$ of \eqref{eq defCCR} defining $CCR(H,\sigma)$,
i.e.\ it is a representation of the canonical commutation relations 
on $\FC_{nc}$. The state $\gamma_{nc}$ is now given by the vector 
$\Phi \in \FC_{nc}$, i.e.\
  \begin{equation*}
  \gamma_{nc}(X) = \langle \Phi, \Pi_{nc}(X)\Phi \rangle_{nc}, \ \ \ \ X \in CCR(H,\sigma).
  \end{equation*}
The triple $(\FC_{nc}, \Pi_{nc}, \Phi)$ is the 
GNS-triple corresponding to the state $\gamma_{nc}$, cf.\ \cite{Pet}.
The algebra of observables for the electromagnetic field in 
the Fock state $\gamma_{nc}$ is modelled by the von Neumann 
algebra $\WC_{nc}$ generated by $\{W_{nc}(f);\ f \in H_{nc}\}$, 
which is just all bounded operators on $\FC_{nc}$. 

\textbf{Remark.} We can reduce the case of a non Fock quasifree 
state to a Fock state by doubling the space $H$ to $H\oplus H$.
We can embed the algebra of canonical commutation relations over
$H$ into the algebra of canonical commutation relations over 
$H\oplus H$ and view the state on $CCR(H,\sigma)$ as the 
restriction of a Fock state on this bigger algebra (cf.\ \cite{Pet}). 
In this way we get representations on a doubled up Fock space. 
Then the algebra of observables is not the whole algebra of 
bounded operators but a true subalgebra.  

The dilation of the semigroup $T_t$ of diagram \eqref{diag dildiag} 
serves as our starting point. We change it by replacing the vacuum 
state $\phi=\gamma_{00}$ on the side channel by the Fock state 
$\gamma_{nc}$ described above. The dilation diagram then changes to
   \begin{equation}\label{diag dildiagsqueezed}\begin{CD}
     \BC @>T^{nc}_t>> \BC              \\
     @V{\Id \ten \I\ten\I}VV        @AA{\Id \ten\phi\ten \gamma_{nc}}A      \\
     \BC\ten\WC^f\ten\WC^s_{nc} @>\hat{T}^{nc}_t>> \BC\ten\WC^f\ten \WC^s_{nc}           \\
  \end{CD}\end{equation}
Coupling the quantum system to a field in another state than the
vacuum has changed its reduced dynamics to $T^{nc}_t$. Changing 
the representation space of the algebra of canonical commutation
relations from $\FC = \FC_{00}$ to $\FC_{nc}$ also means that we 
have to describe the joint evolution of the system and (the two channels 
in) the field in this representation. Making sense of the 
group $\hat{T}^{nc}_t$ will be our main concern for the 
remainder of this section. 
   
For all $f \in H$ the family of operators 
$\big\{W_{nc}(tf)\big\}_{t \in \BB{R}}$
forms a one-parameter group, continuous in the strong operator
topology. Therefore it follows from Stone's theorem that 
for all $f \in H$ there exists a selfadjoint $B_{nc}(f)$ 
such that
  \begin{equation*}
  W_{nc}(tf) = \exp\big(itB_{nc}(f)\big).
  \end{equation*}  
The operators $B_{nc}(f)$ are called \emph{field operators}.
The domain of the operator $B_{nc}(f_k)\ldots$ $B_{nc}(f_1)$
contains $\DC$ for every $f_1,\ldots f_k \in H$ and
$k \in \BB{N}$ (cf.\ \cite{Pet}). For $f,g \in H$ and $t \in \BB{R}$
it follows from the Weyl relation that on the domain $\DC$
  \begin{equation}\label{eq properties fields}\begin{split}
  &1.\ \ \   B_{nc}(tf) = tB_{nc}(f),          \\
  &2.\ \ \   B_{nc}(f+g) = B_{nc}(f) + B_{nc}(g),          \\
  &3.\ \ \   [B_{nc}(f), B_{nc}(g)] = 2 i\sigma(f,g).
  \end{split}\end{equation}

Let $H_0$ be the real Hilbert space $\{f\in H;\ f = (f_1,0)\}$. From 
(\ref{eq properties fields}.3) it immediately follows that the 
family of operators $\{B_{nc}(f); f \in H_0\}$ is commutative. 
Using spectral theory, they can be realised as random variables
on a single measure space. If the field 
described by the algebra $CCR(H,\sigma)$ is 
in the Fock state $\gamma_{nc}$, 
then the joint characteristic function of the 
random variables $B_{nc}(f_1), B_{nc}(f_2),\ldots,B_{nc}(f_k)$ 
is for $t_1,\ldots,t_k \in \BB{R}$ given by
  \begin{equation*}\begin{split}
  &\Big\langle\Phi,\exp\big(it_1B_{nc}(f_1)\big)\exp\big(it_2B_{nc}(f_2)\big)\ldots
  \exp\big(it_kB_{nc}(f_k)\big)\Phi\Big\rangle = \\
  &\Big\langle\Phi,\exp\big(i\sum_{i=1}^k t_i B_{nc}(f_i)\big)\Phi\Big\rangle = 
  \gamma_{nc}\Big(W\big(\sum_{i=1}^k t_i f_i\big)\Big) = 
  \exp\Big(-\frac{1}{2}\sum_{i,j=1}^kt_it_j\alpha_{nc}(f_i,f_j)\Big),
  \end{split}\end{equation*}
i.e.\ their joint distribution is Gaussian with covariance matrix 
$\alpha_{nc}(f_i,f_j)$. In a similar way it can be 
shown that the family $\{B_{nc}(J_0f);\ f\in H_0\}$ is commutative 
and the joint distribution of the random variables 
$B_{nc}(J_0f_1), \ldots B_{nc}(J_0f_k)$ is Gaussian with covariance matrix
$\alpha_{nc}(J_0f_i,J_0f_j)$. The Gaussianity of these fields,
the covariance matrix and the condition $|c|^2 = n^2 + n$ are exactly 
the defining properties of a squeezed vacuum state in the quantum 
optics literature, cf.\ \cite{GaZ}.   

\begin{de}\label{def Anc and A0}
Fix $n\in \BB{R}$ and $c \in \BB{C}$ such that $|c|^2 = n^2+n$. 
On the domain $\DC \subset \FC_{nc}$ we define \emph{creation} and
\emph{annihilation} operators by
  \begin{equation*}\begin{split}
  &A_{nc}^*(f) := \frac{1}{2}\big(B_{nc}(f) - iB_{nc}(J_{nc}f)\big), \ \ \
  A_{nc}(f) := \frac{1}{2}\big(B_{nc}(f) + iB_{nc}(J_{nc}f)\big), \ \ \ f\in H, \\
  &A_0^*(f) := \frac{1}{2}\big(B_{nc}(f) - iB_{nc}(J_0f)\big), \ \ \ \ \ \
  A_0(f) := \frac{1}{2}\big(B_{nc}(f) + iB_{nc}(J_0f)\big), \ \ \ \ \ f\in H.
  \end{split}\end{equation*}
\end{de}
It immediately follows from equation 
(\ref{eq properties fields}.3) that these 
operators satisfy the following commutation 
relations
$[A_0(f), A_0(g)] = [A^*_0(f), A^*_0(g)] = 
[A_{nc}(f), A_{nc}(g)] = [A^*_{nc}(f),A^*_{nc}(g)]$ $= 0,\
[A_{nc}(f), A^*_{nc}(g)] = \langle f,g\rangle_{nc}$ and 
$[A_0(f), A^*_0(g)] = \langle f,g\rangle_{00}$ 
for all $f,g \in H$.
Moreover, it is a standard result (cf.\ \cite{Pet}) that for Fock states
$A_{nc}(f)\Phi = 0,\ f\in H$. Furthermore, we can build up the 
symmetric Fock space by acting with creation operators on the vacuum. 
From all these properties it easily follows that for all $h,f,g\in H$
  \begin{equation*}
  A_{nc}(f)e(g) = \langle f, g\rangle_{nc}e(g),\ \ \ \mbox{and}\ \ \ 
  \big\langle e(h), A^*_{nc}(f)e(g) \big\rangle_{\FC_{nc}} = 
  \langle h, f\rangle_{nc}\big\langle e(h),e(g)\big\rangle_{\FC_{nc}}, 
  \end{equation*}
i.e.\ $A_{nc}(f)$ and $A^*_{nc}(f)$ satisfy the relations of 
section \ref{sec qsc}. This means we can define  
stochastic integrals with respect to $A_{nc}$ and 
$A_{nc}^*$.

Define the following (non-atomic) projection valued measure $\xi$ on the direct 
sum Hilbert space $\HC = L^2(\BB{R}) \oplus H_{nc}$ consisting of a copy of 
$L^2(\BB{R})$ for the forward channel and a copy of $H_{nc}$ for the 
side channel by 
  \begin{equation*}
  \xi(I):\ L^2(\BB{R}) \oplus H_{nc} \to L^2(\BB{R})\oplus H_{nc}:\ 
  g\oplus\begin{pmatrix}f_1 \\ f_2 \end{pmatrix} \mapsto 
  g\chi_I\oplus \begin{pmatrix}f_1\chi_I \\ f_2\chi_I \end{pmatrix},
  \end{equation*}
for all Borel subsets $I$ of $\BB{R}$. Here $\chi_I$ denotes the 
indicator function of the set $I$. Define $\xi$-martingales  
by 
  \begin{equation*}\begin{split}
  &m^f:\ \BB{R}_+ \to \HC:\ t \mapsto m^f_t := \chi_{[0,t]} \oplus 
  \begin{pmatrix}0 \\ 0\end{pmatrix}, \\
  &m^s:\ \BB{R}_+ \to \HC:\ t\mapsto m^s_t :=
  0 \oplus \begin{pmatrix}\chi_{[0,t]} \\ 0 \end{pmatrix}.
  \end{split}\end{equation*}
The measure $\langle\langle m^s,m^s\rangle\rangle$ is then given 
by $\langle\langle m^s,m^s\rangle\rangle\big([0,t]\big) = 
\langle m^s_t, m^s_t\rangle_{nc} = (2n+1+2a)t$. 
For $A_{nc}\big(\frac{m^s_t}{\sqrt{2n+1+2a}}\big)$ and 
$A^*_{nc}\big(\frac{m^s_t}{\sqrt{2n+1+2a}}\big)$ we introduce the 
shorthand notation $A_s(t)$ and $A_s^*(t)$, respectively. Note that 
for stochastic integrals with respect to $A_s(t)$ and $A_s^*(t)$ 
we find the Hudson-Parthasarathy It\^o table. We denote
$A_0(m^s_t)$ and $A^*_0(m^s_t)$ more compactly by $A_0(t)$ 
and $A^*_0(t)$. The following lemma enables the definition of 
stochastic integrals with respect to $A_0(t)$ and $A^*_0(t)$. 

\begin{lem}\label{lem lincomb}
Let $n \in \BB{R}$ and $c \in \BB{C}$ such that $n(n+1) = |c|^2$.
Then for all $t \ge 0$ we can write $A_0(t)$ and $A^*_0(t)$ as 
the following linear combinations of $A_s(t)$ and $A_s^*(t)$
  \begin{equation*}\begin{split}
  &A_0(t) = \frac{n+c}{\sqrt{2n+1+2a}}A_s^*(t) + \frac{n+1+c}{\sqrt{2n+1+2a}}A_s(t), \\
  &A^*_0(t) = \frac{n+\overline{c}}{\sqrt{2n+1+2a}}A_s(t) + 
  \frac{n+1+\overline{c}}{\sqrt{2n+1+2a}}A_s^*(t),
  \end{split}\end{equation*} 
where $a$ is the real part of $c$. 
\end{lem}
\begin{proof}
From Defintion \ref{def Anc and A0} and $J_{nc} = J_0Q_{nc}$ it follows 
that for all $f \in H$
  \begin{equation*}\begin{split}
  A_0(f) &= \frac{1}{2}\big(B_{nc}(f)+iB_{nc}(J_0f)\big) = 
    \frac{1}{2}\big(B_{nc}(f)+iB_{nc}(J_{nc}Q_{nc}^{-1}f)\big) \\
    &=\frac{1}{2}\big(A^*_{nc}(f)+A_{nc}(f)-A^*_{nc}(Q_{nc}^{-1}f)+ A_{nc}(Q_{nc}^{-1}f)\big).
  \end{split}\end{equation*}
Using $J_{nc} = J_0Q_{nc}$ and
  \begin{equation*}
  Q_{nc}^{-1} = \begin{pmatrix}\frac{1 + 4b^2}{2n +1 +2a} & -2b \\ -2b & 2n +1 +2a\end{pmatrix},
  \end{equation*}
we find for the $\xi$-martingale $m_t$
  \begin{equation*}\begin{split}
  Q_{nc}^{-1}m_t &= 
  \begin{pmatrix}\frac{1 + 4b^2}{2n +1 +2a}\chi_{[0,t]} \\ -2b\chi_{[0,t]}\end{pmatrix} = 
  \frac{1}{2n+1+2a} \begin{pmatrix}\chi_{[0,t]}\\ 0\end{pmatrix} - 
  \frac{2b}{2n+1+2a} J_{nc}\begin{pmatrix}\chi_{[0,t]}\\ 0\end{pmatrix}   \\
  &=\Big(\frac{1}{2n+1+2a} - \frac{2b}{2n+1+2a} J_{nc}\Big)m_t.
  \end{split}\end{equation*}
From Definition \ref{def Anc and A0} and equation \eqref{eq properties fields}
we see that $A_{nc}(J_{nc}f) = - iA_{nc}(f)$ and 
$A^*_{nc}(J_{nc}f) = iA^*_{nc}(f)$ for all $f \in H$. Therefore 
it follows that
  \begin{equation*}\begin{split}
  &A_0(t) = 
  \frac{1}{2}\Bigg(A^*_{nc}(m_t)+A_{nc}(m_t)-
  \Big(\frac{1}{2n+1+2a} - \frac{2bi}{2n+1+2a}\Big)A^*_{nc}(m_t)\ + \\
  &\Big(\frac{1}{2n+1+2a} + \frac{2bi}{2n+1+2a}\Big)A_{nc}(m_t)\Bigg)
  = 
  \frac{n+c}{2n+1+2a}A_{nc}^*(m_t) + 
  \frac{n + 1 +c}{2n+1+2a}A_{nc}(m_t) 
  = \\
  &\frac{n+c}{\sqrt{2n+1+2a}}A_s^*(t) + 
  \frac{n + 1 +c}{\sqrt{2n+1+2a}}A_s(t). 
  \end{split}\end{equation*}
\end{proof}

Clearly, we now define for all stochastically integrable
processes $L_t$ stochastic integrals $L_tdA_0(t)$ and
$L_tdA_0^*(t)$ by $L_t\frac{n+c}{\sqrt{2n+1+2a}}dA_s^*(t) +
L_t\frac{n+1+c}{\sqrt{2n+1+2a}}dA_s(t)$ and  
$L_t\frac{n+\overline{c}}{\sqrt{2n+1+2a}}dA_s(t) +
L_t\frac{n+1+\overline{c}}{\sqrt{2n+1+2a}}dA_s^*(t)$, respectively.
Using the Hudson-Parthasarathy It\^o table it easily 
follows that the calculus of these stochastic integrals is 
given by the \emph{squeezed noise It\^o table} \cite{GaZ}, 
\cite{HHKKR}:
\begin{center}
{\large \begin{tabular} {l|lll}
$dM_1\backslash dM_2$ & $dA_0^*(t)$ & $dA_0(t)$ \\
\hline 
$dA_0^*(t)$ & $\overline{c}dt$ & $ndt$ \\
$dA_0(t)$ & $(n+1)dt$ & $cdt$ 
\end{tabular} }
\end{center}

We are now in a position to explain the construction of 
$\hat{T}^{nc}_t$ in the dilation diagram 
\eqref{diag dildiagsqueezed}. The free evolution of the
side channel is again given by the unitary group $S_t$, the 
second quantization of the left shift $s(t)$ on $H_{nc}$, i.e.
  \begin{equation*}
  s(t)\begin{pmatrix} f_1\\f_2 \end{pmatrix} = 
  \begin{pmatrix} f_1(\cdot + t)\\f_2(\cdot + t) \end{pmatrix}, \ \ \ \ \
  \begin{pmatrix} f_1\\f_2 \end{pmatrix} \in H_{nc}.
  \end{equation*}
In the Heisenberg picture the free evolution on $\WC_{nc}$ is then 
given by $X \mapsto S_t^*XS_t$. The system $\BC$ and field 
together form a closed system, thus their joint evolution is 
given by a one-parameter group $\hat{U}_t$ of unitaries, leading
to a Heisenberg picture evolution 
$\hat{T}^{nc}_t := \mbox{Ad}[\hat{U}_t]$ on $\BC\ten\WC^f\ten\WC^s_{nc}$. 
The group $\hat{U}_t$ is a perturbation of the free evolution. 
As in the vacuum case of section \ref{sec dilation}, 
we let this perturbation be given by the cocycle of 
unitaries $U_t := (S_{-t}\ten S_{-t})\hat{U}_t$. The stochastic differential 
equation \eqref{eq HuP} that was satisfied by the cocycle $U_t$  
when the side channel was still in the vacuum state is now changed.
The quantum noise of equation \eqref{eq quantumnoise} takes the
form
  \begin{equation*}
  d\beta_t = -i\big(V_fdA_f(t) - V_f^*dA_f(t) + 
  V_sdA^*_0(t) - V_s^*dA_0(t)\big), \ \ \ \ \ \ \beta_0 = 0,
  \end{equation*}  
in the squeezed noise representation. If the field is in the
vacuum state the operators $A_0(t)$ and $A_s(t)$ coincide. 
$A_0(t)$ should be interpreted as the annihilation
operator of a photon in the side channel and $A_s(t)$ should be 
interpreted as the annihilation operator of a squeezed 
excitation of in the side channel, i.e.\ a quasiparticle consisting 
out of many photons. Using Lemma \ref{lem lincomb} we find
  \begin{equation*}\begin{split}
  id\beta_t ={} &V_fdA^*_f(t) -  V_f^*dA_f(t)+
   \Big(\frac{n+1 +\overline{c}}{\sqrt{2n+1+2a}}V_s
    -\frac{n+c}{\sqrt{2n+1+2a}}V_s^*\Big)dA_s^*(t)\ - \\
    &\Big(\frac{n+1 +c}{\sqrt{2n+1+2a}}V_s^*
    -\frac{n+\overline{c}}{\sqrt{2n+1+2a}}V_s\Big)dA_s(t).
  \end{split}\end{equation*}
Define
  \begin{equation}\label{eq defVnc}
  V_{nc} :=
        \frac{n+1 +\overline{c}}{\sqrt{2n+1+2a}}V_s -\frac{n+c}{\sqrt{2n+1+2a}}V_s^*,
  \end{equation}
then the quantum stochastic differential 
equation for the cocycle $U_t$ is given by
  \begin{equation}\label{eq HuPsqueez}\begin{split}
  &dU_t = 
  \{V_fdA^*_f(t)- V_f^*dA_f(t) + V_{nc}dA_s^*(t) - 
  V_{nc}^*dA_s(t)-\frac{1}{2}(V_{nc}^*V_{nc}+ V_f^*V_f)dt\}U_t, \\ 
  &U_0 = \I.
  \end{split}\end{equation}
In a similar way as in section \ref{sec qsc} this leads to the 
Lindblad operator for the semigroup 
$T^{nc}_t = \exp(tL_{nc})$:
  \begin{equation*}
  L_{nc}(X) = V_f^*XV_f - \frac{1}{2}\{V_f^*V_f, X\} +
  V_{nc}^*XV_{nc} - \frac{1}{2}\{V_{nc}^*V_{nc}, X\}, \ \ \ \ \ X \in \BC.
  \end{equation*}

\section{Control with squeezing}\label{sec con+squeez}

Note that the operator $V_{nc}$ of equation \eqref{eq defVnc} 
for strongly squeezed fields, i.e.\ $n$ and $c$ are big, 
is very close to being skew-selfadjoint. Therefore for strongly 
squeezed fields the dilation is very close to being essentially 
commutative. In this section we exploit this idea and  
control the skew-selfadjoint part of $V_{nc}$.

Write again $V = V_R + iV_I$ with $V_R$ and $V_I$
the selfadjoint operators of equation \eqref{eq defVRI}.
We will again use for $X\in \{R,I\}$ and $\sigma\in\{f,s\}$
the notation $V_X^\sigma := \kappa_\sigma V_X$.
Furthermore we introduce: 
  \begin{equation*}
  W_R := \frac{V^s_R}{\sqrt{2n+1+2a}} \ \ \mbox{and} \ \ 
  W_I := \frac{V^s_I + i(n+c)V_s^* - i (n+\bar{c})V_s}{\sqrt{2n+1+2a}},
  \end{equation*}
i.e.\ $V_{nc} = W_R + iW_I$ with $W_R$ and $W_I$ selfadjoint.
Defining $Y^\sigma_R(t) := i\big(A_\sigma^*(t) - A_\sigma(t)\big)$ and 
$Y^\sigma_I(t) := A_\sigma^*(t) + A_\sigma(t),\ \sigma\in \{f,s\}$   
equation \eqref{eq HuPsqueez}, i.e.\ the laser is off, becomes
  \begin{equation*}\begin{split}
  &dU_t = \Big\{iV_I^fdY^f_I - i V^f_RdY^f_R + iW_IdY^s_I(t) -iW_RdY^s_R(t) -  
  \frac{1}{2}\big(V_f^*V_f + V_{nc}^*V_{nc})dt\Big\}U_t,\\
  &U_0 = \I.
  \end{split}\end{equation*}

Using a homodyne detection scheme we can observe the 
quadratures $X_\phi(t) := e^{-i\phi}A_0(t) +e^{i\phi}A^*_0(t)$
for $\phi \in [0,2\pi)$. With the help of Lemma \ref{lem lincomb} 
this can be written as
  \begin{equation*}
  X_\phi(t) = \frac{e^{-i\phi}(n+c)+e^{i\phi}(n+1+\overline{c})}{\sqrt{2n+1+2a}}A_s^*(t) + 
  \frac{e^{-i\phi}(n+1+c)+e^{i\phi}(n+\overline{c})}{\sqrt{2n+1+2a}}A_s(t).
  \end{equation*}
For simplicity we assume that $c$ is real, i.e.\ $c=a$. 
Note that the variance of $X_0$ has increased due to the squeezing,
while the variance of $X_{\frac{\pi}{2}}$ has decreased. 
Therefore we choose to observe 
$Y_t := X_0(t) = \sqrt{2n+1+2a}Y^s_I(t)$.
   
The Belavkin equation for observing $Y_t$ when the laser is 
still off, follows from equation \eqref{eq Belavkin} (cf.~\cite{WiM3}) 
  \begin{equation}\label{eq Belsqueez}
  d\rho_\bullet^t = L_{nc}(\rho_\bullet^t)dt + 
  \frac{i[W_I,\rho_\bullet^t] + \{W_R,\rho_\bullet^t\} - 2\mbox{Tr}
  \big(\rho_\bullet^tW_R\big)\rho_\bullet^t}{\sqrt{2n+1+2a}}
  \Big(dY_t - 
  2\mbox{Tr}\big(\rho_\bullet^tV^s_R\big)dt\Big).
  \end{equation}
Note that the observed process $Y_t$ is a   
drift, represented by the term $2\mbox{Tr}\big(\rho_\bullet^tV^s_R\big)dt$ plus 
an amplified Wiener process, i.e.\ amplified up to a variance of $(2n+1+2a)t$.
Through the drift term we gain information on the state 
of the two-level system. However, for strong squeezing, i.e.\
$n$ and $a$ big, this information gets lost in the noise
of the amplified Wiener process. In the limit for squeezing
to infinity, the measurement scheme is again non-informative,
just as in the essentially commutative case.
  
We run a control scheme as in section \ref{sec con-squeez} only
now based on the observation of $Y_t$. We correct with the
control unitary given by
  \begin{equation*}
  U^\tau_c = \exp\Big(-i \frac{\Delta(\tau)W_I}{\sqrt{2n+1+2a}}\Big),
  \end{equation*}
where $\Delta(\tau) := Y_\tau-Y_0$. Note that for $c$ real, 
i.e.\ $c=a$ we have
  \begin{equation*}
  W_I = \frac{i\kappa_s\sqrt{2n+1+2a}}{2} 
  \begin{pmatrix}0&1\\ -1&0\end{pmatrix}= \sqrt{2n+1+2a}V_I^s,
  \end{equation*}
i.e.\ we can realise this control unitary by applying 
a laser pulse determined by $h(t) = 
\frac{\kappa_s\Delta(\tau)}{2\kappa_f}\delta_\tau(t)$
for $0 \le t < 2\tau$. The control unitary satisfies
the following quantum stochastic differential equation
  \begin{equation*}\begin{split}
  &dU^\tau_c = \{-iV_I^sdY_\tau - 
    \frac{2n+1+2a}{2}{V_I^s}^2d\tau\}U^\tau_c = U^\tau_c\{-iV_I^sdY_\tau - 
    \frac{2n+1+2a}{2}{V^s_I}^2d\tau\}, \\ 
    &U^0_c =\I.
  \end{split}\end{equation*}
The state after control is again given by 
$\rho^\tau_\bullet := U^\tau_c\tilde{\rho}^\tau_\bullet {U^\tau_c}^*$ where 
$\tilde{\rho}^\tau_\bullet$ is given by the Belavkin 
equation \eqref{eq Belsqueez}. We use the notation below 
Theorem \ref{Itorules} with 
$Z_1 = U^\tau_c,\, Z_2 = \tilde{\rho}^\tau_\bullet$ and $Z_3 = U^{\tau*}_c$. 
For infinitesimal 
$\tau$ evaluated at $\tau =0$, this leads to equation 
\eqref{eq controlrho}, i.e.\
  \begin{equation*}
  d\rho^\tau_\bullet\Big|_{\tau=0} = 
  \big([1]+[2]+[3]+[12]+[13]+[23]+[123]\big)\Big|_{\tau=0}.
  \end{equation*}  
Again $[123] =0$ and further $([1]+[3]+[13])|_{\tau=0} = 
L_{W_I}(\rho^0)d\tau + i[\rho^0, V_I^s]dY_0$.
Furthermore  we have
  \begin{equation*}\begin{split}
  &[2]\Big|_{\tau=0} = L_{nc}(\rho^0)d\tau + 
  \Big(i[V^s_I,\rho^0] + \frac{\{V^s_R,\rho^0\} - 2\mbox{Tr}
  \big(\rho^0V^s_R\big)\rho^0}{2n+1+2a}\Big)
  \Big(dY_0 - 
  2\mbox{Tr}(\rho^0 V^s_R)d\tau\Big), \\
  &\big([12]+[23]\big)\Big|_{\tau = 0} = -2L_{W_I}(\rho^0)d\tau
 -i\big[W_I, \{W_R,\rho^0\}\big]d\tau +
 2\mbox{Tr}\big(\rho^0W_R\big)i[W_I,\rho^0]d\tau. 
  \end{split}\end{equation*}
A calculation shows that 
  \begin{equation*}
   L_{nc}(\rho^0)-L_{W_I}(\rho^0)
  -i\big[W_I, \{W_R,\rho^0\}\big]=  
  L_{V_f}(\rho^0) + L_{W_R}(\rho^0) 
  - \frac{i}{2}[W_IW_R + W_RW_I, \rho^0].
  \end{equation*}
Since $W_IW_R + W_RW_I =0$ for real $c$ and 
$\mbox{Tr}\big(\rho^0W_R\big)[W_I,\rho^0]= 
\mbox{Tr}(\rho^0 V^s_R)[V^s_I,\rho^0]$ , we get
  \begin{equation*}\begin{split}
  d\rho^\tau_\bullet\Big|_{\tau = 0} = L_{V_f}(\rho^0)d\tau +&
  L_{W_R}(\rho^0)d\tau\ + \\
  &\Big(\frac{\{V^s_R,\rho^0\} - 2\mbox{Tr}
  \big(\rho^0V^s_R\big)\rho^0}{2n+1+2a}\Big)
  \Big(dY_0 - 
  2\mbox{Tr}(\rho^0 V^s_R)d\tau\Big).
  \end{split}\end{equation*}
Since we repeat the control every $\tau$ time units with $\tau$ very small, i.e.\
we take $\tau$ infinitesimal, this leads to the following stochastic time evolution
for the density matrix of the two-level atom
  \begin{equation*}
  d\rho^t_\bullet = L_{V_f}(\rho^t_\bullet)dt + 
  \frac{L_{V^s_R}(\rho^t_\bullet)}{2n+1+2a}dt + 
  \Big(\frac{\{V^s_R,\rho^t_\bullet\} - 2\mbox{Tr}
  \big(\rho^t_\bullet V^s_R\big)\rho^t_\bullet}{2n+1+2a}\Big)
  \Big(dY_t - 
  2\mbox{Tr}(\rho^t_\bullet V^s_R)dt\Big).
  \end{equation*}
The first term on the right hand side is again due to the fact that we did not 
measure and correct the forward channel. It is harmless, since we can 
take $\kappa_f$ arbitrarily small. The other two terms converge to $0$ when 
squeezing goes to infinity. Therefore, in the limit, this control scheme 
restores quantum information.  

\newpage
\thispagestyle{plain}
\cleardoublepage

%% file: summary.tex
\pagestyle{plain}
\section*{Summary}
\addtocontents{toc}{\contentsline{chapter}{\numberline{}Summary}{\thepage}}

In this thesis the time evolution of open quantum systems
is studied. We focus on the time evolution of the 
state of the system conditioned on the result of 
some measurement performed continuously in time 
in its environment. The goal is to derive a stochastic 
differential equation for the conditional state in 
which the observed measurement process is a stochastic 
driving term. 

The derivation is carried out within the framework 
of \emph{non-commutative} or \emph{quantum} 
probability theory. In this generalised
probability theory there exists a non-commutative generalisation 
of It\^o's stochastic calculus. The interaction between 
an open system and the quantized electromagnetic field
in the weak coupling limit is governed by a 
stochastic differential equation in this non-commutative 
sense. The stochastic differential equation for the 
conditional state can be derived from it. It is interpreted 
as a non-commutative generalisation of the 
Kushner-Stratonovich filtering equation and is called 
the Belavkin filtering equation.    

In Chapter \ref{ch davies} we study a photon counting
experiment. The open system studied here, is a two-level atom driven 
by a laser. It emits fluorescence photons in its environment,
the electromagnetic field. The emitted photons are detected
continuously in time. We explicitly condition on the 
observed event by sanwiching with the corresponding 
projection in the field. Using the processes introduced 
by Davies \cite{Dav}, this leads to a continuous time 
evolution of the reduced system, i.e.\ the 
two-level atom, interrupted by jumps at the moments at 
which photons are detected. In quantum optics such 
trajectories of the conditioned state are known as  
quantum trajectories \cite{Car}. 

The third chapter focusses on the derivation of the infinitesimal description 
of the time evolution of the conditional state by the 
Belavkin filtering equation. Two separate derivations 
of the filtering equations are given. In the first approach 
we simply differentiate along the trajectories 
for the conditioned state that we obtained in 
Chapter \ref{ch davies}. A diffusive limit of the photon 
counting description enables us to incorporate the situation 
where instead of counting photons in the field we are 
performing a homodyne detection experiment.       
 
The second approach uses the decomposition of 
a von Neumann algebra over its center. The observed process
in the field determines a commutative algebra $\CC$ which is the 
center of its commutant $\AC := \CC'$. This commutant is 
the algebra of observables that are \emph{not demolished} by the observation 
in the field. The central decompostion enables us to decompose
the state restricted to $\AC$ over the possible paths the 
observed process yields. In this way we are able to condition the 
state on the observed path. 

The central decompostion is an existence result. It doesn't 
show how to construct the decompostion. However, using martingale 
techniques familiar from classical filtering theory, it is 
possible to derive a stochastic differential equation for the 
conditioned state which is called the \emph{Belavkin filtering equation}. 
Chapter \ref{ch sseq} concludes with a recipe for the 
derivation of the Belavkin equation for a wide class of 
possible continuous time observations in the field.  

In the fourth and last chapter we are not only interested 
in the time evolution of the state of an open system conditioned
on the result of a measurement in its environment, but 
we also want to use the measurement result for controlling
the state evolution. The objective is to keep the unknown 
state of a qubit fixed in time, i.e.\ to stop the decoherence.
 
We show that in a special case where the interaction with 
the field is such that the qubit only couples to commutative 
noise the above objective can be met. This special case 
is called \emph{essentially commutative}. However in reality
the qubit couples to two non-commutative noises. By putting 
the electromagnetic field in a squeezed state the variance 
of one of the noises increases while the variance of the 
other decreases. In this way we approach the essentially 
commutative case by stronger and stronger squeezing. We 
show that in the limit for the squeezing strength to 
infinity the unknown state of the qubit can again be 
kept fixed.

\cleardoublepage

\section*{Samenvatting}
\addtocontents{toc}{\contentsline{chapter}{\numberline{}Samenvatting}{\thepage}}

In dit proefschrift wordt de tijd evolutie van open
quantum systemen bestudeerd. We concentreren ons op 
de tijd evolutie van de toestand van het systeem 
geconditioneerd op de uitkomst van een meting 
continu in tijd in zijn omgeving. Het doel is de 
afleiding van een stochastische differentiaal vergelijking
voor de geconditioneerde toestand waarin het 
geobserveerde meetproces als een drijvende 
stochastische term voor komt.

De afleiding wordt uitgevoerd binnen het kader van 
\emph{niet-commutatieve} of \emph{quantum} kanstheorie. 
In deze veralgemeniseerde kanstheorie bestaat een 
niet-commutatieve generalisatie van It\^o's stochastische
calculus. De interactie tussen een open systeem en het
gequantiseerde electromagnetische veld in de zwakke 
koppelings limiet wordt beschreven door een stochastische
differentiaal vergelijking in deze niet-commutatieve zin. 
De stochastische differentiaal vergelijking voor de 
geconditioneerde toestand kan eruit worden afgeleid. 
Deze vergelijking wordt ge\"interpreteerd als een 
niet-commutatieve generalisatie van de Kushner-Stratonovich
filter vergelijking en heet de Belavkin filter vergelijking. 

In Hoofdstuk \ref{ch davies} bestuderen we een foton 
detectie experiment. Het bestudeerde open systeem is hier 
een twee niveau systeem gedreven door een laser. Het systeem 
zendt fluorescentie fotonen uit in zijn omgeving, het 
electromagnetische veld. De uitgezonden fotonen worden 
continu in de tijd gedetecteerd. We conditioneren expliciet
op een geobserveerde gebeurtenis door de corresponderende
projecties om de toestand heen te zetten. Gebruikmakend van 
de processen ge\"introduceerd door Davies \cite{Dav} kan 
de tijd evolutie van het twee niveau systeem continu 
in de tijd afgeleid worden. Dit leidt tot een continue 
evolutie tussen de momenten waarop fotonen worden 
gedetecteerd en sprongen op de momenten waarop 
fotonen worden gedetecteerd. In quantum optica staan 
deze trajectorie\"en van de geconditioneerde toestand 
bekend als quantum trajectorie\"en \cite{Car}. 

Het derde hoofdstuk richt zich op de infinitesimale 
beschrijving van de geconditioneerde toestand door
de Belavkin filter vergelijking. De filter 
vergelijkingen worden op twee verschillende manieren 
afgeleid. De eerste methode bestaat uit het eenvoudigweg
differentieren langs een trajectorie van de geconditioneerde
toestand. Een diffusieve limiet van de foton detectie 
beschrijving maakt het mogelijk om ook homodyne detectie 
experimenten in onze beschrijving op te nemen. 

De tweede methode maakt gebruik van de ontbinding van een 
von Neumann algebra over zijn centrum. Het geobserveerde
proces genereert een commutatieve algebra $\CC$ die 
het centrum is van zijn commutant $\AC := \CC'$. Deze 
commutant is de algebra van observabelen die \emph{niet 
gedemoleerd} zijn door de observatie in het veld. De 
centrale ontbinding maakt het mogelijk om de toestand 
beperkt tot $\AC$ te onbinden over de mogelijke 
paden van het geobserveerde proces. Op deze manier 
kunnen we de toestand conditioneren op het 
geobserveerde pad. 
  
De centrale ontbinding is een existentie resultaat. Het
laat niet zien hoe de ontbinding geconstrueerd kan worden. 
Gebruikmakend van martingaal technieken, bekend uit 
klassieke filter theorie, is het mogelijk om een 
stochastische differentiaal vergelijking voor de 
geconditioneerde toestand af te leiden, de Belavkin 
filter vergelijking. Hoofdstuk 3 sluit af met een 
recept voor de afleiding van de Belavkin vergelijking 
voor een grote klasse van mogelijke continue tijd 
observaties in het veld. 

In het vierde en laatste hoofdstuk zijn we niet alleen 
ge\"interesseerd in de tijd evolutie van de geconditioneerde
toestand van een open systeem, maar we willen het 
resultaat van de continue tijd observatie ook gebruiken 
om de toestand van het systeem te controleren. Het 
doel is om een onbekende toestand van een qubit 
vast te houden in de tijd, dat wil zeggen de decoherentie
tegen gaan. 

We vinden dat dit mogelijk is in het geval dat de 
interactie met het veld zodanig is dat de qubit 
slechts koppelt met commutatieve ruis. Dit speciale 
geval heet \emph{essentieel commutatief}. In werkelijkheid 
koppelt de qubit echter met twee onderling niet-commutatieve
ruizen. Door het electromagnetische veld in een samengedrukte
of gesqueezde toestand te nemen, kunnen we de variantie van 
een van de twee ruizen verkleinen terwijl de variantie van de 
andere wordt vergroot. Op deze manier kunnen we door sterker en sterker 
squeezen het essentieel commutatieve geval beter en 
beter benaderen. We laten zien dat in de limiet 
voor de squeezing naar oneindig de onbekende toestand 
van de qubit weer vast kan worden gehouden.

\cleardoublepage
\newpage

\section*{Curriculum Vitae}
\addtocontents{toc}{\contentsline{chapter}{\numberline{}Curriculum Vitae}{\thepage}}

Op 11 oktober 1976 werd ik, Luc Bouten, als tweede in het 
gezin van Piet Bouten en Mia Bouten-van Lier geboren te Helden. 
Ik doorliep de lagere en middelbare school te Panningen.
Op de middelbare school, het Bouwens van der Boije college, 
werd mijn interesse voor wiskunde en natuurkunde gewekt. In 
1995 begon ik aan de studie natuurkunde aan de universiteit
te Nijmegen. Na het eerste studiejaar werd het zwaartepunt 
verlegd naar de studie wiskunde die ik vervolgens begin 2000 
voltooide. In 2002 volgde de voltooiing van de studie 
natuurkunde. Tot mijn leraren behoorden Hans Maassen, 
Gert Heckman, Arnoud van Rooij, Ted Janssen, Thomas Rijken 
en Hubert Knops. Tijdens de studie was ik actief in 
een aantal universitaire raden en voorzag ik gedeeltelijk in mijn 
levensonderhoud door middel van het geven van werkcolleges.

Tegen het einde van mijn studie periode was ik voornamelijk
geinteresseerd geraakt in quantum mechanica. In het bijzonder de 
mathematische zijde van de theorie, te weten spectraal theorie,
kanstheorie en groepentheorie. Dit leidde in april 2000 tot de keuze 
promotie onderzoek te doen in het gebied van de quantum kanstheorie 
onder supervisie van Hans Maassen. Het promotie onderzoek bestond 
uit het toepassen van niet-commutatieve stochastische 
analyse op problemen afkomstig uit de quantum optica. Bijzondere 
aandacht ging hierbij uit naar Belavkin's quantum filter 
theorie. Kennis over quantum optica werd opgedaan tijdens 
een bezoek van drie maanden aan Howard Carmichael in Oregon VS.
Aansluitend aan de voltooiing van het promotie onderzoek in 
april 2004 ben ik een aantal maanden post-doc geweest bij 
Slava Belavkin in Nottingham UK.

\newpage